%% file: main.tex
\documentclass[preprint,twocolumn]{aastex63}

\usepackage{float, bm, graphicx, amsmath, morefloats}
\usepackage[caption=false]{subfig}
\shorttitle{Low Luminosity Type IIP Supernovae}
\shortauthors{Das et al.}

\usepackage{float,graphicx,amsmath,tabularx,booktabs,natbib,threeparttable}
\usepackage{float, bm, graphicx, amsmath, morefloats}
\usepackage[caption=false]{subfig}
\usepackage{float,graphicx,amsmath,tabularx,booktabs,natbib,threeparttable}
\usepackage{subfig}

\newcommand{\lpipe}{\textsc{lpipe}}

\usepackage{comment}
\definecolor{dark-red}{rgb}{0.4,0.15,0.15}
\definecolor{dark-blue}{rgb}{0.15,0.15,0.4}
\definecolor{medium-blue}{rgb}{0,0,0.5}
\hypersetup{
    colorlinks, linkcolor={dark-red},
    citecolor={dark-blue}, urlcolor={medium-blue}
}

\newcommand{\beqa}{\begin{eqnarray}} 
\newcommand{\eeqa}{\end{eqnarray}}

\newcommand{\bsub}{\begin{subequations}}
\newcommand{\esub}{\end{subequations}}
\newcommand{\beal}{\begin{align}}
\newcommand{\ealn}{\end{align}}

\newcommand{\msun}{M$_{\sun}$}

\newcommand{\Msun}{{\ensuremath{\mathrm{M}_{\odot}}}}

\graphicspath{{./}{figures/}}

\begin{document}

\title{Low-Luminosity Type IIP Supernovae from the Zwicky Transient Facility Census of the Local Universe. I: Luminosity Function, Volumetric Rate}

\author[0000-0001-8372-997X]{Kaustav~K.~Das}\thanks{E-mail: kdas@astro.caltech.edu}\affiliation{Cahill Center for Astrophysics, California Institute of Technology, MC 249-17, 
1200 E California Boulevard, Pasadena, CA, 91125, USA}

\author[0000-0002-5619-4938]{Mansi~M.~Kasliwal}
\affiliation{Cahill Center for Astrophysics, 
California Institute of Technology, MC 249-17, 
1200 E California Boulevard, Pasadena, CA, 91125, USA}

\author[0000-0002-4223-103X]{Christoffer Fremling}
\affil{Caltech Optical Observatories, California Institute of Technology, Pasadena, CA 91125, USA}

\author[0000-0003-1546-6615]{Jesper Sollerman}
\affiliation{The Oskar Klein Centre, Department of Astronomy, Stockholm University, AlbaNova, SE-10691 Stockholm, Sweden}

\author[0000-0001-8472-1996]{Daniel A.~Perley}
\affiliation{Astrophysics Research Institute, Liverpool John Moores University, IC2,  Liverpool L3 5RF, UK}

\author[0000-0002-8989-0542]{Kishalay De}
\affiliation{MIT-Kavli Institute for Astrophysics and Space Research
77 Massachusetts Ave. Cambridge, MA 02139, USA}

\author[0000-0003-0484-3331]{Anastasios Tzanidakis}
\affiliation{Department of Astronomy and the DiRAC Institute, University of Washington, 3910 15th Avenue NE, Seattle, WA 98195, USA}

\author{Tawny Sit}
\affiliation{Department of Astronomy, The Ohio State University, Columbus, OH 43210, USA}

\author[0000-0001-5855-5939]{Scott Adams}
\affiliation{Cahill Center for Astrophysics, 
California Institute of Technology, MC 249-17, 
1200 E California Boulevard, Pasadena, CA, 91125, USA}

\author[0000-0003-3768-7515]{Shreya Anand}
\affiliation{Kavli Institute for Particle Astrophysics and Cosmology, Stanford University, Stanford, CA 94305-4085, USA}

\author{Tomas Ahumuda}
\affiliation{Cahill Center for Astrophysics, 
California Institute of Technology, MC 249-17, 
1200 E California Boulevard, Pasadena, CA, 91125, USA}

\author{Igor Andreoni}
\affiliation{Department of Physics and Astronomy, University of North Carolina at Chapel Hill, Chapel Hill, NC 27599-3255, USA}

\author{Seán Brennan}
\affiliation{The Oskar Klein Centre, Department of Astronomy, Stockholm University, AlbaNova, SE-10691 Stockholm, Sweden}

\author{Thomas Brink}
\affiliation{Department of Astronomy, University of California, Berkeley, CA 94720-3411, USA}

\author[0000-0002-0786-7307]{Rachel J. Bruch}
\affil{Department of Particle Physics and Astrophysics, Weizmann Institute of Science, 234 Herzl St, 76100 Rehovot, Israel}

\author{Ping Chen}
\affiliation{Department of Particle Physics and Astrophysics, Weizmann Institute of Science, 234 Herzl St, 76100 Rehovot, Israel}

\author{Matthew R. Chu}
\affiliation{Department of Astronomy, University of California, Berkeley, CA 94720-3411, USA}

\author[0000-0002-6877-7655]{David O. Cook}
\affiliation{IPAC, California Institute of Technology, 1200 E. California Blvd, Pasadena, CA 91125, USA}

\author{Sofia Covarrubias}
\affiliation{Cahill Center for Astrophysics, 
California Institute of Technology, MC 249-17, 
1200 E California Boulevard, Pasadena, CA, 91125, USA}

\author{Aishwarya Dahiwale}
\affiliation{Cahill Center for Astrophysics, 
California Institute of Technology, MC 249-17, 
1200 E California Boulevard, Pasadena, CA, 91125, USA}

\author{Nicholas Earley}
\affiliation{Cahill Center for Astrophysics, 
California Institute of Technology, MC 249-17, 
1200 E California Boulevard, Pasadena, CA, 91125, USA}

\author[0000-0002-9017-3567]{Anna Y. Q.~Ho}
\affiliation{Department of Astronomy, Cornell University, Ithaca, NY 14853, USA}

\author{Avishay Gal-Yam}
\affiliation{Department of Particle Physics and Astrophysics, Weizmann Institute of Science, 234 Herzl St, 76100 Rehovot, Israel}

\author{Anjasha Gangopadhyay}
\affiliation{The Oskar Klein Centre, Department of Astronomy, Stockholm University, AlbaNova, SE-10691 Stockholm, Sweden}

\author{Erica Hammerstein}
\affiliation{Department of Astronomy, University of Maryland, College Park, MD 20742, USA}

\author{K-Ryan Hinds}
\affiliation{Astrophysics Research Institute, Liverpool John Moores University, IC2,  Liverpool L3 5RF, UK}

\author{Viraj Karambelkar}
\affiliation{Cahill Center for Astrophysics, 
California Institute of Technology, MC 249-17, 
1200 E California Boulevard, Pasadena, CA, 91125, USA}

\author{Yihan Kong}
\affiliation{Institute for Astronomy, University of Hawaii at Manoa, Honolulu}

\author[0000-0001-5390-8563]{S.~R.~Kulkarni}
\affiliation{Cahill Center for Astrophysics, 
California Institute of Technology, MC 249-17, 
1200 E California Boulevard, Pasadena, CA, 91125, USA}

\author[0009-0003-6181-4526]{Theophile Jegou du Laz}
\affiliation{Cahill Center for Astrophysics, 
California Institute of Technology, MC 249-17, 
1200 E California Boulevard, Pasadena, CA, 91125, USA}

\author[0000-0002-7866-4531]{Chang Liu}
\affil{Department of Physics and Astronomy, Northwestern University, 2145 Sheridan Rd, Evanston, IL 60208, USA}
\affil{Center for Interdisciplinary Exploration and Research in Astrophysics (CIERA), 1800 Sherman Ave., Evanston, IL 60201, USA}

\author{William Meynardie}
\affiliation{Cahill Center for Astrophysics, 
California Institute of Technology, MC 249-17, 
1200 E California Boulevard, Pasadena, CA, 91125, USA}

\author[0000-0001-9515-478X]{Adam A.~Miller}
\affil{Department of Physics and Astronomy, Northwestern University, 2145 Sheridan Rd, Evanston, IL 60208, USA}
\affil{Center for Interdisciplinary Exploration and Research in Astrophysics (CIERA), 1800 Sherman Ave., Evanston, IL 60201, USA}
\affil{NSF-Simons AI Institute for the Sky (SkAI), 172 E. Chestnut St., Chicago, IL 60611, USA}

\author[0000-0002-7501-5579]{Guy Nir}
\affiliation{Department of Astronomy, University of California, Berkeley, CA 94720-3411, USA}

\author{Kishore C. Patra}
\affiliation{Department of Astronomy, University of California, Berkeley, CA 94720-3411, USA}

\author[0000-0002-8041-8559]{Priscila J. Pessi}
\affiliation{The Oskar Klein Centre, Department of Astronomy, Stockholm University, AlbaNova, SE-10691 Stockholm, Sweden}

\author{R. Michael Rich}
\affiliation{Department of Physics \& Astronomy, Univ. of California Los Angeles, 430 Portola Plaza, Los Angeles, CA 90095-1547, USA}

\author[0000-0002-5683-2389]{Nabeel~Rehemtulla}
\affiliation{Department of Physics and Astronomy, Northwestern University, 2145 Sheridan Rd, Evanston, IL 60208, USA}
\affiliation{Center for Interdisciplinary Exploration and Research in Astrophysics (CIERA), 1800 Sherman Ave., Evanston, IL 60201, USA}
\affiliation{NSF-Simons AI Institute for the Sky (SkAI), 172 E. Chestnut St., Chicago, IL 60611, USA}

\author[0000-0003-4725-4481]{Sam Rose}
\affiliation{Cahill Center for Astrophysics, 
California Institute of Technology, MC 249-17, 
1200 E California Boulevard, Pasadena, CA, 91125, USA}

\author[0000-0001-7648-4142]{Ben Rusholme}
\affiliation{IPAC, California Institute of Technology, 1200 E. California
             Blvd, Pasadena, CA 91125, USA}

\author[0000-0001-6797-1889]{Steve Schulze}
\affiliation{Center for Interdisciplinary Exploration and Research in Astrophysics (CIERA), 1800 Sherman Ave., Evanston, IL 60201, USA}

\author{Yashvi Sharma}
\affiliation{Cahill Center for Astrophysics, 
California Institute of Technology, MC 249-17, 
1200 E California Boulevard, Pasadena, CA, 91125, USA}

\author[0000-0003-2091-622X]{Avinash Singh}
\affiliation{The Oskar Klein Centre, Department of Astronomy, Stockholm University, AlbaNova, SE-10691 Stockholm, Sweden}

\author[0000-0001-7062-9726]{Roger Smith}
\affiliation{Caltech Optical Observatories, California Institute of Technology, Pasadena, CA  91125, USA}

\author{Robert Stein}
\affiliation{Cahill Center for Astrophysics, 
California Institute of Technology, MC 249-17, 
1200 E California Boulevard, Pasadena, CA, 91125, USA}

\author{Milan Sharma Mandigo-Stoba}
\affiliation{Department of Physics \& Astronomy, Univ. of California Los Angeles, 430 Portola Plaza, Los Angeles, CA 90095-1547, USA}

\author{Nora L. Strotjohann}
\affil{Department of Particle Physics and Astrophysics, Weizmann Institute of Science, 234 Herzl St,
76100 Rehovot, Israel}

\author[0000-0003-3658-6026]{Yu-Jing Qin}
\affiliation{Cahill Center for Astrophysics, 
California Institute of Technology, MC 249-17, 
1200 E California Boulevard, Pasadena, CA, 91125, USA}

\author{Jacob Wise}
\affiliation{Astrophysics Research Institute, Liverpool John Moores University, IC2,  Liverpool L3 5RF, UK}

\author[0000-0002-9998-6732]{Avery Wold}
\affiliation{IPAC, California Institute of Technology, 1200 E. California Blvd, Pasadena, CA 91125, USA}

\author[0000-0003-1710-9339]{Lin~Yan}
\affil{Caltech Optical Observatories, California Institute of Technology, Pasadena, CA 91125, USA}

\author{Yi Yang}
\affiliation{Department of Physics, Tsinghua University, Beijing, 100084, China}

\author[0000-0001-6747-8509]{Yuhan Yao}
\affiliation{Department of Astronomy, University of California, Berkeley, CA 94720-3411, USA}

\author[0000-0001-8985-2493]{Erez Zimmerman}
\affil{Department of Particle Physics and Astrophysics, Weizmann Institute of Science, 234 Herzl St,
76100 Rehovot, Israel}

\begin{abstract}
We present the luminosity function and volumetric rate of a sample of Type IIP supernovae (SNe) from the Zwicky Transient Facility Census of the Local Universe survey (CLU). This is the largest sample of Type IIP SNe from a systematic volume-limited survey to-date. The final sample includes 330 Type IIP SNe and 36 low-luminosity Type II (LLIIP) SNe with $M_{\textrm{r,peak}}>-16$ mag, which triples the literature sample of LLIIP SNe. The fraction of LLIIP SNe is $19^{+3}_{-4}\%$ of the total CLU Type IIP SNe population ($8^{+1}_{-2}\%$ of all core-collapse SNe).  This implies that while LLIIP SNe likely represent the fate of core-collapse SNe of $8-12$ \Msun\ progenitors, they alone cannot account for the fate of all massive stars in this mass range. To derive an absolute rate, we estimate the ZTF pipeline efficiency as a function of the apparent magnitude and the local surface brightness. We derive a volumetric rate of $(3.9_{-0.4}^{+0.4}) \times 10^{4}\ \textrm{Gpc}^{-3}\ \textrm{yr}^{-1}$ for Type IIP SNe and  $(7.3_{-0.6}^{+0.6})  \times 10^{3}\ \textrm{Gpc}^{-3}\ \textrm{yr}^{-1}$ for LLIIP SNe. Now that the rate of LLIIP SNe is robustly derived, the unresolved discrepancy between core-collapse SN rates and star-formation rates cannot be explained by LLIIP SNe alone. \\\\

\end{abstract}

\keywords{}

\newpage

\section{Introduction} \label{sec:intro}
%\textcolor{black}{re-structure intro after deciding scope of paper.}

%\textcolor{black}{Current goal: Main focus on LLIIP SNe. To quadruple the no. of published LLIIP SNe, understand their rates, and extend II luminosity function to the fainter end. For a fair comparison, understand the properties in the context of the entire CLU Type II population. Lightcurve modeling and nebular spectroscopy in Part II and Part III}

Core-collapse supernovae (CCSNe), the explosive endpoints of massive stars, drive chemical evolution in galaxies, provide impetus for the formation of new generations of stars, and leave behind neutron star or black hole remnants. Despite the discovery of over 10,000 SNe, the fate of stars at the low-mass end of CCSNe, with progenitor initial masses of $\approx 8$--12 \Msun, is not well understood \citep[e.g., see][]{Janka2012,Sukhbold2016}. Exploring this mass range is important, because it probes the boundary between whether a star has a large enough mass to form a neutron star from its central core or, instead, leaves behind a white dwarf. Also, it is uncertain whether stars in this mass range undergo iron core-collapse from Red Supergiants (RSG) or if electron capture in super-Asymptotic Giant Branch (sAGB) stars drives the SN explosions \citep{Nomoto1984, Jerkstrand2018, Leung2020, Kozyreva2021}. %sAGB stars are both the rarest of the low/intermediate mass stars, and also the most common of the stars on the more massive side of the boundary. Hence, if they do indeed produce ECSNe, then they may make a signiﬁcant contribution to the overall SN rate.

% nice theory review: https://arxiv.org/pdf/2005.02420

Stars in the $8–12$\ \Msun\ range exhibit a distinctly different core structure compared to those with higher progenitor mass, characterized by significantly lower compactness \citep{Sukhbold2016}. The evolution of such stars is more complex to model than at $>12$ \Msun\ due to the development of thermal pulses that are numerically challenging to follow, and highly uncertain late-time mass-loss \citep[e.g.,][]{Miyaji1980, Woosley2015}. The lowest mass CCSNe have steep density gradients outside their degenerate cores, liberate less gravitational binding energy, have lower neutrino luminosities, and ultimately are associated with lower luminosities, explosion energy and nickel yield \citep{Janka2012, Muller2016, Burrows2019, Eldridge2019, Stockinger2020, Burrows2021, Sandoval2021, Barker2022, Burrows2024}.

It is critical to understand this faint end of the core-collapse luminosity function, constrain their rates and couple the stellar evolution initial mass function (IMF) to the known SN populations. Determining the SN rate and tying it to the mass boundary between stars that do and do not explode as SNe is vital for determining the number of neutron stars, galactic chemical evolution and dust evolution models, and the total energy released by SNe into the environment. The CCSN distribution traces the formation of massive stars and the CCSN rate should match the massive star formation rate using the known star-formation rate density and IMF, given our current understanding of stellar evolution. However, there is an apparent discrepancy between the two absolute rates with the star-formation implying a much higher CCSN rate than is observed. This ongoing debate has been discussed extensively in the literature \citep[e.g.,][]{Horiuchi2011, Botticella2012, Horiuchi2013, Cappellaro2015, Xiao2015, Jencson2019}. One possibility is that this disagreement arises from a missing population of low-luminosity CCSNe.

One such class of faint CCSNe is low-luminosity Type IIP SNe (LLIIP SNe), which likely represent the explosions of $8-12$~\Msun\ progenitor stars. \textcolor{black}{LLIIP SNe exhibit a faint peak luminosity,  defined here as $M_{\textrm{r,peak}} \ge -16$ mag, motivated by prior literature \citep[e.g.,][]{Spiro2014}. Recent studies, based on pre-explosion images of SNe 2005cs, 2008bk, 2018aoq, 2022acko \citep{Maund2005, Li2006, Mattila2008, ONeill2019, VanDyk2023}
also show that SNe with $r$-band peak $>-16$ mag have progenitor mass estimates less than 11 \Msun, while those with brighter peak have progenitor masses $>$12 \msun (see Figure \ref{fig:progenitors} in Appendix \ref{sec:progenitors})}. The nickel mass ($\sim0.005$~\msun) of LLIIP SNe is an order of magnitude lower than that typically found in standard Type II SNe \citep[e.g.,][]{Rodriguez2021}. Additionally, LLIIP SNe are characterized by slow expansion velocities (approximately 1300 $-$ 2500 km~s$^{-1}$ at 50 days after the explosion), indicating low explosion energies. The low energy and low nickel mass could be explained by the explosion of $8-12$ \msun\ progenitors. This is also favored by lightcurve simulations \citep{Pumo2017, Fraser2011}, evolutionary numerical simulations of low mass RSGs \citep{Lisakov2017},  and nebular spectroscopy  \citep{Jerkstrand2018}. A correlation is also observed between the explosion energy, nickel mass and
progenitor mass estimates from pre-SN imaging \citep{Eldridge2019}, where lower-mass progenitors ($<12$ \Msun) are associated with lower nickel mass ($<0.01$~\msun)
and lower explosion energy ($<2\times 10^{50}$ erg s$^{-1}$).

However, there are only about a dozen such objects presented in the literature \citep[][]{Turatto1998, Spiro2014, Jager2020, Muller2020, Reguitti2021, Sheng2021, Valerin2022, Kozyreva2022, Bostroem2023, Teja2024, Dastidar2025}. For a Salpeter IMF \citep{Salpeter1955}, around 50\% of the potential CCSN progenitors reside in the $8-12$ \Msun\ mass range. Given this fact, it is surprising that such SNe are rarely detected. What do half of the stars that undergo core-collapse explode as? It is likely that this deficit of discovery can be explained by the connection between these relatively low-mass progenitors and the occurrence of low-luminosity SNe, which are more difficult to detect and classify.

In this series of three papers, we present the analysis of a sample of 36 LLIIP SNe detected by Zwicky Transient Facility (ZTF) as part of the Census of the Local Universe (CLU) Survey, which roughly triples the number of existing LLIIP SNe in the literature. The focus of the first paper is to estimate their luminosity function and volumetric rates.

To place these results in a broader context, we also present a sample of 330 Type IIP SNe, obtained from the largest systematic volume-limited SN survey to date. The previous Type II SNe sample from a volume-limited sample survey includes 62 Type II SNe from the Lick Observatory Supernova Search \citep[LOSS;][]{Li11}. Other Type II samples from systematics surveys include 50 Type II SNe from the Panoramic Survey Telescope and Rapid Response System \citep[PS1;][]{Sanders2015}, 34 Type IIP SNe from the Sloan Digital Sky Survey II Supernova Survey \citep[SDSS-II SNS;][]{Taylor2014} and 21 Type II SNe from the Caltech Core-collapse SN Project \citep[CCCP;][]{Arcavi2012}.

In Section \ref{sec:sample}, we define the sample selection criteria. The data obtained and the extinction correction method are described in Sections \ref{sec:data} and \ref{sec:extinction}, respectively. In Sections \ref{sec:LF} and \ref{sec:rates}, we estimate the luminosity function and the volumetric rate of LLIIP SNe and compare with the overall CLU Type IIP SN population. We discuss the implications of the measured LLIIP rates in Section \ref{sec:discussion} and conclude in Section \ref{sec:conclusion}. In companion Papers II and III, we will present the analysis of their lightcurves and nebular spectra, respectively.

\section{Sample Selection}

\subsection{Census of the Local Universe}\label{sec:ZTF CLU}

ZTF conducts an optical time-domain survey with the 48-inch Schmidt telescope (P48) situated at Palomar Observatory \citep{Bellm2019, Graham2019, Dekany20}. The ZTF Census of the Local Universe (CLU) experiment aims to build a comprehensive spectroscopic sample of transients in the local Universe. To accomplish this, we match the hosts of these transients with the galaxy catalog from the Census of the Local Universe \citep{Cook2019}. This catalog contains around 234,500 galaxies with established distances, compiled from pre-existing spectroscopic galaxy surveys and the CLU-H$\alpha$ survey (refer to \citealt{Cook2019} for further details).

%\textcolor{black}{since using all and no abs mag cut, not have phase I, II but rather a uniform survey less than 150 Mpc?}

The CLU experiment is designed as a volume-limited SN survey where sources at less than 150 Mpc ($z\leq0.033$) and spatially coincident with (within 30 kpc) or visibly associated with a galaxy in the CLU catalog were assigned for spectroscopic follow-up for classification. For more detailed discussion of the CLU experiment filtering, see \citet{De2020}. The filter is implemented on the Global Relay of Observatories Watching Transients Happen (\texttt{GROWTH}) Marshal \citep{Kasliwal2019} and the \texttt{Fritz} Marshal \citep{van2019, Coughlin2023}, both of which are web portals designed for vetting and coordinating transient follow-up observations.

After passing through the CLU filter, the transients are allocated for spectroscopic classification using: the Spectral Energy Distribution Machine \citep[SEDM;][]{Blagorodnova2018, Kim2022} at the Palomar 60-inch telescope, the Double-Beam Spectrograph \citep[DBSP;][]{Oke1982} at the Palomar 200-inch Hale telescope, and the Low-Resolution Imaging Spectrometer \citep[LRIS;][]{Oke1995}. In this paper, we consider SNe saved to the CLU experiment starting from 1 October 2018 until 1 April 2024. We exclude the targets saved to ZTF in 2020 as the CLU experiment was inactive during that year. A total of 1745 SNe saved to CLU were classified during this interval.

\begin{figure*}
    \centering
    \includegraphics[width=9cm]{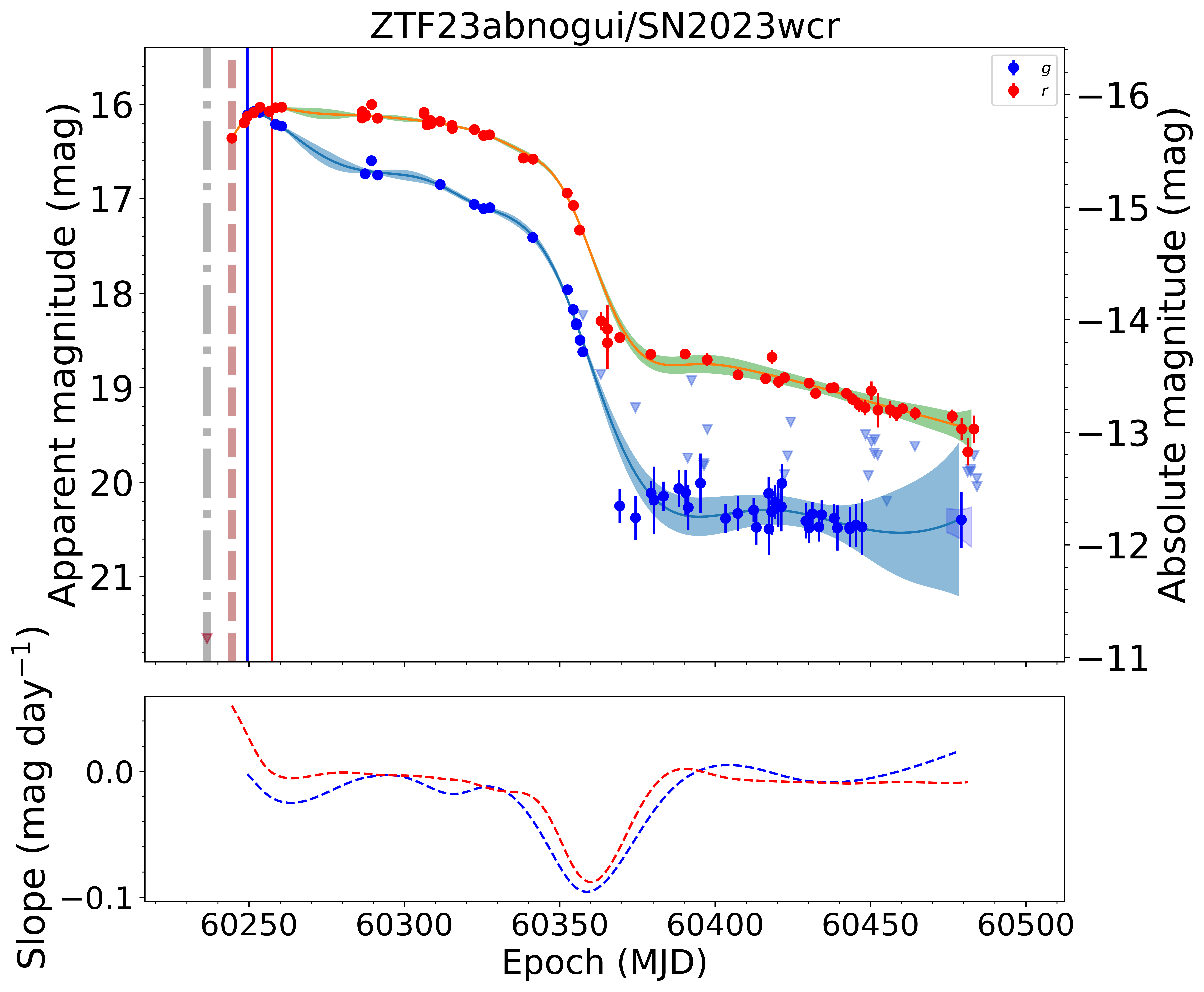}\includegraphics[width=9cm]{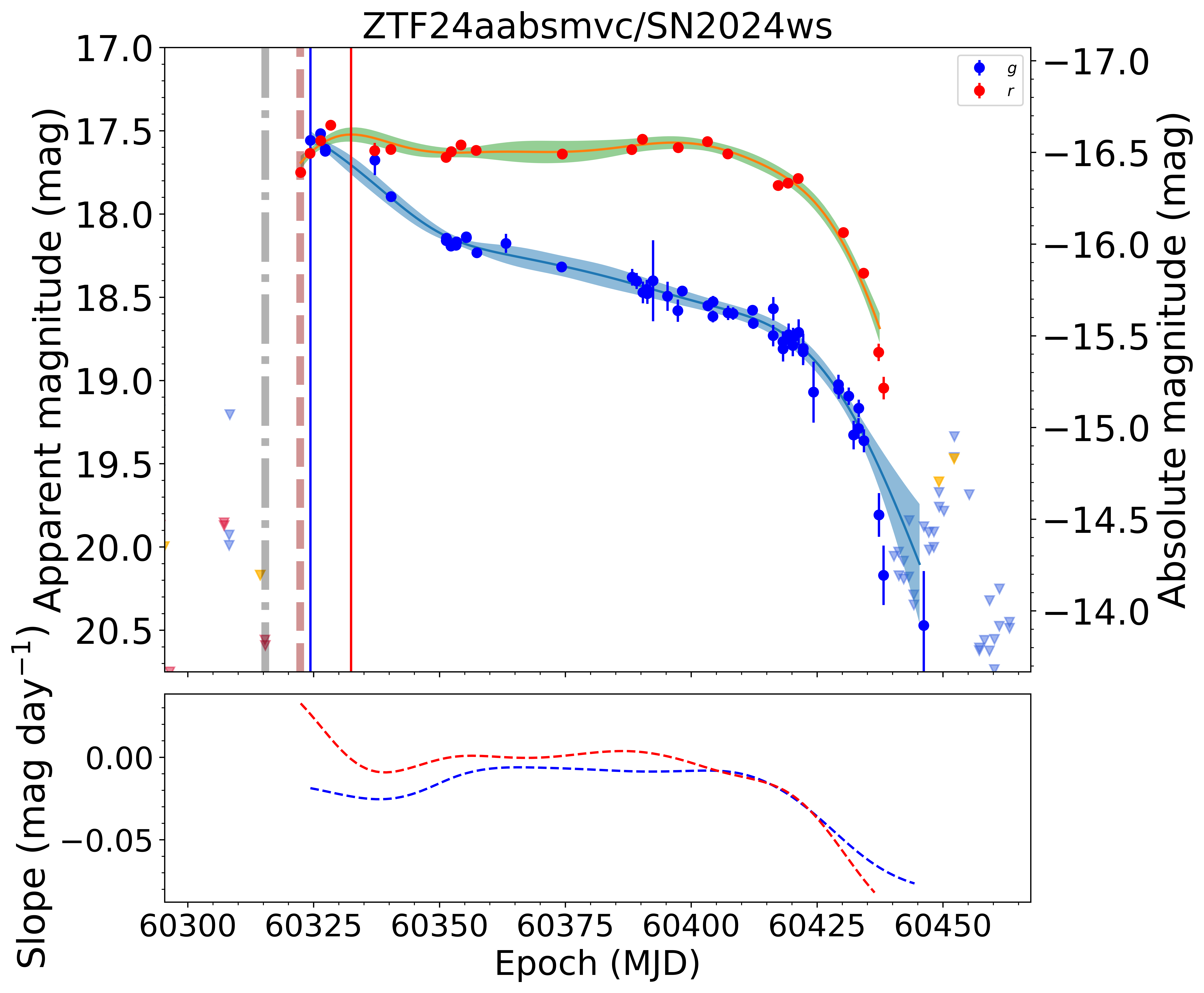}
    \caption{Example of SN light curves and the GP interpolations. The apparent magnitudes have been corrected for Galactic and host extinction. The lower panels show the slope of the light curves, i.e., the derivative of the GP fits, in units of mag per day. The vertical dashed red line shows the epoch of the first detection. The vertical dashed grey line shows the epoch of last non-detection. The vertical solid red and blue lines show the epochs of peak $r$- and $g$-band magnitudes, respectively.}
    \label{fig:GP}
\end{figure*}

\subsection{Sample of LLIIP SNe}\label{sec:sample}

%\textcolor{black}{updated numbers in Table 1; finalize numbers}
We apply the following selection criteria on the ZTF SN sample obtained from the CLU survey:

\begin{enumerate}
    %\item SNe in CLU: The initial sample included 1,665 supernovae identified as part of the CLU experiment from 2018 October
     %1 to 2024 April 1. CLU classification efforts were not operational in 2020, so we exclude the targets saved to ZTF in 2020. 
    
    \item Type II Classification: From the CLU sample, 727 candidates classified as Type II (including subtypes IIP, II, IIL,  II-norm, and II-pec) were selected.
    
    \item Peak Magnitude: A Gaussian process fit was applied to the forced photometry light curves, and the sample was restricted to those with a peak magnitude \( m_{peak_r} < 20 \) mag, resulting in 719 candidates. The peak magnitude and other lightcurve parameters are measured on this GP fit (see Section \ref{section:photdata} for details.). The 20 mag cut is chosen as the pipeline recovery efficiency and classification completeness drops below 80\% for alerts fainter than 20 mag (see Section \ref{sec:LF}).
    
    %\item Number of detections: 
    \item In order to constrain the $r$-band peak, we first require that the candidates have more than 10 detections, reducing the sample to 626. We also require that the candidates pass any one of the following conditions: 
    \begin{enumerate}
        \item Candidates with at least one detection before the peak and one after the peak, and for which the duration between the last non-detection and the first detection (\( t_{first det} - t_{last non-det} \)) was \(\leq 20\) days.
        \item If \( t_{first det} -  t_{last non-det}  \geq 20 \) days, then candidates with more than 10 pre-peak detections and more than 2 post-peak detections, as well as a magnitude difference \( mag_{first det} - mag_{peak} > 1 \) mag, were considered.
        \item For candidates where the first detection corresponds to the peak, the last non-detection had to occur within 10 days of the first detection.
    \end{enumerate}
    394 candidates pass this quality cut.
    \item Plateau Criterion: Candidates were required to show a plateau phase, defined as a duration of $\ge$ 40 days with a drop of less than 1 magnitude, resulting in a sample of 330 candidates.

    %selfnote: ZTF20abeohfn not pass if only ZTF.
    %plateau pass but is it? : ZTF21abcacpa
    %should be removed: no plateau r band: ZTF18abjkryl, ZTF22abrexqa, ZTF22abxlqol, ZTF23aberpzw, ZTF23abjrolf 
    %unclear: ZTF21acgunkr, 
    %for second paper:
    %not good fit : ZTF19aanrrqu, ZTF21abcacpa, ZTF23aajrmfh, ZTF21aaqugxm, ZTF21aavhnpk, ZTF21aavhnpk, ZTF22aakdbia, ZTF22aaolwsd, ZTF23abmoxlu
    %not good bol fit: ZTF19abjrjdw, 

    \item Peak Absolute Magnitude: To obtain the absolute magnitudes, we use the distance modulus calculated from the host redshift assuming a cosmological model with $\Omega_M = 0.3$, $\Omega_\Lambda = 0.7$, and $h = 0.7$. For galaxies closer than 25 Mpc, we correct for the Virgo, Great Attractor, and Shapley supercluster infall based on the NASA Extragalactic Database object page \citep[NED\footnote{https://ned.ipac.caltech.edu/};][]{Helou1991}. We ignore corrections due to peculiar motions for galaxies farther than 25 Mpc. If the peculiar motion of the galaxy is $\sim 300$ km~s$^{-1}$, then the error of the peak magnitude is $\le 0.1$ mag. Table \ref{tab:nearbydistances} lists the distances used for the nearby galaxies and their references. We apply a K-correction of $2.5\times$ log(1 + z) for all the SNe in our sample.  The LLIIP sample is restricted to those with a peak absolute magnitude less luminous than \(-16\) mag, leaving 48 LLIIP SN candidates.
    
    \item Host Extinction Correction: Finally, after correcting for host galaxy extinction (see Section \ref{sec:extinction} for details), 36 LLIIP SNe remained in the final sample.
\end{enumerate}

\begin{figure}
    \centering
    \includegraphics[width=\columnwidth]{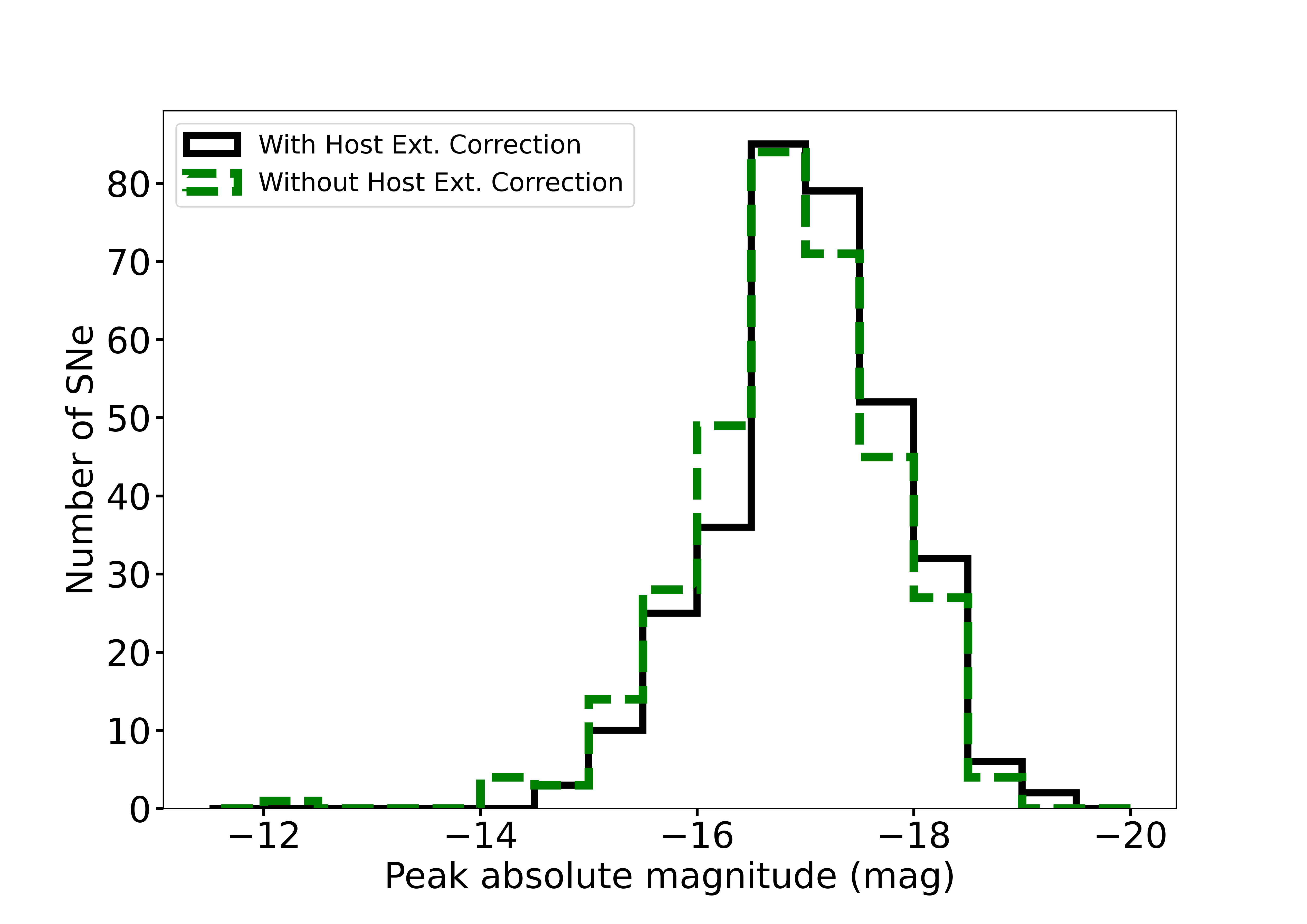}
    \caption{The distribution of the number of SNe in the final ZTF CLU Type IIP sample, consisting of 330 SNe, as a function of peak $r$-band absolute magnitude, with and without host-extinction correction}
    \label{fig:raw}
\end{figure}

The summary of the selection criteria and the sample are provided in Tables \ref{tab:search} and \ref{tab:sample1} respectively. The distribution of the number of SNe as a function of peak absolute magnitude is shown in Figure \ref{fig:raw}. A complementary ZTF experiment is the magnitude-limited Bright Transient Survey \citep[BTS;][]{Fremling2020, Perley2022}, aimed at classifying transients brighter than a peak apparent magnitude of 18.5 mag without any redshift cut. 62 SNe in our sample are also part of the Type II SN sample of the BTS survey (Y. Qin et al., in prep.; K. Hinds et al., in prep.)

%The peak luminosity for the first-peak vs the time above half-maximum is plotted in Figure \ref{fig:lumtime}. 

\begin{table*}[htb!]
    \caption{}
    \centering
    \begin{tabular}{ccc}
     \hline
       Step & Criteria  & \# Candidates \\
     \hline
       %1 & SNe in CLU Phase I+II & 1665 \\
       %\hline
       1 & SNe in CLU classified as Type II  (IIP, II, II-norm, II-pec) & 727\\ 
       \hline
       
       2 & Gaussian process fit to forced photometry lightcurves, $m_{peak_r}$ $<$ 20 mag & 719 \\ 
       \hline
       3 & Number of detections ($N_{pre-peak}$ + $N_{post-peak}$) $>$ 10 \\
       3A  & If first detection is the peak, require that the last non-detection is within 10 days of first detection & \\
       3B & $\geq$1 detection before peak and $\geq$1 detection after peak and $t_{first det} - t_{last non-det}  \leq 20$ days & 394\\
       3C & If $t_{first det} - t_{last non-det} \geq 20$ days, $N_{pre-peak}$ $>10$, $N_{post-peak}$ $>2$, $ mag_{first det} - mag_{peak} >$ 1 mag &  \\
       \hline
       4 & Criteria for plateau$:$ There should be a duration of 40 days with a drop of $<$ 1 mag & 330 \\
       \hline
       5 & LLIIP: Peak absolute magnitude $\geq$ -16 mag (no host extinction) & 48  \\
        & LLIIP: Peak absolute magnitude $\geq$ -16 mag (after host extinction) & 36  \\
     \hline
    \end{tabular}
    \label{tab:search}
\end{table*}

\input{table_sample1_LLIIP}

\section{Data}
\label{sec:data}

In this section, we describe the photometric and spectroscopic data used.

\subsection{Optical photometry}
\label{section:photdata}

We perform forced point-spread function photometry on the ZTF difference images at the location of these SNe using the ZTF forced-photometry service developed by \citet{Masci2019, Masci2023} in $g$ and $r$ bands. For this work, we consider
anything less than a 3$\sigma$ detection an upper limit. We used a Gaussian Process (GP) algorithm\footnote{https://george.readthedocs.io/} \citep{Ambikasaran2015}
to interpolate the photometric measurements and measure the slope. Figure \ref{fig:GP} shows the lightcurves of SNe 2023wcr and 2024wcs as examples. The vertical dashed red line shows the epoch of first detection, and the vertical dashed grey line shows the epoch of last non-detection. The explosion epoch is defined to be midway between the epoch of first detection and last non-detection. The epochs of peak $r$- and $g$-band  magnitudes are shown by the vertical solid red and blue lines. The photometry data, lightcurve plots, and spectroscopy of all SNe in the sample will be made available on WISEREP and Zenodo after publication.

\subsection{Optical spectroscopy}
\label{section:spectra}

For each transient, at least one spectrum is usually obtained close to the peak luminosity to establish a spectroscopic classification as described in Section \ref{sec:ZTF CLU}. We employ the \texttt{SuperNova Identification} \citep[\texttt{SNID};][]{Blondin2007} code for these classifications. For spectra that show host galaxy contamination, we utilize \texttt{superfit} \citep{Howell2005}. % for more accurate classification. Additional spectral follow-up is conducted as part of the ZTF survey efforts. We primarily utilize the Double Beam Spectrograph \citep[DBSP;][]{Oke1982} on the Palomar 200-inch Hale telescope and the Spectral Energy Distribution Machine \citep[SEDM;][]{Blagorodnova2018} on the Palomar 60-inch telescope for this purpose. 
The reduction of DBSP spectra follows the procedures outlined in \citet{Bellm2016} and \citet{Roberson2022}, while the SEDM data reductions are detailed in \citet{Rigault2019}. The LRIS spectra were reduced using the automated \lpipe{} \citep{Perley2019} pipeline.

We also make use of spectra from the Alhambra Faint Object Spectrograph and Camera at the Nordic Optical Telescope \citep[NOT;][]{Djupvik2010} and from the Spectrograph for the Rapid Acquisition of Transients \citep[SPRAT;][]{Piascik2014} on the Liverpool Telescope. The NOT data reduction utilized the PyNOT\footnote{\href{https://github.com/jkrogager/PyNOT}{https://github.com/jkrogager/PyNOT}} and PypeIt \citep{Prochaska2020a} pipelines. The SPRAT data is processed using a pipeline based on FrodoSpec \citep{Barnsley2012}. %For late-time nebular-phase observations, spectra are collected with the Low-Resolution Imaging Spectrometer \citep[LRIS;][]{Oke1995} on the Keck I telescope, typically beginning about 30 days after the explosion. The reduction of LRIS spectra is performed using the \lpipe{} \citep{Perley2019} pipeline. 

\section{Extinction Correction}
\label{sec:extinction}

Extinction is divided into two components: the first component represents dust extinction from the Milky Way, while the second component accounts for extinction originating from the host galaxies of the SNe. To correct for Galactic extinction, we employ the reddening maps provided by \citet{Schlafly11}, the extinction law described by \citet{Cardelli1989} and a value of $R_V = 3.1$. 

\begin{figure*}[!htbp]
    \centering
    \includegraphics[width=0.51\textwidth]{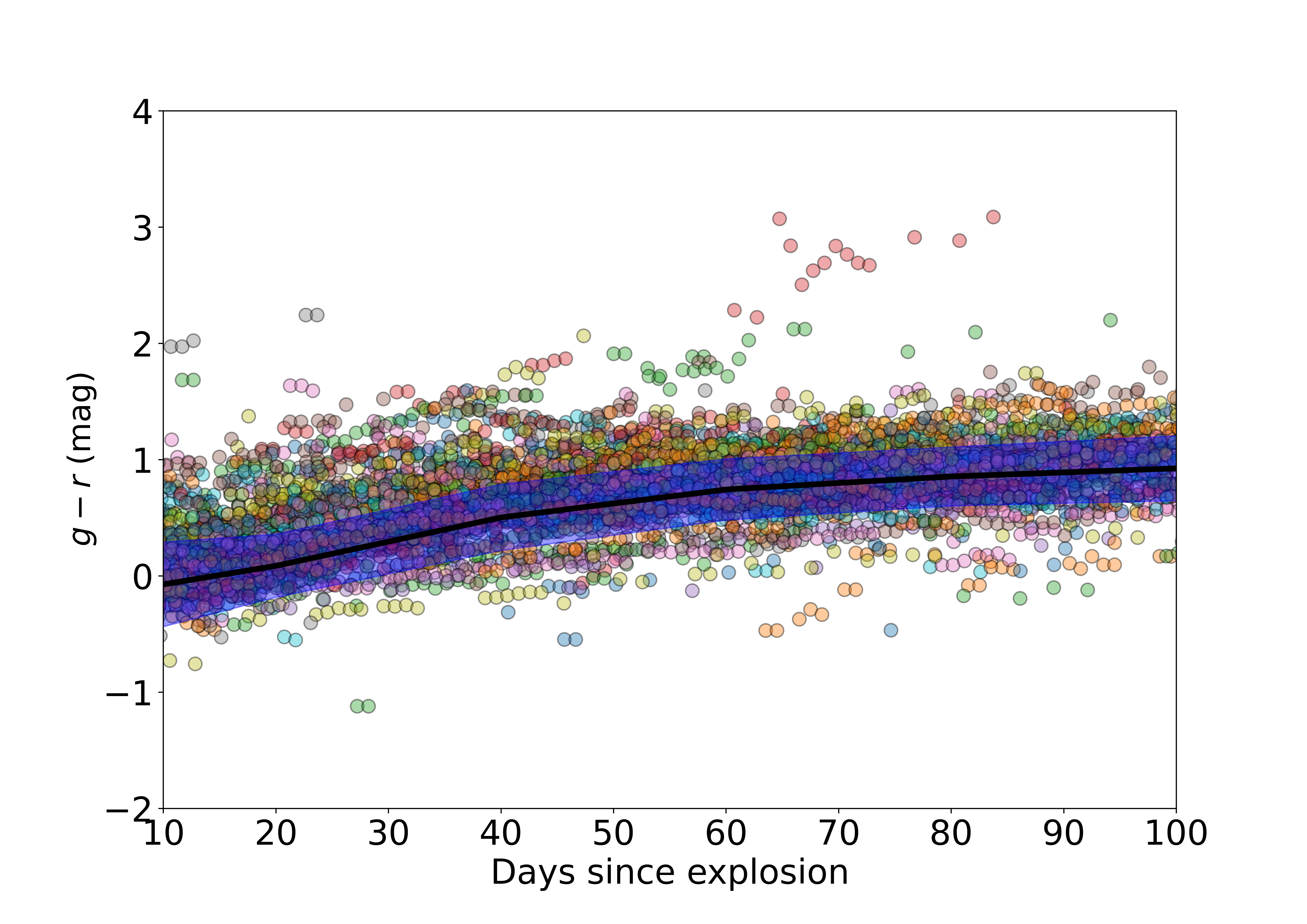}\includegraphics[width=0.51\textwidth]{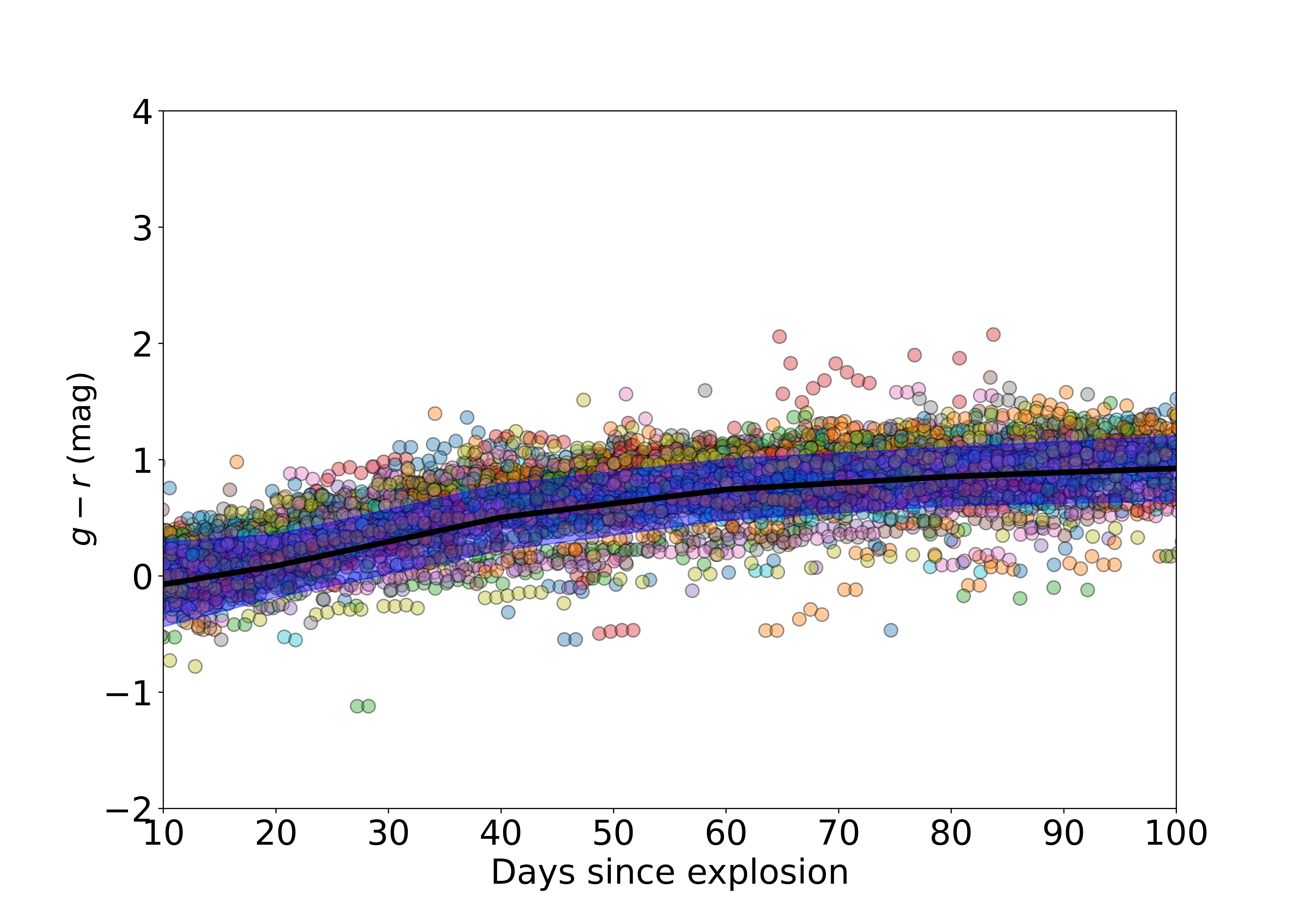}
    \caption{\textit{Left:} The $g-r$ colors of the 
 ZTF CLU Type II SN sample are shown in colored circles. The average color and standard deviation (1$\sigma$ range) of the distribution, after filtering out SNe with \ion{Na}{1} D EW $>$ 1 $\textrm{\AA}$, are shown by the black solid line and blue shaded region, respectively. \textit{Right:} The sample $g-r$ color distribution after host-extinction correction. The color template is available on \href{https://zenodo.org/records/14538857?token=eyJhbGciOiJIUzUxMiJ9.eyJpZCI6ImFlZmFjOTY2LWU5MDMtNDU1My1iNjEwLWVmMDA0NTRjMDBmNSIsImRhdGEiOnt9LCJyYW5kb20iOiJmNGRjMWJiMTJhNmE5ODBmM2VkYzU4YTQ1OGE4NGM5MSJ9.j-VvClhhUEAr0JLwY5sprDydZw1OPGP-LfXjl5h9KRk1PMSCRhR5iM2geD3YgnKoCzkCD1aAFgewgSOAHkoB0A}{Zenodo}}
    \label{fig:hostext}
\end{figure*}

To estimate the host-galaxy extinction, we use the average $g-r$ color curve of the sample. First, we measure the \ion{Na}{1} D absorption lines of the host environment \citep{Poznanski2012, Stritzingetr2018} when a high SNR SN spectrum is available. To measure the template $g-r$ color, we discard objects with \ion{Na}{1} D EW $>$ 1 $\textrm{\AA}$. The 1$\sigma$ scatter is 0.35 mag. The average color and standard deviation (1$\sigma$ range) are shown by the black solid line and blue shaded region, respectively, in Figure \ref{fig:hostext}. This template is available on \href{https://zenodo.org/records/14538857?token=eyJhbGciOiJIUzUxMiJ9.eyJpZCI6ImFlZmFjOTY2LWU5MDMtNDU1My1iNjEwLWVmMDA0NTRjMDBmNSIsImRhdGEiOnt9LCJyYW5kb20iOiJmNGRjMWJiMTJhNmE5ODBmM2VkYzU4YTQ1OGE4NGM5MSJ9.j-VvClhhUEAr0JLwY5sprDydZw1OPGP-LfXjl5h9KRk1PMSCRhR5iM2geD3YgnKoCzkCD1aAFgewgSOAHkoB0A}{Zenodo}. We then measure the $A_\textrm{V,host}$ assuming that the SN color exceeding the 1$\sigma$ $g - r$ template is caused by host extinction.  80 out of the 330 Type IIP SNe has $A_\textrm{V,host} > 0$ . The sample color curve after this correction is shown in the right panel of Figure \ref{fig:hostext}. The remaining 1$\sigma$ scatter in the $g - r$ color is likely due to intrinsic properties of the SNe, which could be dominated by photospheric temperature differences \citep[e.g., see][]{deJaeger2018}. Table \ref{tab:sample1} lists the measured $A_\textrm{V,host}$ values.

\begin{figure}
    \centering
    \includegraphics[width=\columnwidth]{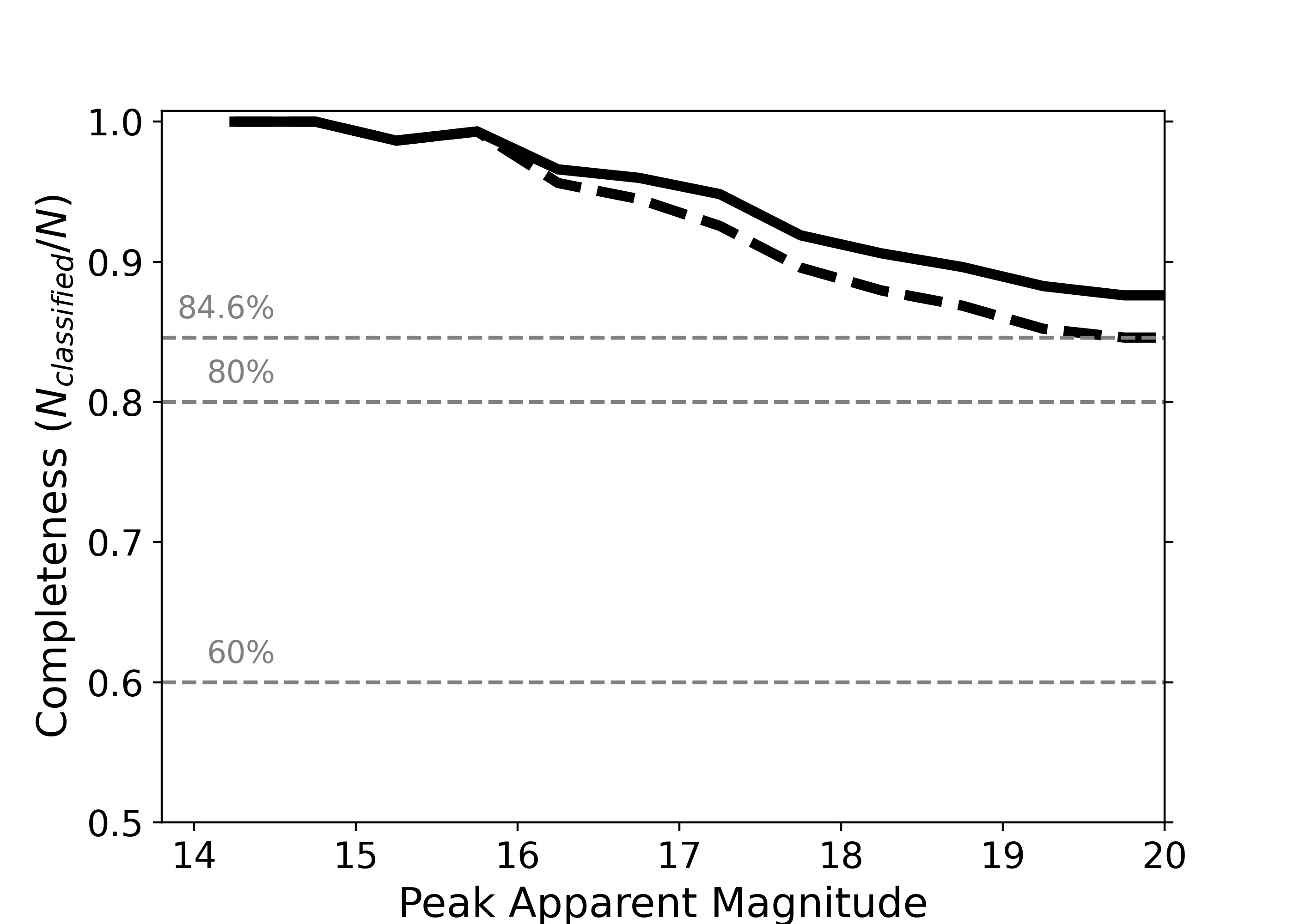}
    \caption{The cumulative distribution of the spectroscopic completeness of the ZTF CLU survey (with 2064 saved transients) as a function of peak $r$-band apparent magnitude. The dashed line accounts for the entire sample, whereas the solid black line omits 79 transients that are probably spurious. As a conservative lower limit, we consider the completeness depicted by the solid dashed line.}
    \label{fig:completeness}
\end{figure}

\begin{figure*}
    \centering
    \includegraphics[width=12cm]{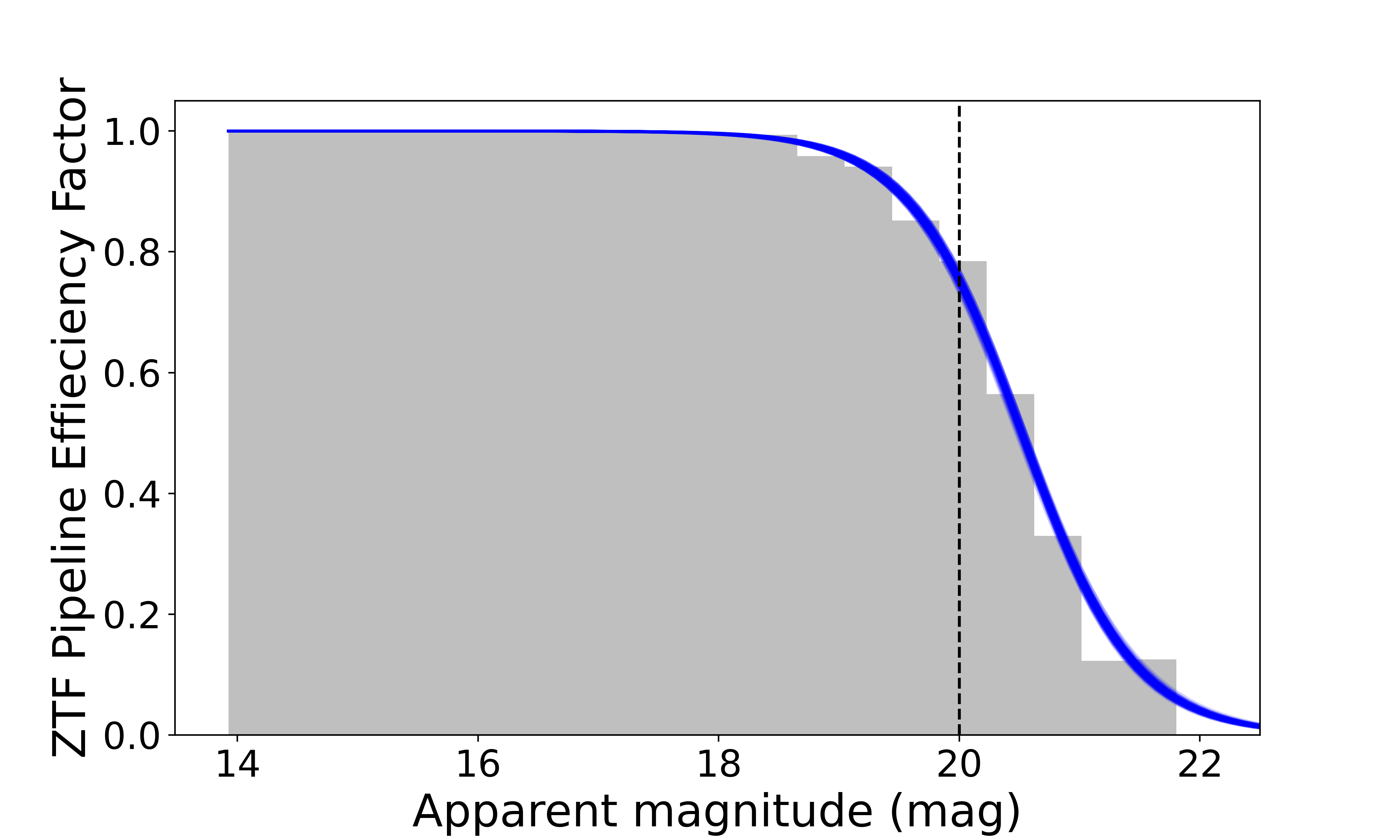} \includegraphics[width=12cm]{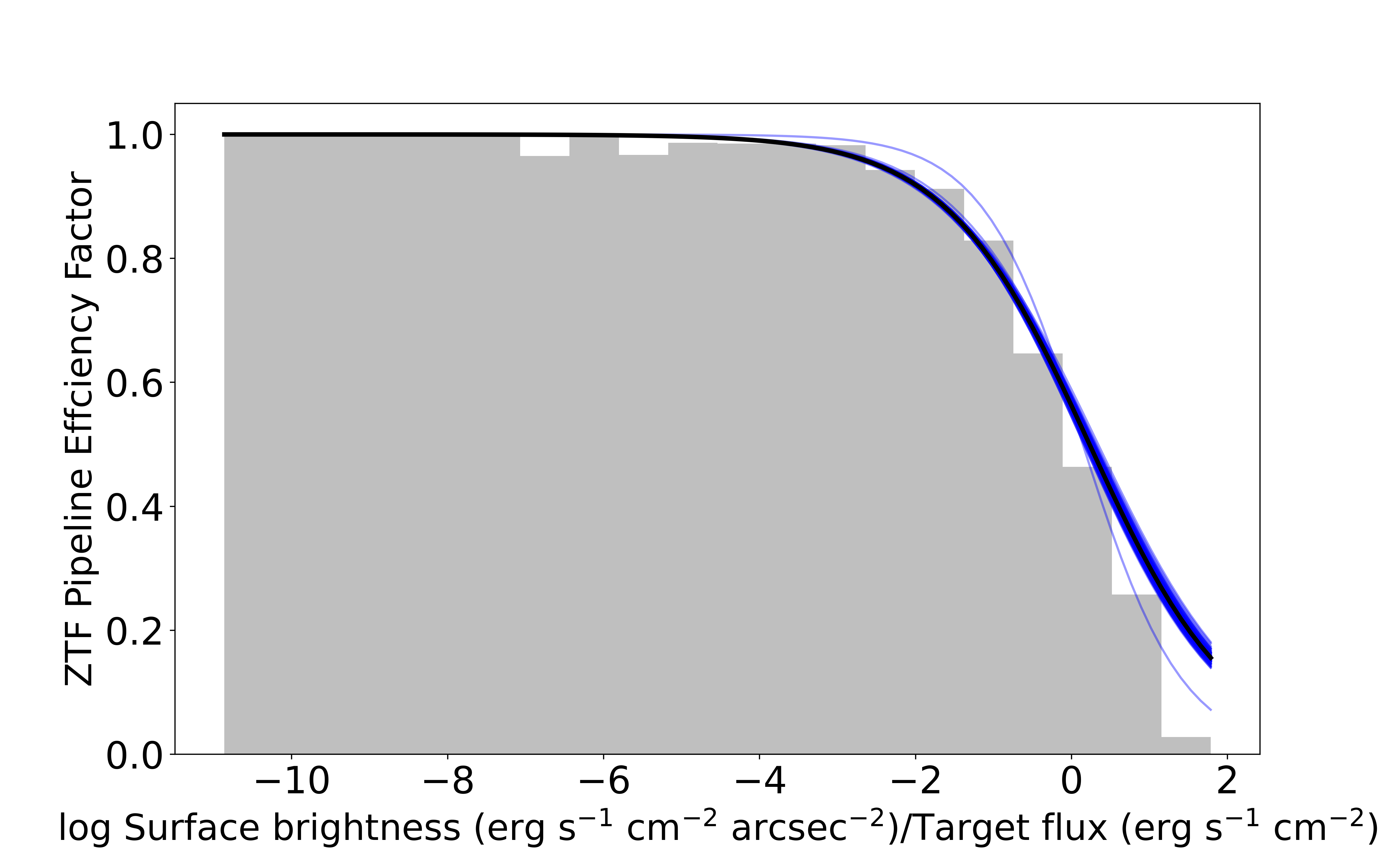}
    \caption{The ZTF pipeline recovery efficiency as a function of apparent magnitude (\textit{upper}) and as a function of the ratio of local surface brightness and target flux (\textit{lower}). The best-fit parameters of the logistic function is listed in Appendix \ref{sec:pipeline}. The data required to reproduce the plot is available on \href{https://zenodo.org/records/14538857?token=eyJhbGciOiJIUzUxMiJ9.eyJpZCI6ImFlZmFjOTY2LWU5MDMtNDU1My1iNjEwLWVmMDA0NTRjMDBmNSIsImRhdGEiOnt9LCJyYW5kb20iOiJmNGRjMWJiMTJhNmE5ODBmM2VkYzU4YTQ1OGE4NGM5MSJ9.j-VvClhhUEAr0JLwY5sprDydZw1OPGP-LfXjl5h9KRk1PMSCRhR5iM2geD3YgnKoCzkCD1aAFgewgSOAHkoB0A}{Zenodo}.}
    \label{fig:pipeline}
\end{figure*}

\begin{figure*}
    \centering
    \includegraphics[width=10cm]{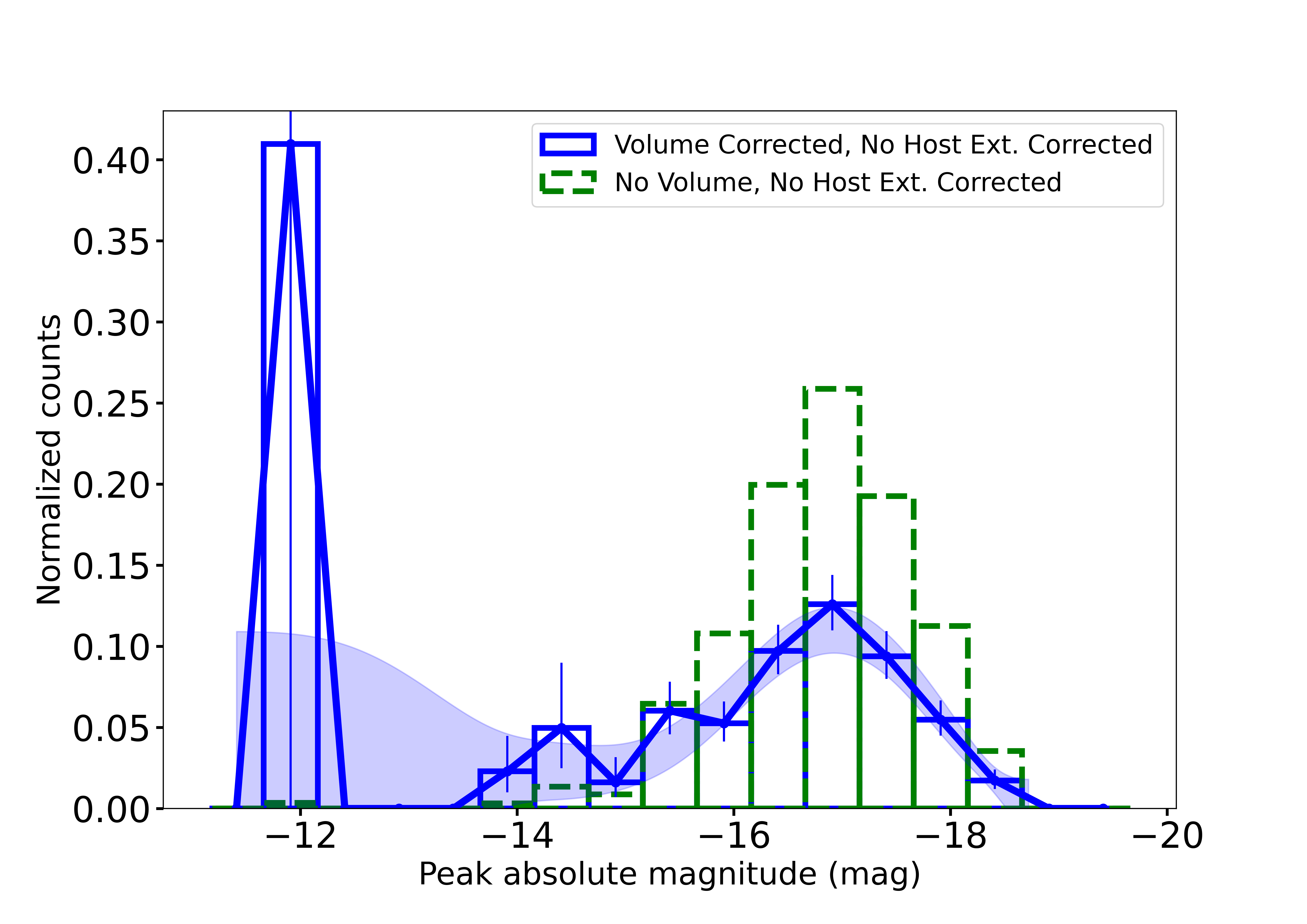}\includegraphics[width=10cm]{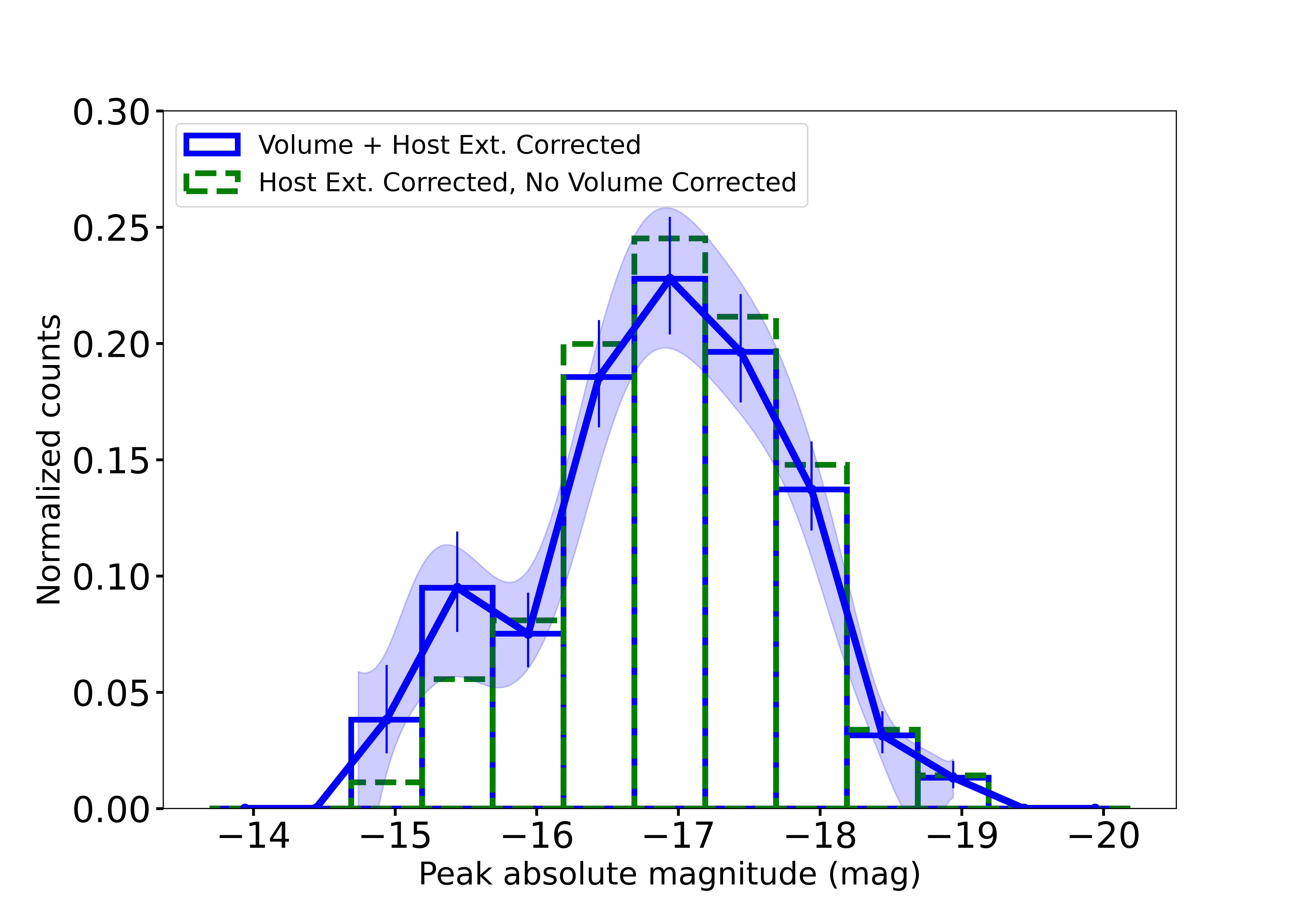}
    \caption{The peak $r$-band absolute magnitude distribution of the sample of 330 Type IIP SNe in CLU with Poisson errorbars before host-extinction correction (left) and after host-extinction correction (right). The volume-corrected luminosity function is shown in solid blue, while the raw distribution is shown in dashed green. GP fit is shown in blue. The luminosity function data is available in \href{https://zenodo.org/records/14538857?token=eyJhbGciOiJIUzUxMiJ9.eyJpZCI6ImFlZmFjOTY2LWU5MDMtNDU1My1iNjEwLWVmMDA0NTRjMDBmNSIsImRhdGEiOnt9LCJyYW5kb20iOiJmNGRjMWJiMTJhNmE5ODBmM2VkYzU4YTQ1OGE4NGM5MSJ9.j-VvClhhUEAr0JLwY5sprDydZw1OPGP-LfXjl5h9KRk1PMSCRhR5iM2geD3YgnKoCzkCD1aAFgewgSOAHkoB0A}{Zenodo}.}
    \label{fig:lumfunc}
\end{figure*}

\section{Luminosity function}
\label{sec:LF}

 With the use of our systematic and controlled ZTF CLU event sample, we determine a luminosity function for Type IIP and LLIIP SNe. The CLU LLIIP sample nearly \textcolor{black}{triples} the current number of published LLIIP SNe in the literature. Our final sample comprises 330 Type IIP SNe, with 36 ($\approx$ 11\%) LLIIP SNe ($M_{r} > -16$ mag).

We use the $r$-band photometry to measure the SN II luminosity function and volumetric rate. A volume correction is applied to the luminosity function by weighting each object by $V_{max} = 1/D^{3}_{max}$, where $D_{max}$ ($\propto 10^{\frac{1}{5}(20-M_r, peak)}$) is the farthest distance at which a transient can be detected given a limiting magnitude of 20 mag. % We note that $M_\textrm{r,peak}$ is the peak absolute magnitude of the SN before extinction correction.
For sources detectable past 150 Mpc, $D_{max}$ was set to the maximum CLU experiment distance of 150 Mpc. 

We also correct for the CLU spectroscopic completeness and the ZTF pipeline recovery efficiency factor as they are dependent on the apparent magnitude.
Of the total of 2064 transients that are saved in CLU, 1745 (84.6\%) transients are classified. The overall completeness of the CLU survey as a function of peak apparent magnitude is indicated by the dashed black line in Figure~\ref{fig:completeness}. There are 79 transients for which we are uncertain whether they are real transients on the basis of their spectroscopic and photometric properties. If we exclude these, the overall completeness is $\sim$ 87\% (solid black line).  Also, we note that the long-lived nature and relative ease of identifying the H$\alpha$ P-Cygni feature make Type IIP events easier to classify. So the Type IIP SNe completeness is likely to be higher than the solid black line. As a conservative estimate, for the rest of the paper, we use the spectroscopic completeness depicted by the dashed black line.

%The conservative completeness as a function of the apparent magnitude is indicated by the black line, which accounts for a cumulative completeness of $\sim$81\%. The completeness without considering these doubtful transients is shown in the dashed black line, which results in a net completeness of $\sim$85\%.

The ZTF image subtraction pipeline has two potential sources of inefficiency relevant to this calculation. In each science image, the pipeline actively masks pixels affected by quality criteria, such as saturation from high brightness, cosmic rays, and defective pixels. This dynamic masking does not have a significant effect as Type IIP SNe have long durations ($\sim$100 days) and are eventually detected by the pipeline. A significant issue relates to the decreased efficiency of the image subtraction algorithm for faint alerts and on bright galaxy backgrounds. Here, we model the efficiency of the ZTF pipeline as a function of the alert flux and the ratio of background surface brightness to target flux. To estimate this, we check whether alert photometry exists for a given epoch for which there is a $\ge5\sigma$ forced photometry detection. The local host surface brightness is extracted using PS1 $r$-band images with a 3$"$ radius aperture. To estimate the conditional probability of the ZTF pipeline efficiency, we consider the binary data, \(X\), which represents whether the ZTF pipeline successfully recovers a transient in alerts (\(X = 1\)) or not (\(X = 0\)). We model \(X\) using the Bernoulli distribution:
\[
X \sim \text{Bern}(p),
\]
where \( p \) is parameterized by a logistic function dependent on the ratio  (\( r \)) of the host surface brightness and the target flux:
\[
p(r) = \frac{1}{1 + \exp(a(r - c))},
\]
with \( a \) and \( c \) representing the model parameters that need to be determined. The logistic function was chosen because it smoothly transitions from 1 to 0 and captures the pipeline efficiency behavior as a function of \( r \). We then estimate the model parameters using the  \texttt{emcee} package \citep{Foreman-Mackey13}. The best-fit values of $a$ and $c$ are  $a=1.12^{+0.02}_{-0.02}$ and $c=0.21^{+0.03}_{-0.03}$ respectively. As expected, the recovery efficiency decreases as this ratio increases. The best-fit efficiency is $\sim50\%$ when the host surface brightness and target fluxes are comparable.

We also measure the inefficiency as a function of the apparent magnitude alone, using \[
p(m) = \frac{1}{1 + \exp(b(m - d))},
\]
where  \( m \) is the alert apparent magnitude. The best-fit values obtained are $b=2.28^{+0.03}_{-0.04}$ mag$^{-1}$ and $d=20.57^{+0.01}_{-0.01}$ mag. The best fits, and corner plots of the MCMC fit are shown in Figure \ref{fig:pipeline} and Appendix \ref{sec:pipeline}, respectively. The best-fit efficiency decreases with an increase in target alert apparent magnitude and is $\approx$ 80\% when the alert magnitude is 20 mag.

\begin{figure}
    \centering
    \includegraphics[width=9cm]{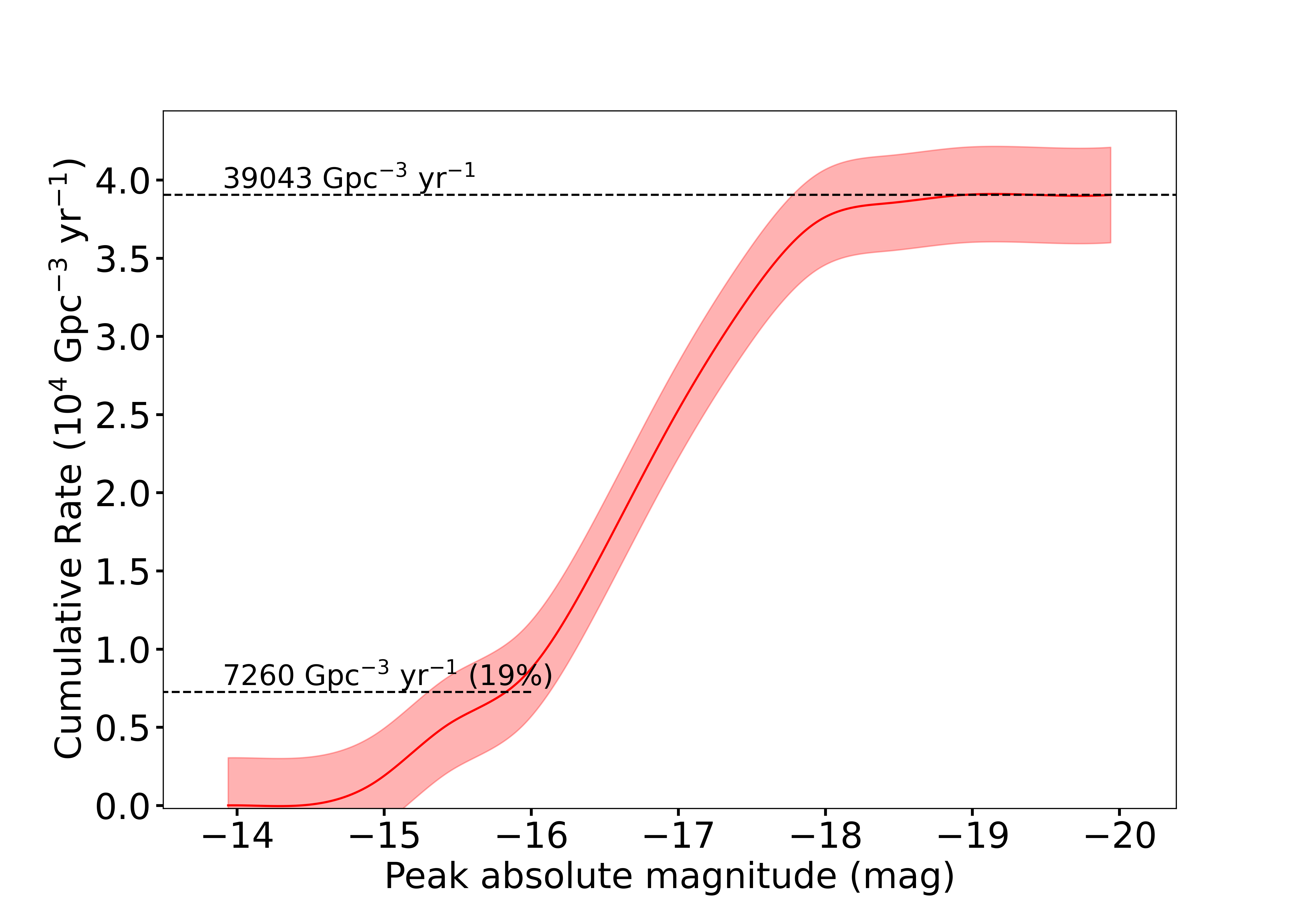}
    \caption{The cumulative distribution of the luminosity function of the ZTF CLU Type IIP sample. The IIP and LLIIP volumetric rates are shown by the horizontal dashed lines.}
    \label{fig:cumrate}
\end{figure}

We then weight each SN by the corresponding classification completeness factor and the pipeline efficiency factor. The luminosity function thus obtained is shown in Figure \ref{fig:lumfunc} in blue. The luminosity function without volume correction is shown by the green dashed lines in Figure \ref{fig:lumfunc}.  After volume-, incompleteness- and pipeline-efficiency correction,  the fraction of LLIIP SNe is $19^{+3}_{-4}\%$ of the entire Type II SN population (see Figure \ref{fig:cumrate}). The fraction of Type IIP SNe fainter than $-15$ mag and $-14.5$ are $<5\%$ and $<1.5\%$ respectively.

\begin{figure}
    \centering
    \includegraphics[width=0.49\textwidth]{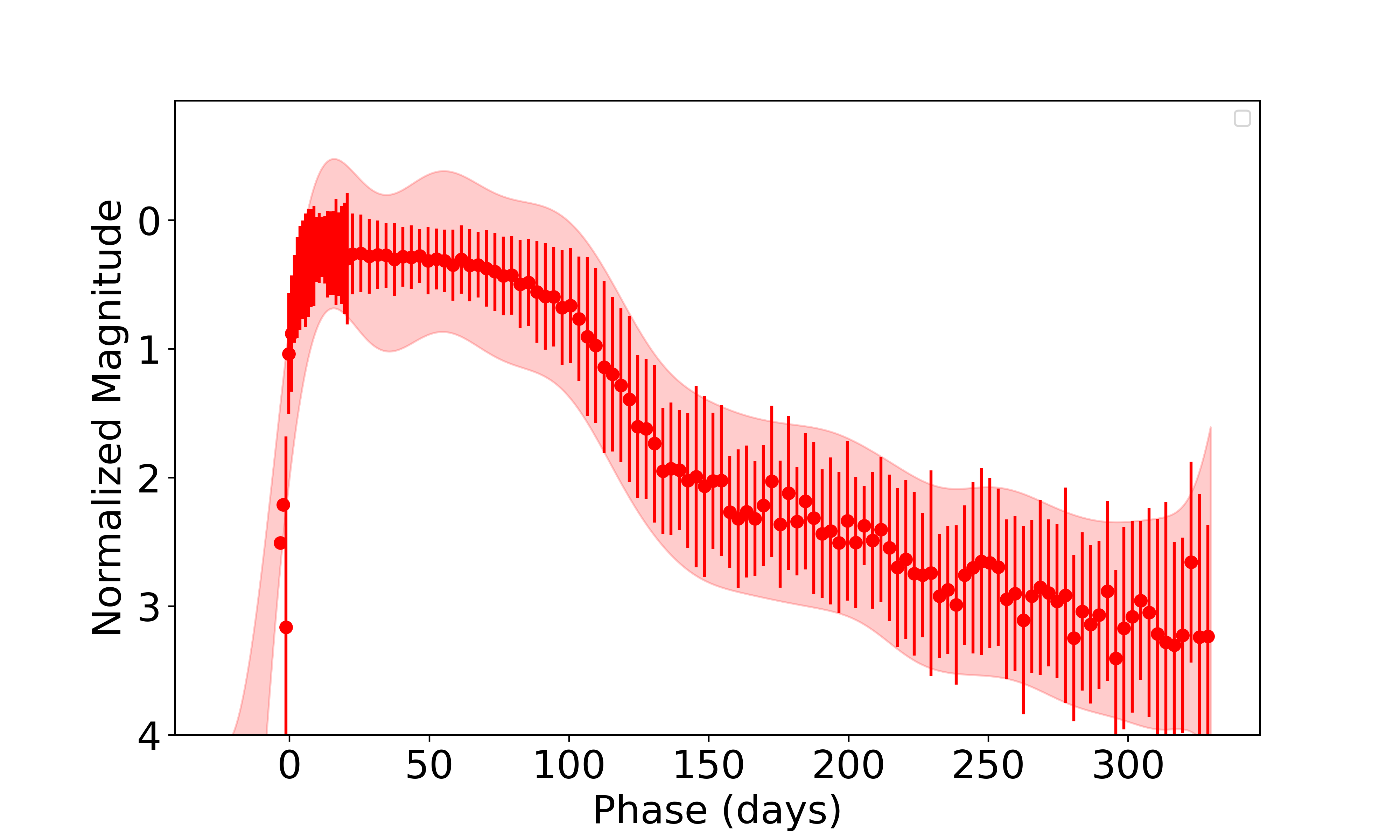}
    \caption{The $r$-band lightcurve template for Type IIP SNe. The solid points depict the mean lightcurve magnitudes of the ZTF CLU Type IIP sample. The line and shaded regions show the $1\sigma$ uncertainties derived from the GP fit.}
    \label{fig:template}
\end{figure}

\begin{figure*}
    \centering
    \includegraphics[width=0.49\textwidth]{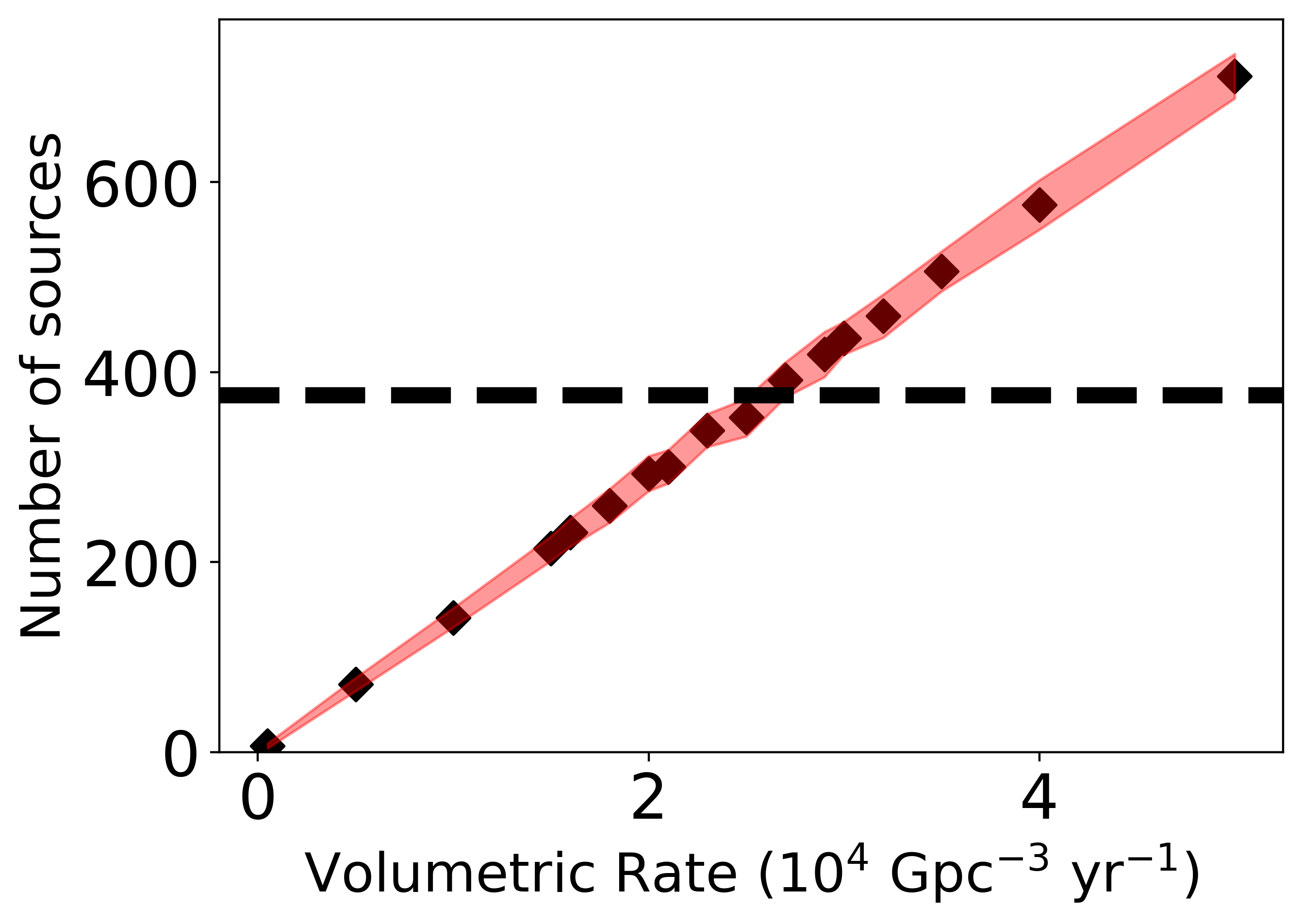}\includegraphics[width=0.49\textwidth]{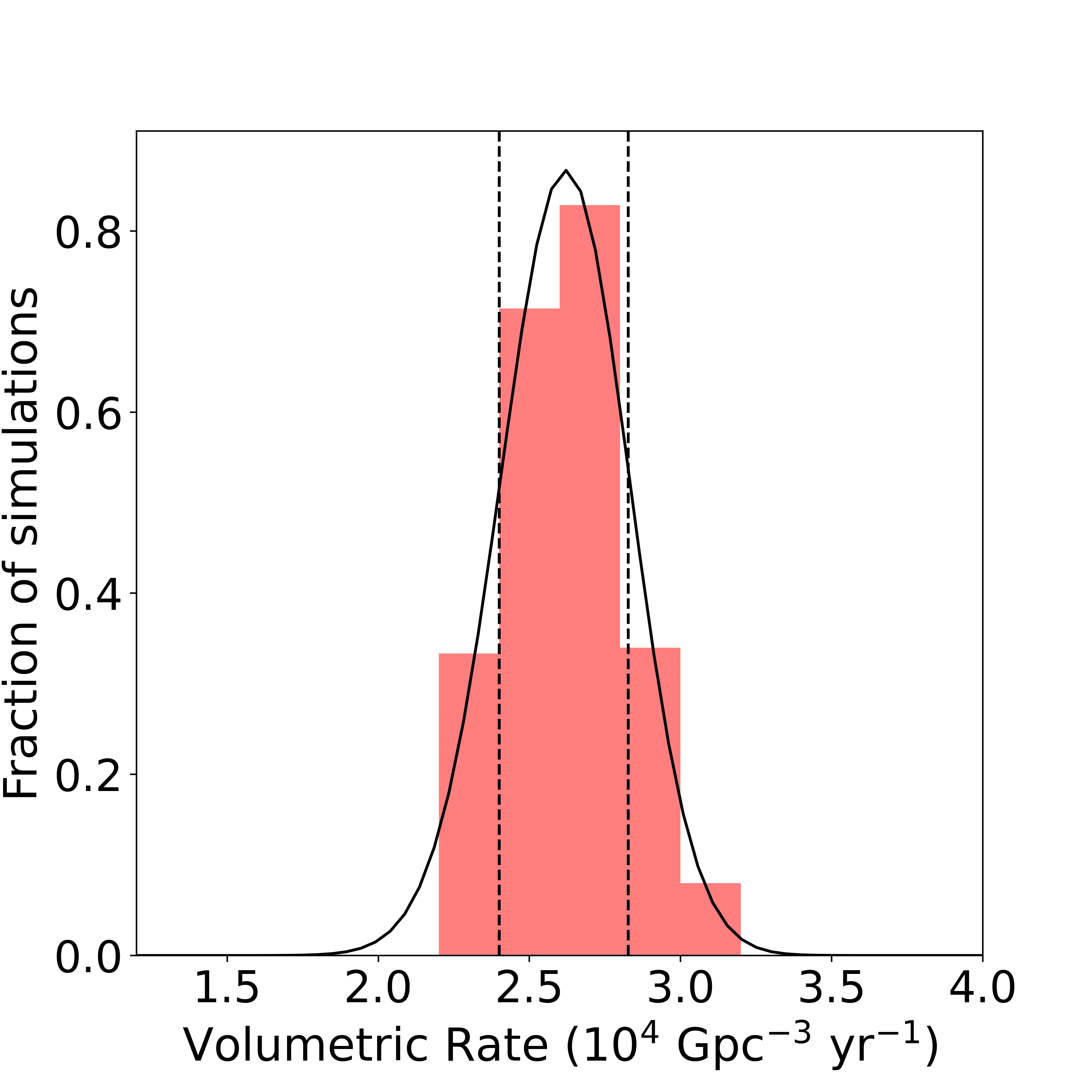}
    \caption{\textit{Left:} The plot shows the number of LLIIP SNe that satisfy the filtering criteria as a function of the volumetric rate. The data points and the shaded region indicate the mean and standard deviation of detected transients, derived from 30 iterations of the simulated survey for each input volumetric rate. The dashed black line represents the observed number of Type IIP SNe in the ZTF CLU sample. \textit{Right:} The fraction of simulations yielding the observed number of transients (within a 10\% margin) is plotted against the volumetric rate.}
\label{fig:rates}
\end{figure*}

\section{The Volumetric rate of Type IIP SNe using skysurvey}
\label{sec:rates}
%Read:https://iopscience.iop.org/article/10.1088/0004-637X/757/1/70

First, we create a template $r$-band lightcurve by compiling the photometry data for all the Type IIP SNe in the sample and apply Gaussian Regression processing (Figure \ref{fig:template}). We use the \texttt{skysurvey} code\footnote{https://github.com/MickaelRigault/skysurvey} to estimate the volumetric rate. \texttt{skysurvey} simulates light curves as they would be observed based on a light curve template (using \texttt{sncosmo}) and survey plan. The survey plan is constructed using the actual pointing schedule of ZTF, detailing observation times, the filters used, and the sky brightness. Additionally, the simulation incorporates data on CCD outlines to account for data losses caused by chip gaps.

For the selection of simulated SNe II from \texttt{skysurvey}, we use the forced photometry-based selection criteria outlined in Section \ref{sec:sample}. We require at least two 5$\sigma$ detections exceeding 20 mag in brightness. These criteria guarantee detection by the ZTF alert system, allowing the transient to pass the CLU filter, with a distance $<150$ Mpc to ensure appropriate CLU filter saving and spectroscopic follow-up. Additionally, the lightcurve quality cuts described in Section \ref{sec:sample} are employed to consistently filter the \texttt{skysurvey} sample.

We generate ZTF lightcurves of Type IIP SNe for a range of input volumetric rates \textcolor{black}{($0.1$–$10.0 \times 10^{4}\ \textrm{Gpc}^{-3}\ \textrm{yr}^{-1}$)  for} the \texttt{skysurvey} simulation up to a distance of 150 Mpc, according to the luminosity function in Figure \ref{fig:lumfunc}. At each rate, 30 survey plan simulations were conducted. For these simulations, we documented the number of transients that successfully passed the filtering process. The detection estimate for each rate was made by computing the mean number of transients detected from the simulations, with the standard deviation serving as the error estimate. Figure \ref{fig:rates} shows the expected number of detected transients along with the fraction of simulations that yield our observed sample size within an error margin of 10\% for a given input volumetric rate.

The fraction of simulations matching our sample size follows a skewed Gaussian distribution, which was used to estimate the rate and its uncertainty based on a 68\% confidence interval. This analysis yields a volumetric rate of \( (2.6_{-0.3}^{+0.3}) \times 10^{4}\ \textrm{Gpc}^{-3}\ \textrm{yr}^{-1} \).

This raw \texttt{skysurvey} rate already accounts for survey conditions such as weather and cadence as we used the real observing history of ZTF to simulate the light curves as they would have been observed. We now apply corrections for spectroscopic incompleteness, pipeline recovery efficiency and galaxy catalog incompleteness. As shown in Section \ref{sec:LF}, the CLU experiment achieved an overall completeness rate of $\approx$ 84\%. 

\begin{figure}
    \centering
    \includegraphics[width=\columnwidth]{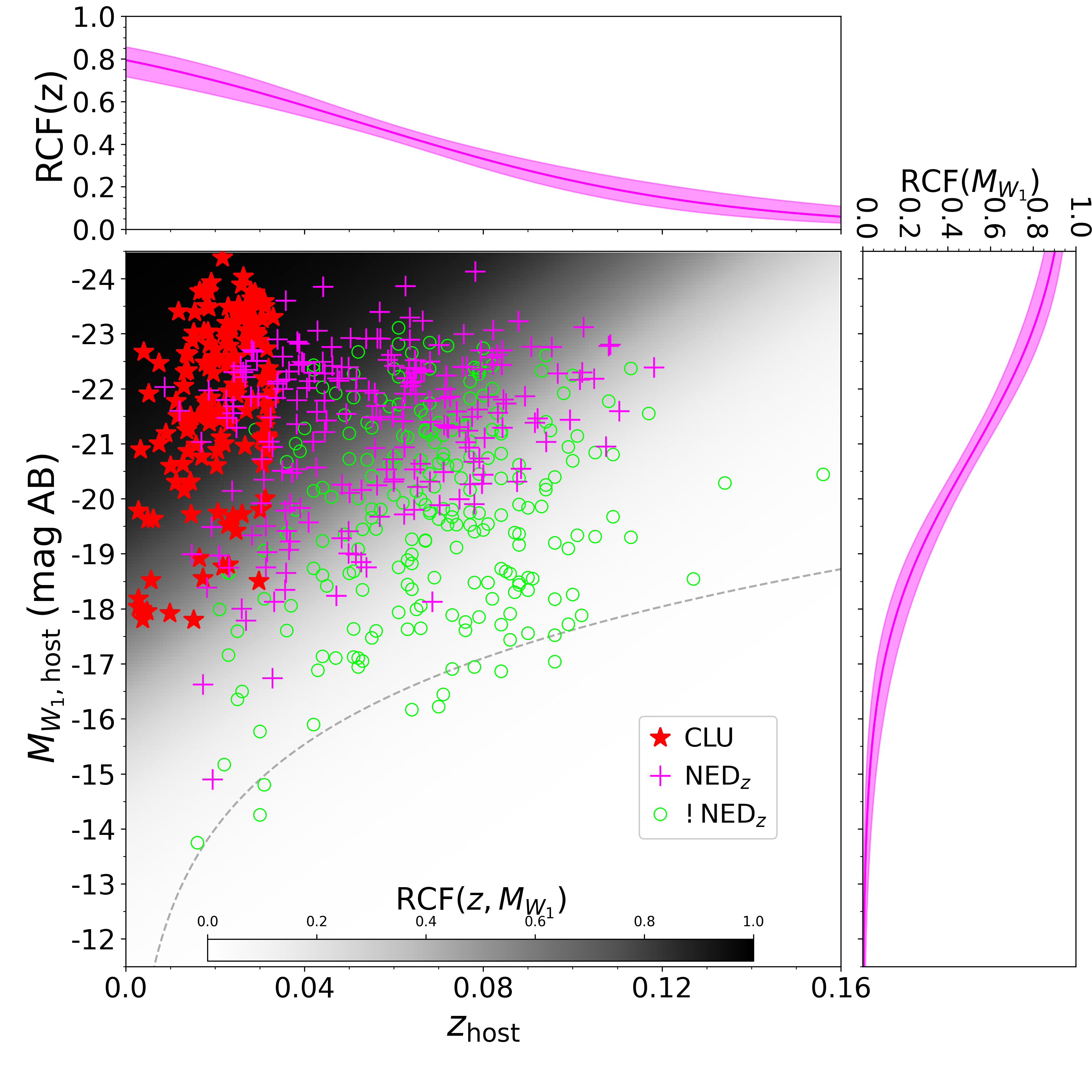}
    \caption{The redshift completeness factor of the galaxy catalogs as a function of the host redshift and $W1$-band magnitude, measured in \citet{Fremling2020}. The red stars depict the hosts of the SNe of the CLU Type IIP sample.}
    \label{fig:rcf}
\end{figure}

In the CLU experiment, our methodology restricts the identification of SNe to those in galaxies where a spectroscopic redshift has been established. Therefore, we must correct the \texttt{skysurvey} rate of LLIIP events for galaxy catalog incompleteness. To correct for the incompleteness of the galaxy catalog, we used the redshift completeness factor derived from the ZTF Bright Transient Survey \citep{Fremling2020}. We use the redshifts and WISE 3.36 $\mu m$ absolute magnitudes (MW1) of the host galaxies of our sample and weight each event by 1/RCF(z,MW1) to estimate the effect of galaxy catalog incompleteness (see Figure \ref{fig:rcf}). This leads to an underestimate of the volumetric rate by $\approx$ 19\%.

Finally, as mentioned earlier, the ZTF image subtraction pipeline has a reduced efficiency of the image subtraction algorithm on bright galaxy backgrounds and fainter targets. We weight each event by the inverse of the Pipeline Efficiency Factor (1/PEF(M, host)). This leads to an underestimate of the rate by $\approx$ 6\%.

Applying the corrections described above, we derive a corrected rate of $(3.9_{-0.4}^{+0.4}) \times 10^{4}\ \textrm{Gpc}^{-3}\ \textrm{yr}^{-1}$ for Type IIP SNe and  $(7.3_{-0.6}^{+0.6})  \times 10^{3}\ \textrm{Gpc}^{-3}\ \textrm{yr}^{-1}$ for LLIIP SNe.

\begin{figure*}
    \centering
    \includegraphics[width=0.35\textwidth]{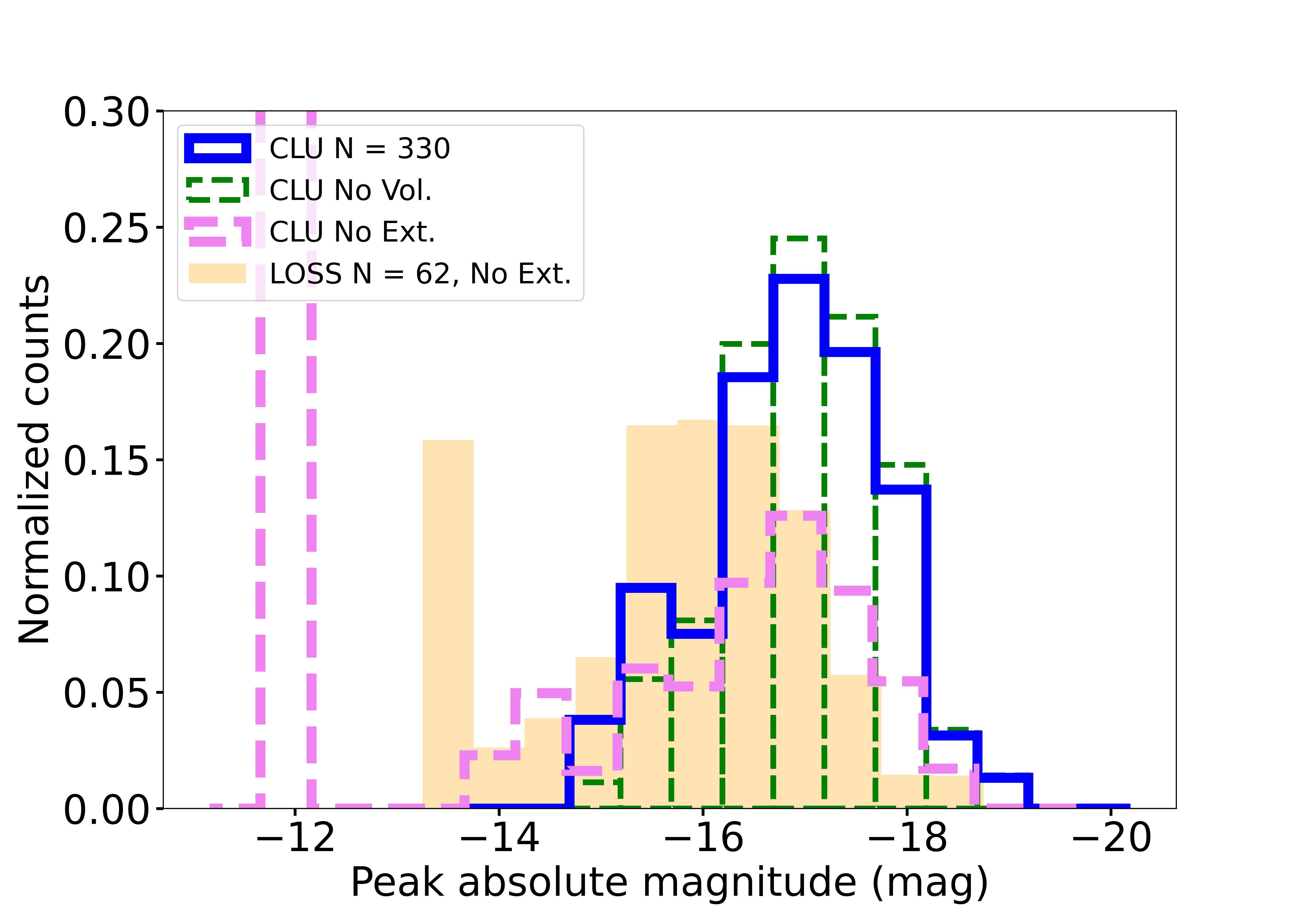}\includegraphics[width=0.35\textwidth]{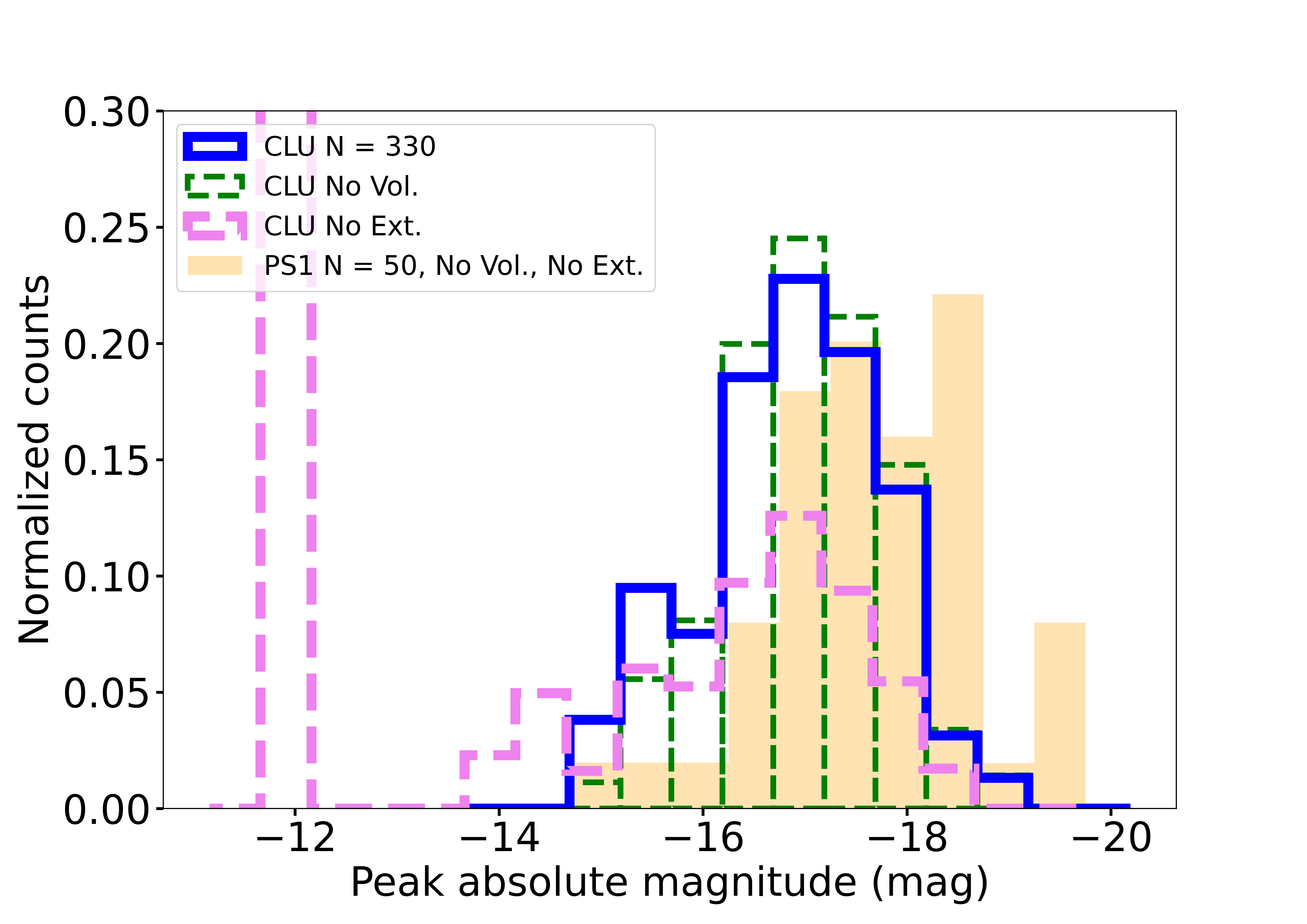}\includegraphics[width=0.35\textwidth]{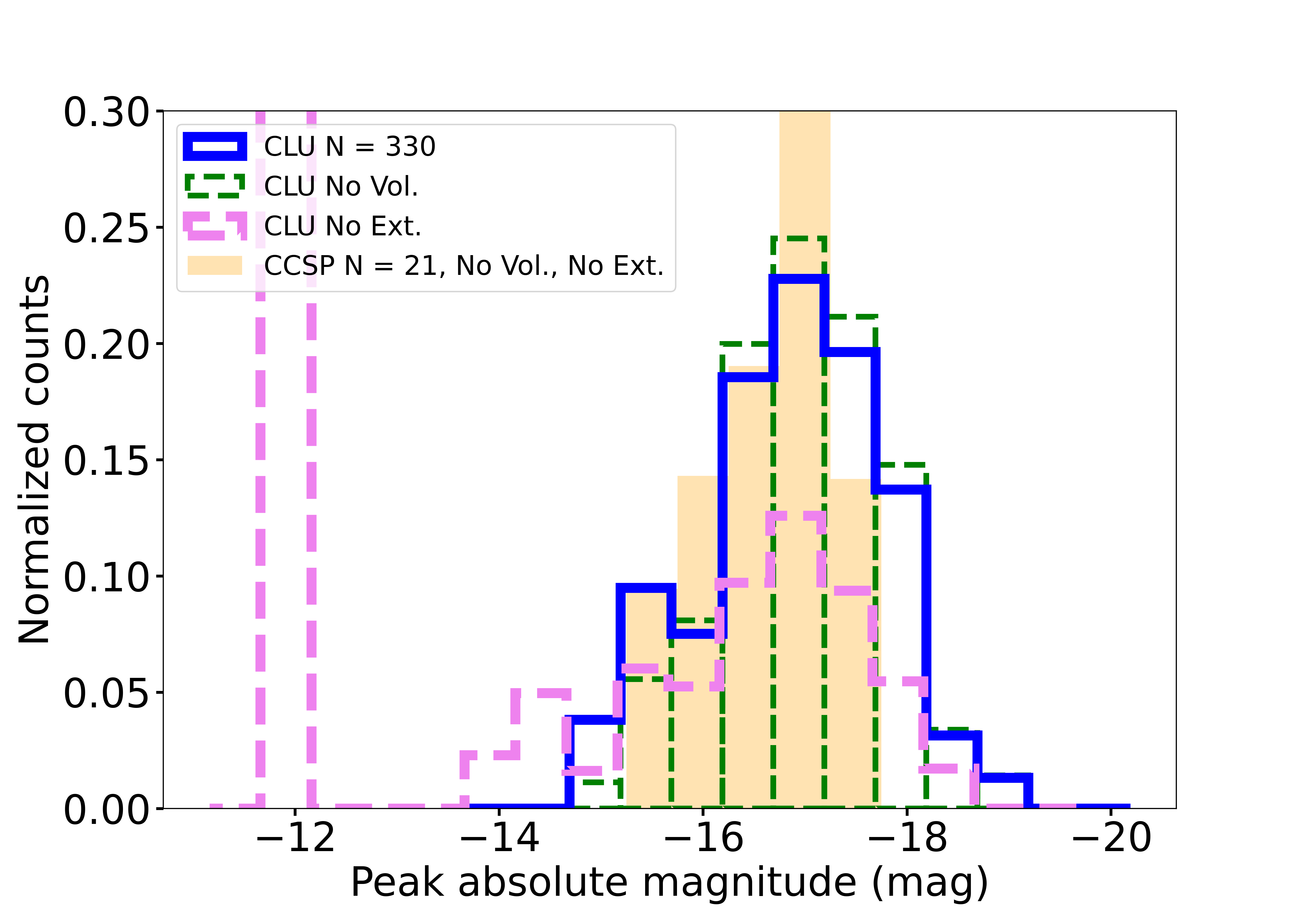}
    
    \includegraphics[width=0.35\textwidth]{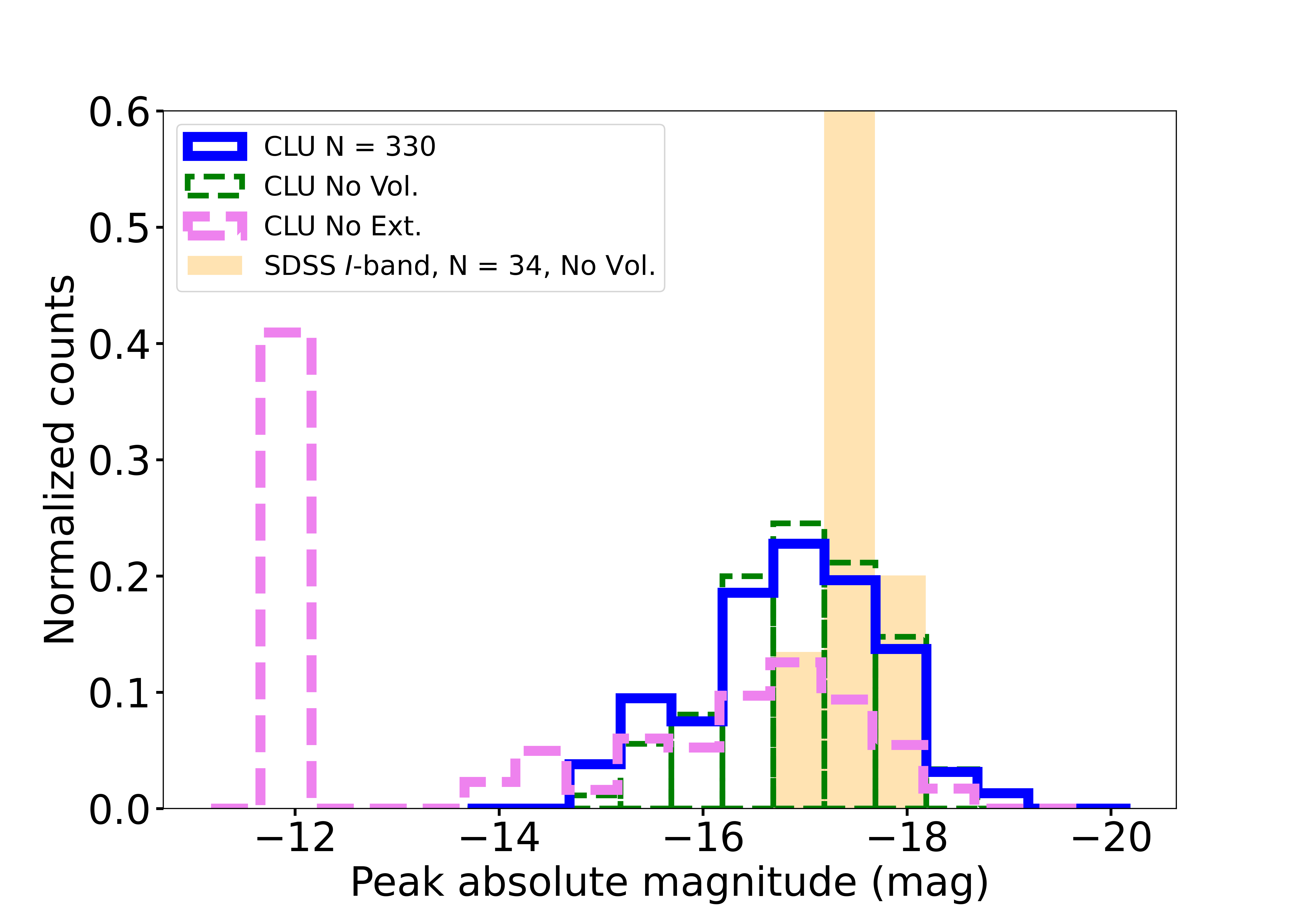}\includegraphics[width=0.35\textwidth]{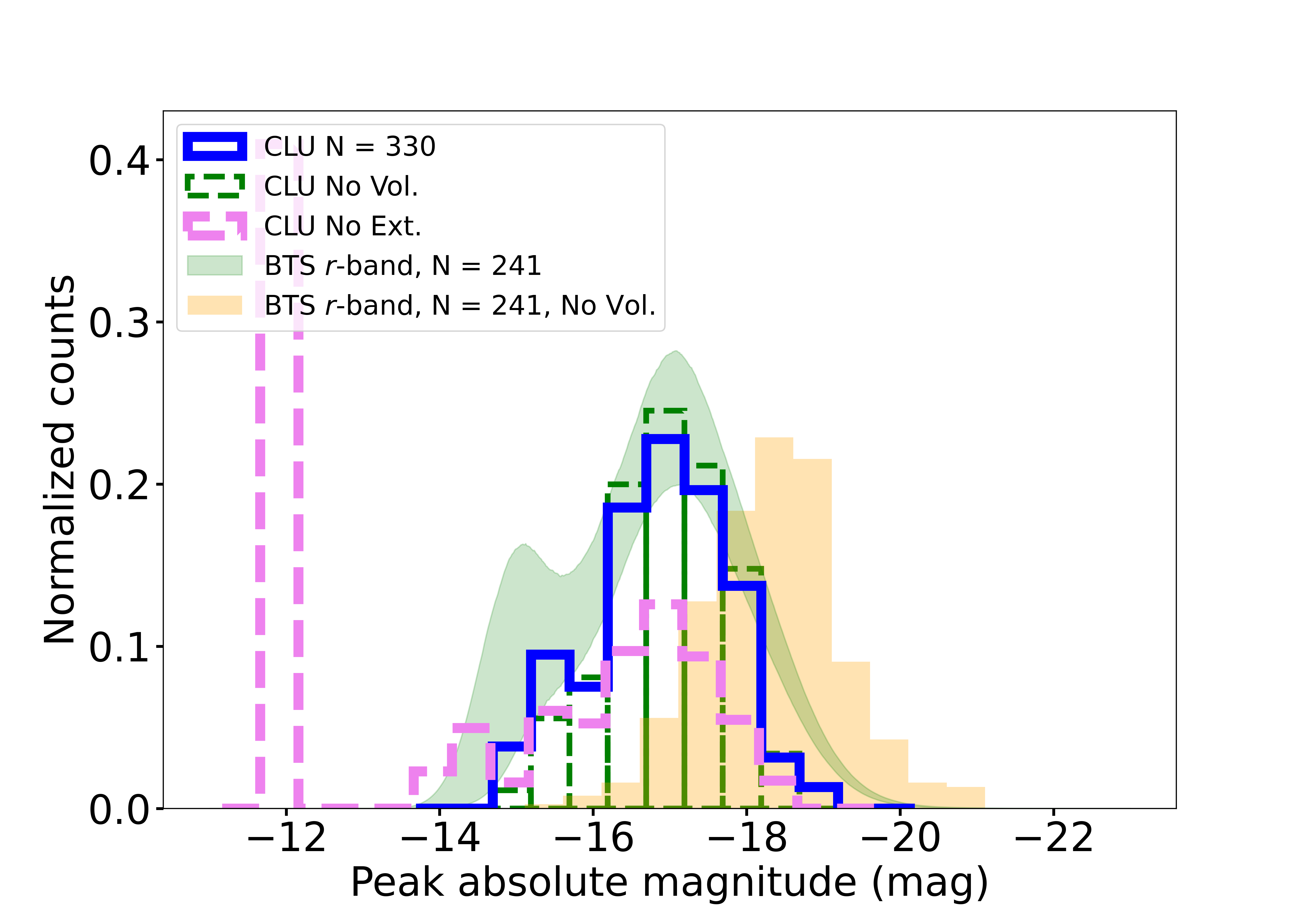}
    \caption{Comparison of the ZTF CLU Type II SNe luminosity function in $r$-band with the LOSS ($r$-band), PS1 ($r$-band), CCCP ($r$-band), SDSS ($I$-band), BTS ($r$-band) surveys (in shaded orange or green). The blue, green, and purple line depicts the volume and host-extinction corrected, only host-extinction corrected and only volume corrected distribution of the CLU Type IIP SNe sample, respectively.}
    \label{fig:literature}
\end{figure*}

\section{Discussion}
\label{sec:discussion}
\subsection{Comparison to the literature}
\label{subsec:lit}

The ZTF CLU sample of 330 SNe is the largest sample of Type IIP SNe from a systematic volume-limited survey to date. In Figure~\ref{fig:literature}, we compare the luminosity function with the $r$-band Type II Luminosity functions from previous systematic SN surveys $-$ 62 Type II SNe from the Lick Observatory Supernova Survey Search \citep[LOSS;][]{Li11}, 21 Type II SNe from the Caltech Core-Collapse  Project \citep[CCCP;][]{Arcavi2012}, 50 Type II SNe from the Panoramic Survey Telescope and Rapid Response System \citep[PS1;][]{Sanders2015}, 34 Type IIP SNe from the Sloan Digital Sky Survey II Supernova Survey \citep[SDSS-II SNS;][]{Andrea2010, Taylor2014}. We also compare with the sample of 241 Type II SNe from the ZTF BTS used to analyze the rise-time properties of Type II SNe (K. Hinds et al., in prep.). In comparison with the LOSS sample, which is corrected for luminosity bias, we show the volume- and host-extinction corrected distribution in solid black and the volume-corrected but not host-extinction-corrected distribution in dashed violet. Our distribution without host extinction correction matches better with the LOSS luminosity function.  In the comparison with the PS1 and CCCP samples, the CLU sample in dashed green is not corrected for luminosity bias. We note that the PS1 $r$-band sample was part of a flux-limited survey and the observed distribution follows what one would typically expect without a volume correction.  The CCCP experiment was also a flux-limited SN experiment. The observed CCCP distribution falls close to the mean and the fainter end of the dispersion of the ZTF CLU sample without any volume correction. The SDSS Type IIP SN sample from \citet{Andrea2010}, which shows the $I$-band magnitude at 50 days post-explosion, does not have any SN fainter than $-16$ mag. The volume-corrected distribution of the peak $r$-band magnitude of the BTS Type II SN sample (K. Hinds et al., in prep.) in shaded green closely resembles the ZTF CLU distribution, with a higher fraction of fainter SNe in ZTF CLU sample without any volume correction (in orange), consistent with the experiment design.  %\textcolor{black}{TBD: replace with volume corrected BTS data}

We obtained a LLIIP fraction of $19^{+3}_{-4}\%$ of the host-extinction corrected sample of CLU Type II SNe. The fraction of LLII SNe is $\sim$ 45\% for the LOSS volume-corrected luminosity function without any host-extinction correction. The fraction of LLIIP in the raw PS1 and CCCP sample distribution is $\sim3\%$ and $\sim9\%$, respectively.

The volumetric rate for Type II SNe from the LOSS survey was $(4.4^{+1.4}_{-1.4}) \times 10^{4}\ \textrm{Gpc}^{-3} \textrm{yr}^{-1}$. Based on the LOSS survey statistics \citep{Li11}, Type IIP SNe are 70\% of all Type II SNe, thus the Type IIP SN rate is $\sim$ $(3.1^{+1.0}_{-1.0}) \times 10^{4}\  \textrm{Gpc}^{-3} \textrm{yr}^{-1}$, which is consistent with the volumetric rate measured with the CLU sample. The Bright Transient Survey \citep[BTS;][]{Perley2022} and the  Sloan Digital Sky Survey II Supernova Survey \citep[SDSS-II SNS;][]{Taylor2014} measured an overall core-collapse SN rate. BTS measured a CCSN rate of $(10.1^{+5.3}_{-3.5}) \times 10^{4}\ \textrm{Gpc}^{-3}\ \textrm{yr}^{-1}$, and SDSS-II SNS measured a CCSN rate of $(10.6^{+1.9}_{-1.9}) \times 10^{4}\ \textrm{Gpc}^{-3}\ \textrm{yr}^{-1}$. If we assume Type II SNe constitute 57\% of all CCSNe and Type IIP SNe are 70\% of all Type II SNe \citep{Li11}, the Type IIP SN rates in BTS and SDSS-II would be $(4.0^{+2.1}_{-1.4}) \times 10^{4}\ \textrm{Gpc}^{-3}\ \textrm{yr}^{-1}$ and $(4.2^{+0.8}_{-0.8}) \times 10^{4}\ \textrm{Gpc}^{-3}\ \textrm{yr}^{-1}$ respectively, which is consistent with the ZTF CLU volumetric rate.

%We note that unlike CLU and LOSS which probe $z<0.03$, the mean redshift of the Type II distribution in the SDSS and BTS CCSN sample is 0.072 and ? respectively.

%(exclude this sentence?) As a consistency check, we also run \texttt{skysurvey} with a Type Ia SN template, and the rate we recover is consistent to within 10\% with the SN Ia rate of $\sim2.5\times10^4\textrm{Gpc}^{-3} \textrm{yr}^{-1}$ reported in \citet{perley2022}. However, our Type IIP rate is much lower than that derived from the Bright Transient Survey ($\sim$5--get IIP fraction rates). 

\subsection{Can LLIIP SNe account for the fate of all $8-12$\ \Msun stars?}

The fraction of massive stars that explode in this mass range is given by:

\[
\frac{\int_{M_{\text{min}}}^{12} \psi(M) \, dM}
     {\int_{M_{\text{min}}}^{M_{\text{max}}} \psi(M) \, dM},
\]
where $M_{\text{min}}$ is the minimum progenitor mass that explodes, $M_{\text{max}}$ is the maximum progenitor mass of stars that explode, and $\psi(M) \propto M^{-2.35}$ for the Salpeter IMF.
If we assume $M_{\text{max}} = 40$ \msun\ and $M_{\text{min}} = 8$ \msun, $8-12$ \Msun\ progenitors should account for $48\%$ of all massive stars that collapse. If we assume $M_{\text{max}} = 100$ \msun\ and $M_{\text{min}} = 8$ \msun, $8-12$ \msun\ progenitors account for 44\% of all massive stars that explode.  However, even though LLIIP SNe are the main observational candidates for low-mass CCSNe, we find that LLIIP SNe constitutes only $19^{+3}_{-4}\%$ of all Type IIP SNe. If we assume that $40\%$ of all CCSNe are Type IIP SNe \citep{Li11}, LLIIP SNe are only $8^{+1}_{-2}\%$ of all CCSNe. 
The mean local star-formation rate density within $\sim$150 Mpc is $(17^{+7}_{-5}) \times 10^6\ \mathrm{M_\odot\ Gpc^{-3}\ yr^{-1}}$  \citep{Horiuchi2011}. This is also in good agreement with the local cosmic star-formation rate (SFR) of $(18.5 \pm 1.20) \times 10^6\ \mathrm{M_\odot\ Gpc^{-3}\ yr^{-1}}$ \citep{Hopkins2006}. Assuming a SFR to SN rate conversion factor of 0.0088 \citep{Horiuchi2011}, the local SN rate would be $\approx 14 \times \mathrm{10^{4}\ yr^{-1}\ Gpc^{-3}}$. %Assuming that Type IIP SNe are 40\% of all CCSNe \citep{Li11}, the local SN rate contribution from Type IIP SNe should be $\approx 5.6 \times \mathrm{10^{4}\ yr^{-1}\ Gpc^{-3}}$, which is a factor of $1.4^{+0.2}_{-0.1}$, higher than the ZTF CLU Type IIP SNe rate. 
Assuming a Salpeter IMF, the expected SN rate for $8-12$ \Msun\ is $\approx 6.7 \times 10^{4}\ \mathrm{yr^{-1}\ Gpc^{-3}}$, which is more than a factor of $9.2^{+0.8}_{-0.7}$ higher than the ZTF CLU LLIIP SN rate. How can this inconsistency be accounted for?

\textit{Observational bias:} A possible dominant effect is that many SNe are missed because they are too faint for current optical surveys. We can rule out that faint IIP SN ($>-14.5$ mag) account for the missing SNe population as they account for $<1.5\%$ of the entire Type IIP SNe population. However, we note that ZTF CLU is sensitive to only 25 Mpc for potential SNe fainter than $-12$ mag and up to 40 Mpc for potential SNe fainter than $-13$ mag. If Type IIP SNe alone were the fate of all progenitors in this mass range, we require that all Type IIP SN fainter than $\sim -18$ mag originate from this progenitor mass range. The ZAMS mass for the entire sample will be estimated in Papers II and III. The other scenario is that current optical transient surveys like CLU might be missing a significant fraction of low-mass explosions due to extinction  \citep[e.g., see][]{Jencson2019, Fox2021}. \citet{Mattila2012} estimate that locally $18.9^{+19.2}_{-9.5}\%$ of the CCSNe are missed by optical surveys. \citet{Jencson2019} predict an even higher fraction, that $38.5^{+26.0}_{-21.9}\%$ of the CCSNe are missed by optical surveys. \citet{Jencson2019} speculate that the objects SPIRITS16ix and SPIRITS16tn may represent a class of low-energy CCSNe arising preferentially in particularly extinguished environments. SPIRITS16tn is consistent with a LLIIP SN, heavily obscured by $A_V = 7-9$ mag. %If $\sim \textcolor{black}{xx}\%$ of the low-luminosity CCSNe are missed by CLU, the SN rates will be consistent with the predicted IMF and SFR rate. 
Another source of observational bias against LLIIP SNe could be the CLU galaxy catalog, if LLIIP SNe preferentially occurs in small galaxies.

\textit{Explode as other classes of transients:} We know that the majority ($\sim$70\%) of young massive stars live in interacting binary systems and the outer envelopes of $\sim$33\% of massive stars are stripped \citep[e.g., see][]{Sana2012}. Thus, it is possible that a significant fraction of $8-12$\ \Msun\ stars explode as stripped-envelope SNe in binary star systems. These stars with low-mass iron cores produce a low amount of Ni and ejecta mass  \citep{Stockinger2020, Sandoval2021, Burrows2024, Moriya2017, Sawada2022}. Candidates for low nickel mass stripped-envelope SNe include double-peaked Type Ibc SNe such as SN 2021inl \citep{Jacobson2022b, Das2024a}, rapidly evolving Type IIb SNe with a half-life of less than 10 days \citep[e.g.,][Fremling et al., in prep]{Ho2023, Das2023a}, ultra-stripped SNe such as SN 2023zaw \citep{Das2024b, Moore2024}. \citet{Moore2024} measured a volumetric rate of rapidly-evolving SESNe of $\approx (2.5^{+2.5}_{-1.4} \pm 0.9) \times 10^3\ 
 \textrm{Gpc}^{-3}\ \textrm{yr}^{-1}$, which is only $\approx 5\%$ of the Type IIP SN rate. A systematic study of the lowest-nickel mass stripped-envelope SNe in CLU, including their rates and the nickel mass distribution will be explored in a future work. Also, $\sim7-9$ \Msun\ sAGB stars are predicted to undergo electron-capture SNe \citep{Nomoto1984}, which could be the origin of the class of Intermediate Luminosity Red Transients (ILRTs) \citep[e.g., AT 2019abn, NGC 300 2008OT-1, SN 2008S;][]{Botticella2009, Adams2016, Jencson2019b, Rose2024, Valerin2025}. \textcolor{black}{Observations indicate that ILRTs comprise $1-10 \%$ of the overall CCSNe rate \citep{Thompson2009, Cai2021, Karambelkar2023}}.

\textit{Stellar evolution and explosion models:} The third explanation could be our limited theoretical understanding regarding the evolution and fate of $8-12$ \Msun\ progenitors. Stars in this mass range have a qualitatively different core structure than at $>12$ \Msun, with significantly lower compactness \citep{Sukhbold2016}. Their evolution is significantly more complex to model due to degeneracy effects. They develop thermal pulses that are numerically challenging to follow. Due to the steeply falling nature of the IMF, the lower mass limit $M_{\textrm{min}}$ strongly affects the fraction of massive stars that explode. For example, if $M_{\textrm{min}} = 9.7$ \Msun, then the fraction of massive stars $<12$ \Msun\ is $\sim25\%$. To match the observed rate of LLIIP SNe, we require a minimum progenitor mass of \( M_{\textrm{min}} = 11.3^{+0.6}_{-0.7} \, M_\odot \). The low rate of LLIIP SNe can be reasonably explained if a fraction of potential progenitors within this mass range either fail to undergo Fe-core collapse or do not successfully explode. This outcome is plausible if their cores fail to achieve the necessary physical conditions for core collapse. %This is seen in recent simulations of stars in this mass range \citep[e.g., see][]{Burrows2019, Burrows2024}. 
Another possibility in binary stellar evolution is stellar mergers. \citet{Sana2012} predict that more than half of the progenitors of Type II SNe are merged stars or binary mass gainers. Thus, if single stars in the $8-12$ \Msun\ merge, then we might instead observe the SN explosions of these progenitors that have a higher mass and undergo more luminous explosions \citep{Zapartas2019, Zapartas2021}.

It is likely that one or more of the above factors play a role in explaining the fate of all the lowest-mass stars that undergo core collapse.

%\subsection{Comparison to the star-formation rates}

%The CCSN distribution traces the formation of massive stars as the interval from massive star formation to core-collapse supernova (CCSN) is brief on astronomical timescale. The CCSNe rate should match the massive star formation rate using the known star formation rate density and initial mass function, given our current understanding of stellar evolution. \citet{Horiuchi2011} note an apparent discrepancy between the two absolute rates with the star formation implying a much higher CCSN rate than is observed. This disagreement could be caused by a very large misunderstanding of stellar evolution, a large change in the initial stellar mass function, or large population of dim or extincted CCSNe that evade detection as a function of redshift that preserves the observed redshift dependence. \citet{Botticella2012} conclude that the star formation rate derived from H$\alpha$ is too small by a factor of two, which would exacerbate further the discrepancy.

%very large misunderstanding of stellar evolution, a large change in the initial stellar mass function, extincted CCSNe that evade

%aim to correct for missing dim CC SNe, an SN luminosity function that is complete to dimmer CC SNe must be adopted. All catalog SNR should be considered lower limits since they are derived from simple counting of likely incomplete SN discoveries.

%Neither host nor Galactic extinction corrections have been applied to the previous catalogs.

\section{Summary}
\label{sec:conclusion}
%\textcolor{black}{In progress..}

In summary, we present the largest sample of 330 Type IIP SNe to derive the luminosity function to-date from a spectroscopic complete volume-limited SN survey. The sample is particularly critical to understand the low-luminosity population of CCSNe, with the sample of 36 LLIIP SNe ($M_{\textrm{r,peak}} \ge -16$ mag) tripling the sample of LLIIP SNe in the literature. The key takeaways from the analysis are:

\begin{enumerate}  

\item The luminosity function peaks at $-17$ mag and does not show a significant population $>-14.5$ mag, accounting for $<$ 1.5 \% of all Type IIP SNe. The fraction of LLIIP SNe is $19^{+3}_{-4}\%$ of the Type IIP SNe population and $8^{+1}_{-2}\%$ of the CCSN population (see Figure \ref{fig:lumfunc}).

\item We model the efficiency of the ZTF subtraction pipeline as a function of the alert flux and the ratio of background surface brightness to target flux. As expected, the recovery efficiency decreases as this ratio increases. The best-fit efficiency is $\sim50\%$ when the host surface brightness and target fluxes are comparable. Similarly, the best-fit efficiency decreases with an increase in target alert apparent magnitude and is $\approx$ 80\% when the target magnitude is 20 mag (see Figure \ref{fig:pipeline}).  

\item We derive a volumetric rate of  $(3.9_{-0.4}^{+0.4}) \times 10^{4}\ \textrm{Gpc}^{-3}\ \textrm{yr}^{-1}$ for Type IIP SNe and  $(7.3_{-0.6}^{+0.6})  \times 10^{3}\ \textrm{Gpc}^{-3}\ \textrm{yr}^{-1}$ for LLIIP SNe (see Figures \ref{fig:rates}, \ref{fig:cumrate}).

\item The expected SN rate for $8-12$ \Msun\ is $\approx 6.7 \times 10^{4}\ \mathrm{yr^{-1}\ Gpc^{-3}}$, which is more than a factor of $9.2^{+0.8}_{-0.7}$ higher than the ZTF CLU LLIIP SNe rate. While LLIIP SNe represent the explosions of the lowest massive stars that explode, they cannot account for all $8-12$ \Msun\ progenitors. %Now that the rate of LLIIP SNe is robustly derived, \textcolor{black}{this discrepancy must invoke dust-obscured SNe or fallback events.}

%\item We can rule out that there is a missing population of LLIIP SNe, and we need to invoke dust obscured transients or failed SNe to explain the low SN rates compared to the local star-formation rate. 

\end{enumerate}

The robust LLIIP rate measured in this work shows a significant discrepancy between the calculated and predicted volumetric rates of SNe in the low mass-end ($8-12$ \Msun) of core-collapse SNe. Future deep surveys such as the Legacy Survey of Space and Time \citep[LSST;][]{Ivezic08} in synergy with high-cadenced surveys of ZTF could reveal a hidden population of $\geq -13$ mag CCSNe. To enhance the efficiency of spectroscopic completeness for faint SNe, we will conduct the ZTF Complete Astronomical Transient Survey within 150 Mpc \citep[CATS150;][]{Das2025a}. CATS150 will utilize the high efficiency of the Next Generation Palomar Spectrograph (NGPS) on the Palomar 200-inch Hale Telescope to achieve spectroscopic completeness exceeding 95\% for even the faintest SNe. Also, SN survey in the local universe carried out at longer wavelengths by ground-based surveys such as the Wide-field Infrared Transient Explorer \citep[WINTER;][]{Lourie2020}, Dynamic REd All-sky Monitoring Survey \citep[DREAMS;][]{Soon2020}, Prime-focus Infrared Microlensing Experiment \citep[PRIME;][]{Kondo2023}, Cryoscope and space-based telescopes such as the \texttt{Nancy Grace Roman Space Telescope} will be critical in constraining the missing SN population due to dust obscuration.

%To further understand their progenitor properties, we will present the analysis of the lightcurve properties and nebular spectroscopy sample of the CLU LLIIP SNe sample in Paper II and Paper III, respectively.

\section{Data Availability}
All the photometric and spectroscopic data of the SNe in the sample will be made available on WISEREP and Zenodo after publication. The meta-data of the sample, extinction template, and pipeline recovery efficiency are available as a machine-readable tables on \href{https://zenodo.org/records/14538857?token=eyJhbGciOiJIUzUxMiJ9.eyJpZCI6ImFlZmFjOTY2LWU5MDMtNDU1My1iNjEwLWVmMDA0NTRjMDBmNSIsImRhdGEiOnt9LCJyYW5kb20iOiJmNGRjMWJiMTJhNmE5ODBmM2VkYzU4YTQ1OGE4NGM5MSJ9.j-VvClhhUEAr0JLwY5sprDydZw1OPGP-LfXjl5h9KRk1PMSCRhR5iM2geD3YgnKoCzkCD1aAFgewgSOAHkoB0A}{Zenodo}.

\section{Acknowldegement}

%KKD acknowledges support from the Schwartz Reisman Collaborative Science Program, which is supported by the Gerald Schwartz and Heather Reisman Foundation.

We thank the anonymous referee for their constructive feedback, which helped improve the quality of this manuscript. 

Based on observations obtained with the Samuel Oschin Telescope 48-inch and the 60-inch Telescope at the Palomar Observatory as part of the Zwicky Transient Facility project. ZTF is supported by the National Science Foundation under Grants No. AST-1440341 and AST-2034437 and a collaboration including current partners Caltech, IPAC, the Oskar Klein Center at Stockholm University, the University of Maryland, University of California, Berkeley , the University of Wisconsin at Milwaukee, University of Warwick, Ruhr University Bochum, Cornell University, Northwestern University and Drexel University. Operations are conducted by COO, IPAC, and UW.

%Based on observations obtained with the Samuel Oschin Telescope 48-inch and the 60-inch Telescope at the Palomar Observatory as part of the Zwicky Transient Facility project. ZTF is supported by the National Science Foundation under Grants No. AST-2034437 and a collaboration including current partners Caltech, IPAC, the Oskar Klein Center at Stockholm University, the University of Maryland, University of California, Berkeley , the University of Wisconsin at Milwaukee, University of Warwick, Ruhr University Bochum, Cornell University, Northwestern University and Drexel University. Operations are conducted by COO, IPAC, and UW.

Zwicky Transient Facility access for N.R., A.A.M., S.S., and C.L. was supported by Northwestern University and the Center for Interdisciplinary Exploration and Research in Astrophysics (CIERA).
N.R., C.L.,~and A.A.M.~are supported by DoE award \#DE-SC0025599.

S. Schulze is partially supported by LBNL Subcontract 7707915.

%Based on observations obtained with the Samuel Oschin Telescope 48-inch and the 60-inch Telescope at the Palomar Observatory as part of the Zwicky Transient Facility project. ZTF is supported by the National Science Foundation under Grant No. AST-2034437 and a collaboration including Caltech, IPAC, the Oskar Klein Center at Stockholm University, the University of Maryland, University of California, Berkeley, the University of Wisconsin at Milwaukee, University of Warwick, Ruhr University, Cornell University, Northwestern University and Drexel University. Operations are conducted by COO, IPAC, and UW.

SED Machine is based upon work supported by the National Science Foundation under
Grant No. 1106171. 

The ZTF forced-photometry service was funded under the Heising-Simons Foundation grant \#12540303 (PI: Graham).

The Gordon and Betty Moore Foundation, through both the Data-Driven Investigator Program and a dedicated grant, provided critical funding for SkyPortal .

This research has made use of the NASA/IPAC Extragalactic Database (NED), which is funded by the National Aeronautics and Space Administration and operated by the California Institute of Technology.

The Liverpool Telescope is operated on the island of La Palma by Liverpool John Moores University in the Spanish Observatorio del Roque de los Muchachos of the Instituto de Astrofisica de Canarias with financial support from the UK Science and Technology Facilities Council.

The W. M. Keck Observatory is operated as a scientific partnership among the California Institute of Technology, the University of California and the National Aeronautics and Space Administration. The Observatory was made possible by the generous financial support of the W. M. Keck Foundation. The authors wish to recognize and acknowledge the very significant cultural role and reverence that the summit of Maunakea has always had within the indigenous Hawaiian community.  We are most fortunate to have the opportunity to conduct observations from this mountain.

\bibliography{main}
\bibliographystyle{aasjournal}

\appendix

\section{Progenitor Mass from pre-explosion images Vs SN peak magnitude}
\label{sec:progenitors}
%\textcolor{black}{should keep this section?}
Here, we show the distribution of the Zero-Age Main-Sequence (ZAMS) mass estimated for Type II SNe from pre-explosion images. We can see that all LLIIP SNe have progenitor mass estimates of less than $<11$ \Msun while SNe brighter than $-16$ mag have more massive progenitors. The SNe shown here are: SN 2003gd \citep{Maund2009, Smartt2015}, SN 2005cs \citep{Maund2014b, Smartt2015}, SN 2009md \citep{Fraser2011, Smartt2015}, SN 2006my \citep{Maund2014b, Smartt2015}, SN 2012A \citep{Tomasella2013, Smartt2015}, SN 2013ej \citep{Fraser2014, Smartt2015}, SN 2004et \citep{Crockett2011, Smartt2015}, SN 2008bk \citep{Maund2014b, Smartt2015}, SN 2004A \citep{Maund2014b, Smartt2015}, SN 2012aw\citep{Kochanek2012, Smartt2015}, SN 2009hd \citep{Elias-Rosa2011, Smartt2015}, SN 2009kr \citep{Fraser2010, Smartt2015}, SN 2012ec \citep{Maund2013, Smartt2015}, SN 2018aoq \citep{ONeill2019}, SN 2022acko \citep{Dyk2023}. \textcolor{black}{A positive correlation between progenitor luminosity and $V-$band magnitude at 50 days since explosion is also reported in \citet{Rodriguez2022}.}

\begin{figure}[h!]
    \centering
    \includegraphics[width=\columnwidth]{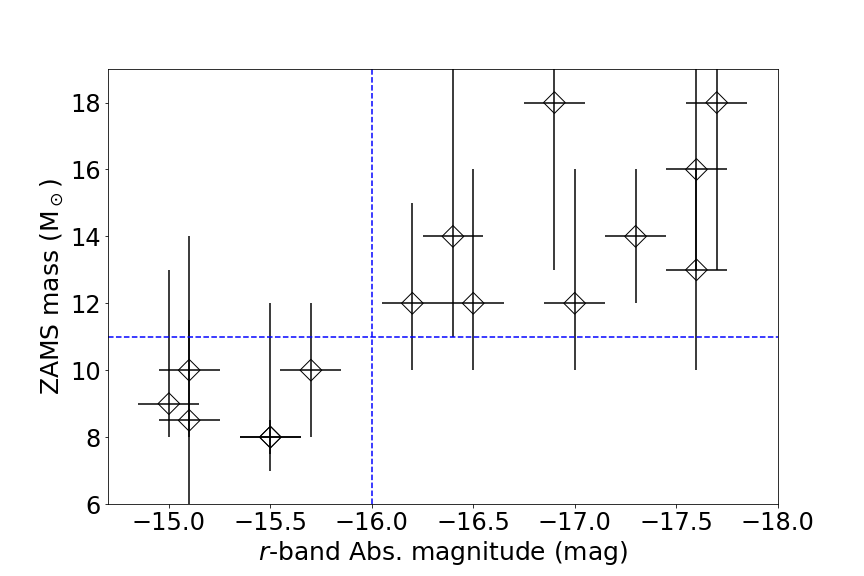}
    \caption{The distribution of the Zero-Age Main-Sequence (ZAMS) mass estimated for Type II SNe from pre-explosion images as a function of the peak $r$-band magnitude.}
    \label{fig:progenitors}
\end{figure}

\section{Distance correction for nearby galaxies}

\renewcommand{\thetable}{\Alph{table}}
\setcounter{table}{0}

Table \ref{tab:nearbydistances} lists the distance measurements used for galaxies closer than $25$ Mpc. We correct for the Virgo, Great Attractor, and Shapley supercluster infall based on the NASA Extragalactic Database object page \citep[NED\footnote{https://ned.ipac.caltech.edu/};][]{Helou1991}.

\begin{table}[h!]
\centering
\caption{Distance measurements used for galaxies $<25$ Mpc.}
\label{tab:nearbydistances}
\begin{threeparttable}
\begin{tabular}{@{}l@{}r@{}}
\hline
Name & Distance (Mpc) \\ \hline
%SN 2018hna/ZTF18acbwaxk$^{a}$  & 12.82  \\
SN 2021gmj/ZTF21aaqgmjt$^{a}$  & 13.10  \\
SN 2022acko/ZTF22abyivoq$^{b}$  & 23.40  \\
SN 2023axu/ZTF23aabngtm$^{c}$  & 13.68  \\
SN 2023hlf/ZTF23aaitpjv$^{d}$  & 9.54   \\
SN 2022aagp/ZTF22abtspsw$^{d}$  & 25.15  \\
SN 2023ijd/ZTF23aajrmfh$^{d}$  & 14.94  \\
SN 2022jzc/ZTF22aakdbia$^{d}$ & 19.64  \\
%ZTF22abtjefa$^{f}$ & 11.10 \\
\hline
\end{tabular}

\begin{tablenotes}
\item[] $^{a}$\citet{Valerin2022}; 
$^{b}$\citet{Dyk2023}; $^{c}$\citet{Shrestha2024}; $^{d}$\citet{Helou1991}. %; $^{f}$ Assuming a peculiar velocity of 300 $\textrm{km\ s^{-1}}$.
\end{tablenotes}

\end{threeparttable}
\end{table}

\section{ZTF Pipeline Recovery Efficiency Fits}
\label{sec:pipeline}

\begin{figure*}[h!]
    \centering
    \includegraphics[width=0.45\textwidth]{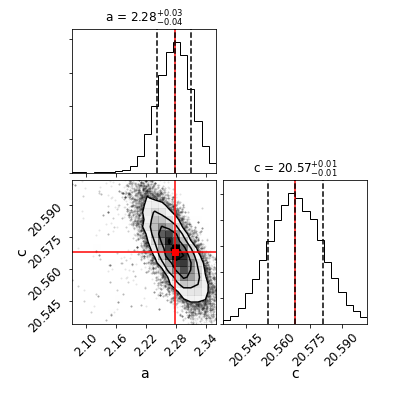}\includegraphics[width=0.45\textwidth]{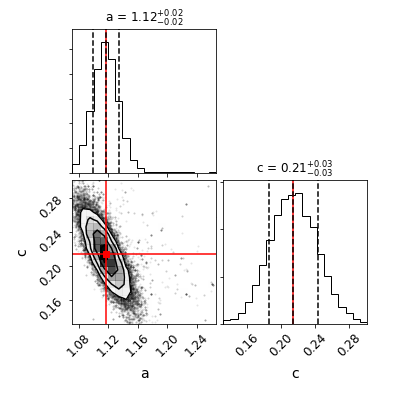}
    \caption{Corner plots for the logistic function fit to the pipeline efficiency as a function of apparent magnitude (\textit{left}) and the ratio of the alert apparent magnitude to the local surface brightness (\textit{right}).}
    \label{fig:corner}
\end{figure*}

The corner plot for the MCMC fit to 
\[
p(m) = \frac{1}{1 + \exp(a(m - c))},
\]
where m is the alert apparent magnitude is shown in Figure \ref{fig:corner}. The best-fit values are $a=2.28^{+0.03}_{-0.04}$ and $c=20.57^{+0.01}_{-0.01}$.

The corner plot for the MCMC fit to 
\[
p(r) = \frac{1}{1 + \exp(a(r - c))},
\]
where $r$ is the ratio of the alert apparent magnitude to the local surface brightness is shown in Figure \ref{fig:corner}. The best-fit values are $a=1.12^{+0.02}_{-0.02}$ and $c=0.21^{+0.03}_{-0.03}$.

\section{Sample of CLU Type II SNe}
\label{sec:appendix_sample}
The full version of the sample summary table is shown in Table \ref{tab:sample}.
\renewcommand{\thetable}{\Alph{table}}
\setcounter{table}{1}
\input{table_sample1_with_continued}
\clearpage

\end{document}

%% file: table_sample1_LLIIP.tex
\begin{table*}
\caption{Summary table of the CLU Type II SN sample. Only the first 45 rows are shown here. The full table is shown in the Appendix \ref{sec:appendix_sample}. 
The peak absolute magnitudes have been measured by assuming Milky Way extinction ($\mathrm{A_{V,MW}}$) and host galaxy extinction 
($\mathrm{A_{V,host}}$) as described in Section \ref{sec:extinction}. The $t_{\rm expl}$ column shows the explosion epoch. The Peak mag$_{\textrm{r}}$ shows the peak $r$-band absolute magnitude. A machine-readable version of the full sample table is available in \href{link}{DOI:10.5281/zenodo.14538857}. }
\label{tab:sample1}
\small
\begin{center}
\begin{tabular}{ccccccccc}
\hline
Name & RA & Dec & Redshift & $t_{\rm exp}$ & 1st detection & Peak mag$_{\textrm{r}}$ & $A_{V,MW}$ & $A_{V,\text{host}}$ \\
 & (hh:mm:ss) & (dd:mm:ss) & & (MJD) & (MJD) & (mag) & (mag) & (mag) \\
\hline

ZTF18abnxfve/AT2018lrz & 22:29:52.72 & +36:43:50.1 & 0.025 & 58347.8 & 58349.3 & -15.94 & 0.30 & 0.00 \\
ZTF18abrzbtb/SN2018ggu & 07:43:04.67 & +50:17:22.2 & 0.019 & 59120.0 & 59120.5 & -15.66 & 0.19 & 0.00 \\
ZTF19aaabzpt/SN2018lab & 06:16:26.51 & -21:22:32.4 & 0.009 & 58479.3 & 58487.3 & -15.81 & 0.24 & 0.71 \\
ZTF19aalycsv/AT2019txj & 10:48:39.19 & +76:48:05.3 & 0.023 & 58543.3 & 58546.2 & -15.27 & 0.07 & 0.00 \\
ZTF19aamwhat/SN2019bzd & 14:47:32.03 & -19:45:57.7 & 0.008 & 58559.7 & 58568.4 & -15.83 & 0.24 & 0.07 \\
ZTF19aatmadu/AT2019esn & 14:51:56.11 & +51:15:51.2 & 0.027 & 58602.4 & 58605.4 & -16.00 & 0.06 & 0.00 \\
ZTF19abalqkq/AT2019khq & 17:50:41.76 & +14:49:26.3 & 0.014 & 58645.9 & 58650.3 & -15.76 & 0.26 & 1.63 \\
ZTF19abctzkc/AT2019tti & 00:18:59.83 & +08:46:28.2 & 0.019 & 58644.4 & 58646.4 & -15.67 & 0.55 & 0.21 \\
ZTF19abejaiy/SN2019krp & 14:07:33.70 & +14:38:03.3 & 0.017 & 58670.2 & 58671.2 & -15.45 & 0.04 & 0.00 \\
ZTF19ablfoqa/AT2019tya & 02:09:38.03 & +01:33:06.7 & 0.032 & 58694.0 & 58694.5 & -15.93 & 0.08 & 0.00 \\
ZTF19abllxfy/AT2019ttl & 21:52:43.47 & +38:56:00.8 & 0.020 & 58682.9 & 58690.3 & -15.36 & 0.88 & 0.00 \\
ZTF19acalxgp/AT2019qiq & 23:44:56.09 & -04:16:33.1 & 0.029 & 58735.4 & 58737.3 & -15.86 & 0.12 & 0.00 \\
ZTF21aabfwwl/SN2021iy & 11:18:31.68 & -06:16:40.5 & 0.014 & 59217.9 & 59219.4 & -15.55 & 0.14 & 0.00 \\
ZTF21aaobkmg/SN2021eui & 19:20:55.80 & +43:07:14.6 & 0.015 & 59273.5 & 59276.5 & -15.34 & 0.28 & 0.00 \\
ZTF21aaqgmjt/SN2021gmj & 10:38:47.27 & +53:30:30.3 & 0.003 & 59292.3 & 59293.2 & -15.02 & 0.06 & 0.00 \\
ZTF21abbomrf/AT2020ghq & 14:45:20.59 & +38:44:18.5 & 0.015 & 59347.8 & 59349.2 & -15.72 & 0.03 & 0.00 \\
ZTF21abglcxm/SN2021qcs & 15:29:22.82 & -12:14:54.4 & 0.011 & 59377.3 & 59378.2 & -15.65 & 0.43 & 0.42 \\
ZTF21acdoyqt/SN2021zgm & 18:35:48.34 & +22:27:45.2 & 0.013 & 59479.2 & 59480.1 & -15.51 & 0.44 & 0.00 \\
ZTF22aaahubo/AT2022cru & 08:23:26.30 & -04:55:06.5 & 0.023 & 59575.4 & 59600.3 & -15.89 & 0.12 & 0.00 \\
ZTF22aakdbia/SN2022jzc & 12:05:28.66 & +50:31:36.8 & 0.002 & 59714.3 & 59715.2 & -14.91 & 0.05 & 0.57 \\
ZTF22aanrqje/SN2022mji & 09:42:54.06 & +31:51:03.6 & 0.004 & 59732.7 & 59741.2 & -15.00 & 0.05 & 0.85 \\
ZTF22aaywnyg/SN2022pru & 11:59:07.65 & +52:41:58.5 & 0.004 & 59787.7 & 59797.2 & -15.42 & 0.07 & 0.00 \\
ZTF22aazmrpx/SN2022raj & 02:03:17.52 & +29:14:04.9 & 0.012 & 59798.4 & 59800.4 & -15.20 & 0.15 & 0.00 \\
ZTF22abssiet/SN2022zmb & 10:38:43.18 & +56:33:14.4 & 0.014 & 59885.5 & 59887.5 & -15.72 & 0.02 & 0.07 \\
ZTF22abvaetz/SN2022aang & 07:59:21.83 & +18:06:40.9 & 0.016 & 59894.5 & 59901.5 & -15.44 & 0.08 & 0.00 \\
ZTF22abyivoq/SN2022acko & 03:19:38.98 & -19:23:42.8 & 0.006 & 59917.8 & 59922.2 & -15.83 & 0.08 & 0.00 \\
ZTF23aaarmtb/SN2023qh & 09:07:15.44 & +37:12:54.8 & 0.024 & 59947.4 & 59957.4 & -15.89 & 0.06 & 0.00 \\
ZTF23aackjhs/SN1995al & 09:50:56.03 & +33:33:11.0 & 0.005 & 59989.8 & 59992.3 & -14.88 & 0.04 & 0.07 \\
ZTF23aagkajy/AT2023gdt & 10:30:10.64 & +43:21:23.4 & 0.014 & 60050.2 & 60051.2 & -15.42 & 0.03 & 0.71 \\
ZTF23aajrmfh/SN2023ijd & 12:36:32.47 & +11:13:19.7 & 0.007 & 60078.2 & 60079.0 & -15.39 & 0.09 & 0.00 \\
ZTF23aamzlzc/AT2023kne & 17:25:19.11 & +58:49:02.6 & 0.028 & 60095.4 & 60097.3 & -15.81 & 0.10 & 0.00 \\
ZTF23aaqwpio/AT2023nca & 16:39:26.36 & +11:12:45.3 & 0.023 & 60129.3 & 60135.2 & -15.61 & 0.14 & 0.00 \\
ZTF23aasyvbf/SN2023nmh & 00:37:38.73 & -04:16:53.2 & 0.020 & 60142.4 & 60144.5 & -15.93 & 0.10 & 0.00 \\
ZTF23abnogui/SN2023wcr & 12:23:31.29 & +74:57:01.3 & 0.005 & 60240.5 & 60244.5 & -15.58 & 0.09 & 0.00 \\
ZTF23abvommm/SN2023acbr & 02:27:03.18 & -09:25:02.3 & 0.016 & 60300.7 & 60305.2 & -15.57 & 0.08 & 0.00 \\
ZTF24aabppgn/SN2024wp & 11:24:39.25 & +14:56:52.9 & 0.014 & 60320.5 & 60325.5 & -15.52 & 0.10 & 0.00 \\
ZTF24aahwfsa/AT2024fas & 10:51:55.36 & +37:35:23.9 & 0.026 & 60393.4 & 60396.4 & -15.80 & 0.04 & 0.00 \\
ZTF24aaplfjd/SN2024jxm & 0:58:01.36 & +30:42:23.84 & 0.016 & 60460.0 & 60460.5 & -16.00 & 0.18 & 0.00 \\

\hline
\end{tabular}
\end{center}
\end{table*}

%% file: table_sample1_with_continued.tex
\begin{table*}
\caption{Summary table of the CLU Type II SN sample. The peak absolute magnitudes have been measured by assuming Milky Way extinction ($\mathrm{A_{V,MW}}$) and host galaxy extinction 
($\mathrm{A_{V,host}}$) as described in Section \ref{sec:extinction}. The $t_{\rm expl}$ column shows the explosion epoch. The Peak mag$_{\textrm{r}}$ shows the peak $r$-band absolute magnitude. A machine-readable version of the full sample table is available in \href{link}{DOI:10.5281/zenodo.14538857}. }
\label{tab:sample}

\small
\begin{center}
\begin{tabular}{ccccccccc}
\hline
Name & RA & Dec & Redshift & $t_{\rm expl}$ & 1st detection & Peak mag$_{\textrm{r}}$ & $A_{V,MW}$ & $A_{V,\text{host}}$ \\
 & (hh:mm:ss) & (dd:mm:ss) & & (MJD) & (MJD) & (mag) & (mag) & (mag) \\
\hline

ZTF18aaegvyd/SN2019env & 09:09:35.63 & +29:58:14.4 & 0.024 & 58606.7 & 58608.2 & -17.03 & 0.07 & 0.00 \\
ZTF18aaqkywr/SN2022giv & 12:23:14.92 & +51:11:28.3 & 0.031 & 59655.9 & 59670.4 & -16.77 & 0.05 & 0.00 \\
ZTF18aaqpowv/SN2024iwm & 11:28:17.83 & +26:58:26.37 & 0.032 & 60443.2 & 60444.2 & -17.42 & 0.05 & 0.00 \\
ZTF18aaszvfn/SN2021iaw & 14:23:07.88 & +50:13:14.9 & 0.027 & 59306.4 & 59307.3 & -16.77 & 0.06 & 0.00 \\
ZTF18aawgrxz/SN2021lmp & 15:04:17.64 & +48:55:10.7 & 0.027 & 59338.9 & 59339.3 & -16.85 & 0.04 & 0.42 \\
ZTF18aaxzlmy/SN2018lrq & 13:34:51.37 & +34:03:20.5 & 0.025 & 58218.8 & 58219.3 & -16.62 & 0.03 & 0.99 \\
ZTF18abbpxik/SN2018cqp & 15:22:50.32 & +08:04:49.1 & 0.032 & 58282.7 & 58283.2 & -18.08 & 0.10 & 0.71 \\
ZTF18abbpyvk/SN2018cqi & 13:57:40.91 & +17:30:29.1 & 0.022 & 58284.7 & 58286.2 & -18.29 & 0.09 & 0.07 \\
ZTF18abcpmwh/SN2018cur & 12:59:09.12 & +37:19:00.1 & 0.015 & 58289.7 & 58291.2 & -17.56 & 0.04 & 0.00 \\
ZTF18abdbysy/SN2018cyg & 15:34:08.48 & +56:41:48.6 & 0.011 & 58294.8 & 58295.2 & -17.08 & 0.05 & 2.05 \\
ZTF18abdfwcy/SN2018cwa & 12:42:43.85 & +34:06:26.3 & 0.014 & 58290.8 & 58295.2 & -17.27 & 0.05 & 0.00 \\
ZTF18abdjmfh/SN2018dct & 16:15:08.80 & +26:33:36.2 & 0.030 & 58298.8 & 58300.3 & -18.72 & 0.12 & 1.27 \\
ZTF18abjkryl/SN2018dzc & 18:23:09.26 & +21:14:56.4 & 0.010 & 58308.4 & 58308.4 & -16.30 & 0.48 & 0.49 \\
ZTF18abjndhh/SN2018ecj & 11:32:43.02 & +62:25:57.9 & 0.013 & 58322.2 & 58324.2 & -17.03 & 0.03 & 1.34 \\
ZTF18abjovdz/SN2018dht & 20:55:19.05 & +00:32:18.6 & 0.024 & 58309.9 & 58312.4 & -18.36 & 0.23 & 1.48 \\
ZTF18abklbam/SN2018elp & 14:31:19.10 & +21:17:24.7 & 0.031 & 58329.2 & 58331.2 & -18.00 & 0.13 & 0.00 \\
ZTF18abmdpwe/SN2018evy & 18:22:38.17 & +15:41:47.6 & 0.018 & 58339.2 & 58340.2 & -18.06 & 0.54 & 0.00 \\
ZTF18abnxfve/SN2018lrz & 22:29:52.72 & +36:43:50.1 & 0.025 & 58347.8 & 58349.3 & -15.94 & 0.30 & 0.00 \\
ZTF18abokyfk/SN2018fif & 00:09:26.55 & +47:21:14.7 & 0.017 & 58350.8 & 58351.4 & -17.12 & 0.30 & 0.00 \\
ZTF18abrpkdw/SN2018lsa & 21:45:07.20 & +15:35:09.0 & 0.031 & 58355.8 & 58357.2 & -16.49 & 0.29 & 0.00 \\
ZTF18abrzbtb/SN2018ggu & 07:43:04.67 & +50:17:22.2 & 0.019 & 59120.0 & 59120.5 & -15.66 & 0.19 & 0.00 \\
ZTF18abvmlow/SN2018lcd & 01:51:10.65 & -03:29:24.9 & 0.017 & 58479.2 & 58483.2 & -18.02 & 0.08 & 0.00 \\
ZTF18abvvmdf/SN2018gts & 16:36:47.39 & +55:44:08.8 & 0.030 & 58373.2 & 58374.2 & -18.02 & 0.05 & 0.92 \\
ZTF18abwisaz/SN2024kan & 22:49:07.02 & +32:47:53.12 & 0.028 & 60454.5 & 60456.4 & -17.45 & 0.36 & 0.00 \\
ZTF18abxbmcl/SN2018hcp & 08:22:57.66 & +27:42:10.5 & 0.020 & 58377.0 & 58378.5 & -19.19 & 0.10 & 2.05 \\
ZTF18abzrgim/SN2018gvt & 23:07:32.54 & +23:00:20.9 & 0.021 & 58386.8 & 58388.3 & -17.49 & 0.63 & 0.00 \\
ZTF18acbvhit/SN2018hle & 03:39:28.11 & -13:07:02.4 & 0.014 & 58418.7 & 58422.4 & -16.80 & 0.15 & 0.21 \\
ZTF18aceqrcy/SN2018joy & 10:19:40.32 & +12:50:37.7 & 0.032 & 58428.0 & 58429.5 & -17.77 & 0.18 & 0.99 \\
ZTF18acrflch/SN2018jgy & 03:06:59.88 & -06:45:14.0 & 0.029 & 58445.8 & 58448.3 & -17.38 & 0.23 & 0.00 \\
ZTF18acrkaks/SN2018jwq & 05:17:54.17 & +08:54:51.3 & 0.030 & 58438.4 & 58439.4 & -16.51 & 0.93 & 0.00 \\
ZTF18acrtvmm/SN2018jfp & 03:17:56.27 & -00:10:10.7 & 0.023 & 58448.8 & 58450.3 & -17.55 & 0.20 & 0.00 \\
ZTF18acsxkov/SN2018kds & 00:10:09.06 & -04:42:33.2 & 0.030 & 58449.7 & 58456.2 & -16.68 & 0.09 & 0.00 \\
ZTF18acurdnh/SN2018jvr & 10:07:20.35 & +04:04:49.2 & 0.029 & 58458.5 & 58461.5 & -16.38 & 0.05 & 0.00 \\
ZTF18acurqaw/SN2018hwm & 08:21:28.20 & +03:09:52.3 & 0.009 & 58432.5 & 58441.4 & -17.05 & 0.07 & 3.39 \\
ZTF18acyybvg/SN2018kkv & 11:34:33.85 & +22:31:09.2 & 0.032 & 58470.5 & 58471.5 & -16.69 & 0.06 & 0.00 \\
ZTF18adachwf/SN2018lev & 15:13:08.16 & +41:15:49.2 & 0.029 & 58472.2 & 58475.5 & -16.94 & 0.06 & 0.14 \\
ZTF18adazblo/SN2018ldu & 09:09:32.49 & +20:24:25.1 & 0.027 & 58476.4 & 58480.4 & -16.67 & 0.11 & 0.00 \\
ZTF18adbclkd/SN2018kpo & 03:40:43.06 & -06:25:23.9 & 0.017 & 58478.3 & 58481.3 & -17.28 & 0.15 & 0.00 \\
ZTF19aaabzpt/SN2018lab & 06:16:26.51 & -21:22:32.4 & 0.009 & 58479.3 & 58487.3 & -15.81 & 0.24 & 0.71 \\
ZTF19aadnxnl/SN2019va & 13:35:14.68 & +44:45:58.6 & 0.009 & 58498.5 & 58502.5 & -16.87 & 0.05 & 0.00 \\
ZTF19aadnxog/SN2019vb & 13:14:23.07 & +30:29:06.6 & 0.020 & 58498.5 & 58503.5 & -18.04 & 0.04 & 0.00 \\
ZTF19aailepg/SN2019amt & 11:17:52.25 & +30:09:23.5 & 0.030 & 58518.4 & 58522.4 & -17.11 & 0.05 & 0.00 \\
ZTF19aajwkbb/SN2019bsw & 10:05:06.10 & -16:24:21.3 & 0.027 & 58502.5 & 58511.4 & -16.75 & 0.14 & 0.00 \\
ZTF19aaklqoi/SN2021adnr & 08:57:54.94 & +20:07:12.6 & 0.031 & 59514.0 & 59518.5 & -17.27 & 0.08 & 1.20 \\
ZTF19aakpxfm/SN2019aor & 07:24:57.85 & -27:31:53.7 & 0.025 & 58476.4 & 58476.4 & -17.94 & 0.83 & 0.00 \\
ZTF19aalycsv/SN2019txj & 10:48:39.19 & +76:48:05.3 & 0.023 & 58543.3 & 58546.2 & -15.27 & 0.07 & 0.00 \\
ZTF19aamftfu/SN2019cag & 02:23:21.52 & +53:23:41.3 & 0.025 & 58551.6 & 58556.1 & -17.30 & 0.53 & 0.00 \\

\hline
\end{tabular}
\end{center}
\end{table*}

\begin{table*}
\ContinuedFloat
\caption{Continued.}
\small
\begin{center}
\begin{tabular}{ccccccccc}
\hline
Name & RA & Dec & Redshift & $t_{\rm expl}$ & 1st detection & Peak mag$_{\textrm{r}}$ & $A_{V,MW}$ & $A_{V,\text{host}}$ \\
 & (hh:mm:ss) & (dd:mm:ss) & & (MJD) & (MJD) & (mag) & (mag) & (mag) \\
\hline

ZTF19aamwhat/SN2019bzd & 14:47:32.03 & -19:45:57.7 & 0.008 & 58559.7 & 58568.4 & -15.83 & 0.24 & 0.07 \\
ZTF19aanfnvl/SN2019crk & 10:10:31.69 & +10:02:37.0 & 0.032 & 58561.7 & 58567.2 & -17.77 & 0.12 & 0.00 \\
ZTF19aanhhal/SN2019cec & 13:41:40.74 & +55:40:10.7 & 0.026 & 58561.9 & 58562.3 & -17.23 & 0.03 & 0.00 \\
ZTF19aaniore/SN2019ceg & 16:27:51.64 & +62:41:32.5 & 0.030 & 58564.5 & 58567.5 & -17.31 & 0.09 & 0.00 \\
ZTF19aanqzle/SN2019cmm & 11:18:07.29 & +75:08:50.8 & 0.023 & 58570.2 & 58572.2 & -16.03 & 0.16 & 0.00 \\
ZTF19aanrrqu/SN2019clp & 12:13:39.83 & +16:07:24.4 & 0.024 & 58569.8 & 58572.3 & -18.02 & 0.10 & 0.00 \\
ZTF19aapafit/SN2019cvz & 16:30:54.08 & +46:35:18.4 & 0.019 & 58576.5 & 58577.4 & -17.05 & 0.04 & 0.00 \\
ZTF19aaqxosb/SN2019dok & 13:46:51.76 & +16:16:53.7 & 0.019 & 58588.3 & 58589.2 & -16.64 & 0.07 & 0.00 \\
ZTF19aarjfqe/SN2019dvd & 12:31:06.89 & +00:27:54.6 & 0.021 & 58592.8 & 58596.3 & -16.93 & 0.06 & 0.00 \\
ZTF19aarykkb/SN2019dzk & 17:13:21.95 & -09:57:52.1 & 0.024 & 58595.4 & 58598.3 & -18.17 & 1.54 & 0.00 \\
ZTF19aatmadu/SN2019esn & 14:51:56.11 & +51:15:51.2 & 0.027 & 58602.4 & 58605.4 & -16.00 & 0.06 & 0.00 \\
ZTF19aauishy/SN2019evl & 13:31:01.26 & +34:09:12.5 & 0.023 & 58609.3 & 58612.2 & -16.45 & 0.03 & 0.00 \\
ZTF19aavrcew/SN2019fyw & 13:07:18.44 & +02:00:11.5 & 0.019 & 58625.2 & 58633.2 & -17.78 & 0.08 & 0.00 \\
ZTF19aavtcjs/SN2019gss & 15:24:44.44 & +68:43:44.3 & 0.022 & 58630.3 & 58633.3 & -17.89 & 0.07 & 2.19 \\
ZTF19aawgxdn/SN2019gmh & 16:31:03.16 & +41:09:14.2 & 0.031 & 58634.3 & 58635.2 & -17.39 & 0.03 & 0.00 \\
ZTF19aaykqsk/SN2019hci & 11:16:03.35 & -00:31:55.9 & 0.026 & 58641.7 & 58643.2 & -17.02 & 0.17 & 0.00 \\
ZTF19aayrosj/SN2019hrb & 20:54:12.08 & +10:33:01.1 & 0.015 & 58642.5 & 58644.5 & -17.64 & 0.26 & 1.70 \\
ZTF19aazudta/SN2019hqm & 17:36:56.18 & +21:06:17.1 & 0.024 & 58648.8 & 58650.3 & -16.98 & 0.23 & 0.00 \\
ZTF19aazyvub/SN2019hnl & 23:43:10.25 & -02:56:58.6 & 0.023 & 58649.5 & 58651.5 & -17.13 & 0.09 & 0.00 \\
ZTF19abaamsd/SN2019ifm & 17:23:41.33 & +52:00:36.1 & 0.024 & 58652.8 & 58653.3 & -17.33 & 0.09 & 0.00 \\
ZTF19abajxet/SN2019hyk & 14:17:57.85 & +26:25:17.6 & 0.015 & 58655.7 & 58657.2 & -17.77 & 0.06 & 0.00 \\
ZTF19abalqkq/SN2019khq & 17:50:41.76 & +14:49:26.3 & 0.014 & 58645.9 & 58650.3 & -15.76 & 0.26 & 1.63 \\
ZTF19abbnamr/SN2019iex & 23:51:03.63 & +20:08:43.7 & 0.014 & 58659.0 & 58660.5 & -17.40 & 0.20 & 0.00 \\
ZTF19abbwfgp/SN2019ikb & 17:13:17.71 & +43:47:03.5 & 0.026 & 58660.8 & 58661.3 & -18.16 & 0.05 & 0.00 \\
ZTF19abctzkc/SN2019tti & 00:18:59.83 & +08:46:28.2 & 0.019 & 58644.4 & 58646.4 & -15.67 & 0.55 & 0.21 \\
ZTF19abejaiy/SN2019krp & 14:07:33.70 & +14:38:03.3 & 0.017 & 58670.2 & 58671.2 & -15.45 & 0.04 & 0.00 \\
ZTF19abjrjdw/SN2019mkr & 17:11:05.78 & +05:51:07.3 & 0.022 & 58687.7 & 58689.2 & -17.56 & 0.43 & 0.00 \\
ZTF19abjsmmv/SN2019mor & 15:38:38.04 & +36:57:31.0 & 0.019 & 58693.3 & 58694.2 & -17.13 & 0.06 & 0.00 \\
ZTF19ablfoqa/SN2019tya & 02:09:38.03 & +01:33:06.7 & 0.032 & 58694.0 & 58694.5 & -15.93 & 0.08 & 0.00 \\
ZTF19abllxfy/SN2019ttl & 21:52:43.47 & +38:56:00.8 & 0.020 & 58682.9 & 58690.3 & -15.36 & 0.88 & 0.00 \\
ZTF19abpyqog/SN2019oba & 19:57:03.47 & +50:11:20.1 & 0.031 & 58708.3 & 58711.2 & -17.22 & 0.40 & 0.00 \\
ZTF19abqhobb/SN2019nvm & 17:25:38.66 & +59:26:48.2 & 0.018 & 58713.7 & 58714.2 & -17.59 & 0.08 & 0.00 \\
ZTF19abqrhvt/SN2019nyk & 00:15:15.20 & -08:11:21.8 & 0.021 & 58713.4 & 58715.4 & -18.09 & 0.10 & 0.00 \\
ZTF19abqrhvy/SN2019odf & 22:48:44.69 & +27:34:18.5 & 0.032 & 58714.9 & 58715.4 & -18.18 & 0.14 & 0.00 \\
ZTF19abrnjwi/SN2019omb & 00:12:39.66 & +05:30:32.0 & 0.028 & 58717.9 & 58719.4 & -17.20 & 0.07 & 0.00 \\
ZTF19abwztsb/SN2019pjs & 18:04:40.34 & +21:38:04.2 & 0.007 & 58731.7 & 58734.2 & -16.23 & 0.29 & 0.00 \\
ZTF19abzmoov/SN2019qba & 15:37:44.70 & +22:25:36.4 & 0.025 & 58735.6 & 58737.1 & -16.68 & 0.12 & 0.00 \\
ZTF19acalxgp/SN2019qiq & 23:44:56.09 & -04:16:33.1 & 0.029 & 58735.4 & 58737.3 & -15.86 & 0.12 & 0.00 \\
ZTF19acblhxc/SN2019rho & 02:12:49.18 & -06:42:05.9 & 0.017 & 58753.4 & 58756.4 & -17.21 & 0.07 & 0.21 \\
ZTF19acbpqlh/SN2019rpn & 21:19:41.19 & +37:31:19.3 & 0.026 & 58709.8 & 58711.2 & -16.28 & 0.46 & 0.00 \\
ZTF19acbwejj/SN2019upq & 14:29:08.13 & +27:27:00.6 & 0.014 & 58754.1 & 58758.1 & -17.80 & 0.05 & 0.00 \\
ZTF19acewuwn/SN2019ssl & 23:20:17.26 & +35:29:35.3 & 0.027 & 58771.7 & 58772.2 & -16.96 & 0.27 & 0.00 \\
ZTF19acftfav/SN2019ssi & 23:30:56.12 & +15:29:29.8 & 0.013 & 58773.7 & 58774.2 & -16.51 & 0.19 & 0.00 \\
ZTF19acgbkzr/SN2019szo & 00:19:56.64 & +15:05:36.1 & 0.026 & 58774.8 & 58775.3 & -16.73 & 0.13 & 0.00 \\
ZTF19acgzwbm/SN2019tba & 04:55:09.84 & -16:09:01.3 & 0.020 & 58770.5 & 58776.5 & -16.34 & 0.18 & 0.00 \\
ZTF19aclobbu/SN2019twk & 02:23:05.28 & +46:52:56.7 & 0.018 & 58787.3 & 58788.3 & -17.63 & 0.49 & 0.00 \\

\hline
\end{tabular}
\end{center}
\end{table*}

\begin{table*}
\ContinuedFloat
\caption{Continued.}
\small
\begin{center}
\begin{tabular}{ccccccccc}
\hline
Name & RA & Dec & Redshift & $t_{\rm expl}$ & 1st detection & Peak mag$_{\textrm{r}}$ & $A_{V,MW}$ & $A_{V,\text{host}}$ \\
 & (hh:mm:ss) & (dd:mm:ss) & & (MJD) & (MJD) & (mag) & (mag) & (mag) \\
\hline

ZTF19acnphay/SN2019ubr & 06:25:52.32 & +64:44:38.4 & 0.014 & 58769.5 & 58772.5 & -16.57 & 0.32 & 2.40 \\
ZTF19acrcxri/SN2019ult & 23:58:47.97 & +14:44:31.3 & 0.027 & 58795.1 & 58797.1 & -18.42 & 0.11 & 0.00 \\
ZTF19acryurj/SN2019ust & 00:54:22.41 & +31:40:12.6 & 0.022 & 58798.2 & 58800.3 & -18.10 & 0.17 & 0.00 \\
ZTF19actnwtn/SN2019vdl & 09:29:31.80 & +44:25:20.0 & 0.025 & 58803.5 & 58804.5 & -17.35 & 0.06 & 0.57 \\
ZTF19actnyae/SN2019vdm & 11:26:24.51 & +22:37:11.0 & 0.032 & 58804.0 & 58805.5 & -16.88 & 0.04 & 0.00 \\
ZTF19acvtrxj/SN2019vjl & 09:49:23.69 & +01:08:46.6 & 0.025 & 58808.5 & 58812.6 & -17.58 & 0.23 & 0.07 \\
ZTF19acwrrvg/SN2019vsr & 02:01:57.73 & +44:48:32.1 & 0.027 & 58816.7 & 58819.2 & -18.66 & 0.21 & 0.00 \\
ZTF19acxgwvo/SN2019wbd & 23:20:35.17 & -00:52:51.0 & 0.015 & 58815.7 & 58820.2 & -17.20 & 0.10 & 0.35 \\
ZTF19acyjjni/SN2019xbm & 13:07:13.99 & +58:08:03.3 & 0.028 & 58823.0 & 58831.6 & -16.94 & 0.04 & 0.42 \\
ZTF19acykzsk/SN2019wqj & 02:11:37.09 & +34:02:28.7 & 0.021 & 58823.2 & 58827.2 & -16.31 & 0.27 & 0.07 \\
ZTF19acytcsg/SN2019wvz & 10:20:28.67 & +50:28:04.5 & 0.032 & 58832.5 & 58833.4 & -17.60 & 0.03 & 0.00 \\
ZTF19adakmbh/SN2019xgi & 21:55:25.49 & +34:30:37.8 & 0.018 & 58834.1 & 58837.1 & -17.90 & 0.46 & 1.63 \\
ZTF20aarenrz/SN2021qyy & 11:44:29.64 & +69:43:45.4 & 0.009 & 59388.8 & 59390.2 & -16.55 & 0.03 & 0.00 \\
ZTF21aaabwem/SN2020aeqx & 13:13:00.02 & +06:10:12.1 & 0.032 & 59211.0 & 59215.5 & -16.79 & 0.10 & 0.00 \\
ZTF21aaagypx/SN2021V & 11:13:07.92 & +05:04:19.3 & 0.027 & 59213.4 & 59216.4 & -17.43 & 0.20 & 0.00 \\
ZTF21aabfwwl/SN2021iy & 11:18:31.68 & -06:16:40.5 & 0.014 & 59217.9 & 59219.4 & -15.55 & 0.14 & 0.00 \\
ZTF21aabygea/SN2021os & 12:02:54.08 & +05:36:53.1 & 0.019 & 59219.5 & 59221.4 & -17.28 & 0.05 & 0.00 \\
ZTF21aaeoqxf/SN2021aek & 11:59:48.02 & -21:23:13.0 & 0.022 & 59225.5 & 59227.5 & -17.61 & 0.13 & 0.00 \\
ZTF21aaeqwov/SN2021htp & 07:43:04.75 & +50:17:19.4 & 0.019 & 59119.0 & 59119.5 & -16.39 & 0.19 & 0.00 \\
ZTF21aafepon/SN2021ass & 01:50:10.12 & +27:38:42.7 & 0.012 & 59230.6 & 59231.1 & -16.24 & 0.23 & 0.07 \\
ZTF21aafkwtk/SN2021apg & 13:41:19.24 & +24:29:43.9 & 0.027 & 59229.5 & 59231.4 & -16.98 & 0.03 & 0.00 \\
ZTF21aagtqna/SN2021brb & 18:05:15.31 & +46:52:56.0 & 0.023 & 59238.5 & 59248.5 & -16.84 & 0.11 & 0.00 \\
ZTF21aahgspm/SN2021cah & 02:48:30.72 & +50:45:36.0 & 0.016 & 59250.7 & 59251.1 & -17.43 & 0.98 & 0.00 \\
ZTF21aaipypa/SN2021cgu & 11:03:03.71 & +05:05:53.4 & 0.025 & 59252.4 & 59253.4 & -18.34 & 0.10 & 0.00 \\
ZTF21aakvroo/SN2021cwe & 15:45:31.20 & +30:09:43.3 & 0.032 & 59257.0 & 59258.5 & -17.11 & 0.09 & 0.00 \\
ZTF21aaluqkp/SN2021dhx & 11:05:10.37 & -15:21:10.1 & 0.025 & 59263.3 & 59264.3 & -16.99 & 0.14 & 0.00 \\
ZTF21aalxurx/SN2021dqs & 13:44:05.74 & +43:04:18.0 & 0.027 & 59259.9 & 59264.4 & -16.33 & 0.03 & 0.00 \\
ZTF21aanjvng/SN2021dvk & 08:07:38.41 & +08:56:24.2 & 0.030 & 59269.3 & 59271.2 & -16.65 & 0.07 & 0.00 \\
ZTF21aantsla/SN2021ech & 12:06:20.42 & +37:00:47.1 & 0.021 & 59274.3 & 59275.2 & -16.44 & 0.06 & 0.00 \\
ZTF21aanzcuj/SN2021enz & 12:07:05.23 & +42:59:18.3 & 0.024 & 59275.9 & 59276.3 & -16.69 & 0.04 & 0.00 \\
ZTF21aaobkmg/SN2021eui & 19:20:55.80 & +43:07:14.6 & 0.015 & 59273.5 & 59276.5 & -15.34 & 0.28 & 0.00 \\
ZTF21aapkcmr/SN2021fnj & 14:23:42.67 & +28:20:45.7 & 0.030 & 59285.4 & 59290.4 & -17.91 & 0.05 & 0.42 \\
ZTF21aapliyn/SN2021foj & 13:45:26.60 & +47:55:05.3 & 0.028 & 59286.3 & 59291.3 & -17.11 & 0.08 & 0.00 \\
ZTF21aaqgmjt/SN2021gmj & 10:38:47.27 & +53:30:30.3 & 0.003 & 59292.3 & 59293.2 & -15.02 & 0.06 & 0.00 \\
ZTF21aaqjmps/SN2021gvm & 13:30:01.22 & +13:24:40.2 & 0.025 & 59292.8 & 59294.3 & -18.10 & 0.07 & 0.00 \\
ZTF21aaqldsj/SN2021hac & 14:12:30.45 & +34:33:00.3 & 0.032 & 59293.3 & 59294.3 & -16.95 & 0.04 & 0.00 \\
ZTF21aaqugxm/SN2021hdt & 11:34:45.73 & +42:57:55.3 & 0.019 & 59299.4 & 59300.3 & -18.12 & 0.07 & 0.00 \\
ZTF21aaqyifh/SN2021hqe & 09:41:30.96 & +10:38:22.8 & 0.019 & 59297.3 & 59301.2 & -16.86 & 0.06 & 0.71 \\
ZTF21aaqyuun/SN2021hkf & 11:44:23.27 & +08:10:41.4 & 0.019 & 59300.3 & 59302.3 & -16.38 & 0.07 & 0.00 \\
ZTF21aardvtn/SN2021htp & 07:43:04.76 & +50:17:19.6 & 0.019 & 59119.0 & 59119.5 & -16.42 & 0.19 & 0.00 \\
ZTF21aasksnl/SN2021mju & 16:41:47.59 & +19:21:53.5 & 0.028 & 59292.0 & 59301.4 & -16.08 & 0.21 & 0.42 \\
ZTF21aavhnpk/SN2021jsf & 20:50:21.44 & +01:08:29.5 & 0.028 & 59313.0 & 59317.4 & -17.33 & 0.29 & 0.00 \\
ZTF21aaxtzzj/SN2021kqj & 11:04:58.16 & +30:01:46.8 & 0.029 & 59331.8 & 59334.3 & -17.72 & 0.07 & 1.41 \\
ZTF21aazhegf/SN2021llp & 09:29:31.17 & +25:33:25.0 & 0.033 & 59339.2 & 59340.2 & -17.75 & 0.09 & 0.00 \\
ZTF21abbomrf/SN2020ghq & 14:45:20.59 & +38:44:18.5 & 0.015 & 59347.8 & 59349.2 & -15.72 & 0.03 & 0.00 \\
ZTF21abcacpa/SN2021mtb & 20:03:44.13 & +49:59:34.7 & 0.027 & 59350.9 & 59353.4 & -16.45 & 0.39 & 0.00 \\
ZTF21abcmzvk/SN2021nli & 14:02:12.66 & -18:45:16.1 & 0.030 & 59354.2 & 59356.2 & -17.62 & 0.25 & 0.00 \\

\hline
\end{tabular}
\end{center}
\end{table*}

\begin{table*}
\ContinuedFloat
\caption{Continued.}
\small
\begin{center}
\begin{tabular}{ccccccccc}
\hline
Name & RA & Dec & Redshift & $t_{\rm expl}$ & 1st detection & Peak mag$_{\textrm{r}}$ & $A_{V,MW}$ & $A_{V,\text{host}}$ \\
 & (hh:mm:ss) & (dd:mm:ss) & & (MJD) & (MJD) & (mag) & (mag) & (mag) \\
\hline

ZTF21abcpbqd/SN2014gz & 14:15:50.78 & +01:52:57.4 & 0.026 & 59359.3 & 59359.3 & -17.72 & 0.13 & 0.49 \\
ZTF21abfiuqf/SN2021pla & 16:05:38.19 & +69:35:41.2 & 0.024 & 59375.4 & 59376.3 & -17.00 & 0.08 & 0.21 \\
ZTF21abfjaxa/SN2021pkh & 12:48:41.97 & +26:25:06.7 & 0.023 & 59372.2 & 59373.2 & -17.43 & 0.03 & 2.05 \\
ZTF21abfjdev/SN2021pqj & 11:05:35.47 & +19:41:22.2 & 0.032 & 59366.2 & 59367.2 & -16.99 & 0.07 & 0.00 \\
ZTF21abfoytp/SN2021pnh & 15:50:50.63 & +22:14:16.1 & 0.031 & 59373.8 & 59376.3 & -16.68 & 0.17 & 0.49 \\
ZTF21abgilzj/SN2021qcr & 17:10:21.60 & -03:13:49.7 & 0.029 & 59289.0 & 59295.4 & -18.12 & 1.14 & 0.00 \\
ZTF21abglcxm/SN2021qcs & 15:29:22.82 & -12:14:54.4 & 0.011 & 59377.3 & 59378.2 & -15.65 & 0.43 & 0.42 \\
ZTF21abhhrpj/SN2021qiu & 21:45:50.08 & +15:11:01.3 & 0.029 & 59380.4 & 59381.4 & -17.88 & 0.24 & 0.00 \\
ZTF21abiblpl/SN2021qzi & 20:45:13.87 & -05:37:09.9 & 0.027 & 59389.4 & 59391.3 & -17.30 & 0.17 & 0.00 \\
ZTF21abioeyq/SN2021rhk & 14:03:02.40 & +08:45:56.4 & 0.023 & 59394.7 & 59395.2 & -17.74 & 0.07 & 0.00 \\
ZTF21abjcjmc/SN2021skn & 16:24:49.00 & +39:44:04.7 & 0.030 & 59398.3 & 59399.2 & -18.02 & 0.03 & 0.00 \\
ZTF21abjcliz/SN2021skm & 16:16:56.05 & +21:48:35.8 & 0.031 & 59371.8 & 59372.3 & -18.31 & 0.22 & 1.34 \\
ZTF21abkajar/SN2021svy & 13:09:21.83 & +30:55:20.5 & 0.017 & 59402.7 & 59403.2 & -17.01 & 0.03 & 0.00 \\
ZTF21ablvzhp/SN2021tiq & 22:36:54.72 & -12:33:41.9 & 0.024 & 59409.4 & 59411.4 & -18.12 & 0.17 & 0.00 \\
ZTF21abnlhxs/SN2021tyw & 23:05:56.45 & +14:21:27.8 & 0.013 & 59417.9 & 59419.4 & -17.85 & 0.63 & 0.00 \\
ZTF21abnudtb/SN2021txr & 22:30:50.30 & +36:33:48.2 & 0.026 & 59416.9 & 59418.4 & -18.10 & 0.34 & 0.00 \\
ZTF21abouuat/SN2021ucg & 22:47:37.67 & +39:52:59.4 & 0.017 & 59420.4 & 59422.4 & -17.56 & 0.31 & 0.00 \\
ZTF21abrluay/SN2021vfh & 01:31:38.36 & +31:59:23.7 & 0.025 & 59432.4 & 59434.4 & -17.01 & 0.12 & 0.00 \\
ZTF21abtephz/SN2021wun & 15:46:31.98 & +25:25:44.6 & 0.023 & 59425.7 & 59427.2 & -16.86 & 0.13 & 0.14 \\
ZTF21abvcxel/SN2021wvw & 03:14:47.39 & +40:15:47.7 & 0.010 & 59449.4 & 59449.4 & -16.36 & 0.78 & 0.00 \\
ZTF21abvcxid/SN2021xat & 02:53:03.00 & +42:51:07.0 & 0.032 & 59449.0 & 59451.5 & -17.11 & 0.25 & 0.00 \\
ZTF21acafqtj/SN2021yok & 07:28:55.48 & +20:35:09.9 & 0.015 & 59466.5 & 59469.5 & -17.11 & 0.14 & 0.00 \\
ZTF21accdiqz/SN2021ywg & 02:58:44.41 & +17:15:48.0 & 0.020 & 59469.4 & 59471.4 & -17.49 & 1.08 & 1.56 \\
ZTF21acceboj/SN2021yyg & 05:16:21.03 & -13:28:39.9 & 0.012 & 59471.0 & 59471.5 & -16.51 & 0.40 & 0.00 \\
ZTF21accwcrh/SN2021zco & 03:39:13.31 & +15:59:04.9 & 0.032 & 59472.9 & 59474.4 & -17.50 & 0.66 & 0.00 \\
ZTF21acdcxaf/SN2021zex & 02:14:05.67 & +05:10:35.5 & 0.031 & 59475.4 & 59476.4 & -17.09 & 0.12 & 0.00 \\
ZTF21acdezwk/SN2021zet & 21:52:13.34 & -23:22:37.0 & 0.032 & 59475.2 & 59477.2 & -18.24 & 0.10 & 0.00 \\
ZTF21acdoyqt/SN2021zgm & 18:35:48.34 & +22:27:45.2 & 0.013 & 59479.2 & 59480.1 & -15.51 & 0.44 & 0.00 \\
ZTF21acelnth/SN2021zzi & 01:34:39.15 & +55:25:12.4 & 0.025 & 59484.9 & 59485.3 & -17.31 & 0.87 & 0.00 \\
ZTF21acfajbc/SN2021aalq & 09:50:40.54 & +47:57:52.4 & 0.025 & 59487.0 & 59488.5 & -18.21 & 0.02 & 0.00 \\
ZTF21acgpjbx/SN2021aaqn & 02:37:58.18 & -01:49:53.2 & 0.028 & 59492.8 & 59494.3 & -17.39 & 0.11 & 0.00 \\
ZTF21acgqhru/SN2021aatd & 00:59:04.17 & -00:12:12.0 & 0.015 & 59492.8 & 59494.3 & -16.63 & 0.07 & 0.00 \\
ZTF21acgrrnl/SN2021aayf & 06:22:08.33 & +50:25:44.7 & 0.018 & 59492.9 & 59496.4 & -16.31 & 0.39 & 0.00 \\
ZTF21acgunkr/SN2021aaxs & 08:33:35.18 & +19:44:30.1 & 0.026 & 59490.5 & 59496.5 & -17.77 & 0.09 & 0.00 \\
ZTF21achkqhi/SN2021abpd & 02:28:52.37 & -05:29:14.2 & 0.031 & 59499.9 & 59501.4 & -17.60 & 0.07 & 0.00 \\
ZTF21achpqlr/SN2021abkm & 18:22:37.63 & +15:42:17.7 & 0.018 & 59496.7 & 59502.2 & -16.84 & 0.54 & 0.00 \\
ZTF21aciiaio/SN2021abqs & 11:21:37.61 & +20:09:02.1 & 0.013 & 59503.0 & 59504.5 & -16.37 & 0.07 & 0.35 \\
ZTF21acissla/SN2021achr & 00:22:23.67 & -19:47:37.0 & 0.025 & 59497.3 & 59503.3 & -17.49 & 0.05 & 1.27 \\
ZTF21acjglei/SN2021acma & 02:30:33.57 & +30:52:17.3 & 0.018 & 59514.8 & 59517.3 & -16.61 & 0.25 & 0.07 \\
ZTF21ackrkqq/SN2021addc & 03:53:38.46 & +37:15:47.3 & 0.019 & 59519.3 & 59521.3 & -16.38 & 1.50 & 0.00 \\
ZTF21aclmgzk/SN2021adxd & 00:39:22.90 & +02:48:14.5 & 0.018 & 59522.3 & 59524.2 & -16.92 & 0.05 & 1.06 \\
ZTF21acpqqgu/SN2021aewn & 10:04:06.68 & +31:11:02.7 & 0.021 & 59534.5 & 59536.5 & -16.63 & 0.07 & 0.21 \\
ZTF21acqxomi/SN2021afud & 09:06:41.11 & -10:00:29.0 & 0.025 & 59545.4 & 59550.4 & -16.69 & 0.22 & 0.00 \\
ZTF22aaacxkp/SN2022abq & 13:22:56.82 & +28:19:08.9 & 0.008 & 59599.2 & 59600.4 & -16.55 & 0.05 & 0.00 \\
ZTF22aaahubo/SN2022cru & 08:23:26.30 & -04:55:06.5 & 0.023 & 59575.4 & 59600.3 & -15.89 & 0.12 & 0.00 \\
ZTF22aaaowlo/SN2022ces & 13:54:14.54 & -01:26:34.6 & 0.024 & 59614.7 & 59623.5 & -16.80 & 0.13 & 0.00 \\
ZTF22aaevwec/SN2022gwg & 13:50:25.69 & +68:33:18.1 & 0.031 & 59675.4 & 59676.4 & -17.67 & 0.05 & 0.28 \\

\hline
\end{tabular}
\end{center}
\end{table*}

\begin{table*}
\ContinuedFloat
\caption{Continued.}
\small
\begin{center}
\begin{tabular}{ccccccccc}
\hline
Name & RA & Dec & Redshift & $t_{\rm expl}$ & 1st detection & Peak mag$_{\textrm{r}}$ & $A_{V,MW}$ & $A_{V,\text{host}}$ \\
 & (hh:mm:ss) & (dd:mm:ss) & & (MJD) & (MJD) & (mag) & (mag) & (mag) \\
\hline

ZTF22aafsqud/SN2022hql & 13:48:06.22 & +12:04:29.8 & 0.023 & 59682.8 & 59683.2 & -17.00 & 0.09 & 0.00 \\
ZTF22aagvgwl/SN2022hss & 12:25:38.39 & +07:11:33.0 & 0.025 & 59684.8 & 59687.4 & -17.97 & 0.07 & 0.00 \\
ZTF22aahhgjh/SN2022ihb & 13:46:13.40 & +23:05:10.9 & 0.030 & 59691.8 & 59693.2 & -17.59 & 0.04 & 0.00 \\
ZTF22aahyqkz/SN2022iob & 19:10:37.07 & +37:39:18.7 & 0.028 & 59684.5 & 59689.4 & -17.62 & 0.48 & 0.92 \\
ZTF22aaijrci/SN2022iyl & 20:58:08.06 & +00:27:10.0 & 0.030 & 59694.5 & 59698.4 & -17.04 & 0.21 & 0.14 \\
ZTF22aajipum/SN2022joe & 14:29:20.63 & -22:56:09.7 & 0.026 & 59704.3 & 59707.3 & -16.54 & 0.27 & 0.14 \\
ZTF22aajutqu/SN2022jux & 08:07:22.25 & +40:23:34.8 & 0.026 & 59711.2 & 59712.2 & -17.56 & 0.16 & 0.00 \\
ZTF22aajuufc/SN2022juw & 08:30:31.46 & +18:12:13.7 & 0.027 & 59711.7 & 59712.2 & -16.98 & 0.10 & 0.00 \\
ZTF22aakdbia/SN2022jzc & 12:05:28.66 & +50:31:36.8 & 0.002 & 59714.3 & 59715.2 & -14.91 & 0.05 & 0.57 \\
ZTF22aakdqqg/SN2022kad & 14:58:43.33 & +11:37:50.8 & 0.020 & 59713.4 & 59714.4 & -17.82 & 0.10 & 0.00 \\
ZTF22aanrqje/SN2022mji & 09:42:54.06 & +31:51:03.6 & 0.004 & 59732.7 & 59741.2 & -15.00 & 0.05 & 0.85 \\
ZTF22aaolwsd/SN2022mxv & 23:51:05.12 & +20:09:08.9 & 0.014 & 59747.4 & 59751.4 & -18.10 & 0.20 & 0.00 \\
ZTF22aapargp/SN2022niw & 15:57:16.80 & +19:28:28.1 & 0.033 & 59752.3 & 59753.3 & -17.25 & 0.10 & 0.00 \\
ZTF22aarskhm/SN2022ohx & 20:46:37.43 & -2:21:50.56 & 0.029 & 59760.9 & 59762.4 & -17.07 & 0.18 & 0.28 \\
ZTF22aarycqo/SN2022ojo & 01:44:35.61 & +37:41:50.7 & 0.019 & 59761.4 & 59765.4 & -19.18 & 0.15 & 1.13 \\
ZTF22aaslyzf/SN2022oor & 15:04:29.79 & +02:20:18.7 & 0.032 & 59767.2 & 59768.2 & -17.05 & 0.14 & 0.00 \\
ZTF22aasojye/SN2022omr & 23:41:41.10 & +50:02:58.9 & 0.023 & 59766.9 & 59768.4 & -16.89 & 0.62 & 0.00 \\
ZTF22aativsd/SN2022ovb & 22:22:29.55 & +36:00:17.9 & 0.018 & 59773.4 & 59774.4 & -18.15 & 0.37 & 0.00 \\
ZTF22aatpwfw/SN2022paf & 22:05:26.42 & -00:31:58.7 & 0.031 & 59774.4 & 59775.4 & -16.93 & 0.28 & 0.00 \\
ZTF22aattfmb/SN2022oyp & 18:05:13.16 & +46:52:49.5 & 0.023 & 59775.8 & 59776.3 & -16.87 & 0.11 & 0.28 \\
ZTF22aaudjgc/SN2022pfx & 21:50:50.87 & -00:50:48.9 & 0.027 & 59775.9 & 59778.4 & -18.03 & 0.28 & 0.07 \\
ZTF22aavbfhz/SN2022phi & 1:20:15.03 & +17:49:56.49 & 0.029 & 59779.4 & 59782.4 & -16.60 & 0.21 & 0.00 \\
ZTF22aavobvq/SN2022prv & 15:40:07.76 & +20:40:31.7 & 0.015 & 59781.7 & 59784.3 & -18.26 & 0.17 & 0.00 \\
ZTF22aaxzzoc/SN2022qhc & 17:16:35.13 & +07:19:44.4 & 0.022 & 59789.3 & 59791.2 & -17.28 & 0.48 & 0.42 \\
ZTF22aaywnyg/SN2022pru & 11:59:07.65 & +52:41:58.5 & 0.004 & 59787.7 & 59797.2 & -15.42 & 0.07 & 0.00 \\
ZTF22aazmrpx/SN2022raj & 02:03:17.52 & +29:14:04.9 & 0.012 & 59798.4 & 59800.4 & -15.20 & 0.15 & 0.00 \\
ZTF22abadzpo/SN2022rfz & 17:22:20.39 & +02:00:58.0 & 0.030 & 59799.8 & 59802.2 & -18.65 & 0.49 & 0.00 \\
ZTF22abbecow/SN2022rqg & 16:56:14.62 & +55:01:10.3 & 0.029 & 59801.7 & 59805.2 & -16.20 & 0.07 & 0.00 \\
ZTF22abfavpu/SN2022tmb & 03:20:33.60 & +37:29:54.8 & 0.019 & 59825.0 & 59825.5 & -17.10 & 1.04 & 0.00 \\
ZTF22abfwxtr/SN2022udq & 00:05:55.84 & +22:29:26.3 & 0.022 & 59834.7 & 59839.2 & -16.96 & 0.20 & 0.00 \\
ZTF22abfxkdm/SN2022ubb & 23:08:58.34 & +12:02:39.0 & 0.016 & 59834.3 & 59839.3 & -16.57 & 0.24 & 0.00 \\
ZTF22abgwgsv/SN2022vpm & 17:22:32.88 & +26:46:04.2 & 0.022 & 59842.7 & 59843.2 & -17.80 & 0.12 & 0.00 \\
ZTF22abhsxph/SN2022vyc & 04:33:10.32 & +76:34:05.3 & 0.026 & 59844.9 & 59846.4 & -17.01 & 0.43 & 0.00 \\
ZTF22abitour/SN2022wbr & 04:29:15.65 & +40:14:46.9 & 0.020 & 59838.0 & 59847.5 & -16.34 & 1.67 & 0.00 \\
ZTF22abkbjsb/SN2022vym & 08:54:01.09 & +18:41:18.1 & 0.015 & 59849.0 & 59853.5 & -16.48 & 0.06 & 0.21 \\
ZTF22abkhrkd/SN2022wol & 01:51:27.78 & +36:03:51.5 & 0.018 & 59853.9 & 59854.3 & -16.02 & 0.21 & 0.00 \\
ZTF22ablnrcv/SN2022xav & 09:39:17.64 & +32:18:38.3 & 0.023 & 59857.0 & 59858.5 & -16.96 & 0.05 & 0.42 \\
ZTF22abmsaxp/SN2022xjk & 02:16:32.50 & -11:20:59.4 & 0.013 & 59860.9 & 59861.4 & -17.15 & 0.10 & 0.00 \\
ZTF22abnujbv/SN2022xus & 06:54:05.13 & +08:34:13.3 & 0.009 & 59869.4 & 59871.4 & -16.42 & 0.60 & 0.00 \\
ZTF22abpxxil/SN2022yma & 03:08:48.32 & -07:02:06.5 & 0.029 & 59871.9 & 59873.3 & -17.16 & 0.20 & 0.49 \\
ZTF22abrexqa/SN2022yyz & 19:07:01.55 & +28:59:50.0 & 0.013 & 59880.1 & 59881.1 & -17.52 & 0.64 & 0.00 \\
ZTF22absqhkw/SN2022zkc & 04:47:58.59 & -16:39:37.0 & 0.032 & 59885.4 & 59887.4 & -17.40 & 0.14 & 0.00 \\
ZTF22abssiet/SN2022zmb & 10:38:43.18 & +56:33:14.4 & 0.014 & 59885.5 & 59887.5 & -15.72 & 0.02 & 0.07 \\
ZTF22abtcsyd/SN2022zxt & 08:40:16.38 & +56:02:36.0 & 0.026 & 59891.5 & 59893.4 & -17.25 & 0.10 & 0.00 \\
ZTF22abtspsw/SN2022aagp & 09:10:41.90 & +07:12:20.3 & 0.005 & 59895.4 & 59897.4 & -17.09 & 0.12 & 0.00 \\
ZTF22abulusd/SN2022aatx & 09:15:15.32 & +11:53:04.6 & 0.017 & 59899.4 & 59902.5 & -17.27 & 0.08 & 0.49 \\
ZTF22abvaetz/SN2022aang & 07:59:21.83 & +18:06:40.9 & 0.016 & 59894.5 & 59901.5 & -15.44 & 0.08 & 0.00 \\

\hline
\end{tabular}
\end{center}
\end{table*}

\begin{table*}
\ContinuedFloat
\caption{Continued.}
\small
\begin{center}
\begin{tabular}{ccccccccc}
\hline
Name & RA & Dec & Redshift & $t_{\rm expl}$ & 1st detection & Peak mag$_{\textrm{r}}$ & $A_{V,MW}$ & $A_{V,\text{host}}$ \\
 & (hh:mm:ss) & (dd:mm:ss) & & (MJD) & (MJD) & (mag) & (mag) & (mag) \\
\hline

ZTF22abxomzd/SN2022acbu & 02:30:43.15 & -02:55:56.9 & 0.019 & 59906.8 & 59910.2 & -18.75 & 0.08 & 3.11 \\
ZTF22abyivoq/SN2022acko & 03:19:38.98 & -19:23:42.8 & 0.006 & 59917.8 & 59922.2 & -15.83 & 0.08 & 0.00 \\
ZTF22abyivxh/SN2022acwe & 02:28:07.20 & -05:43:37.0 & 0.030 & 59917.8 & 59922.2 & -17.22 & 0.07 & 0.00 \\
ZTF22abyohff/SN2022acrl & 11:34:21.03 & +15:39:49.2 & 0.017 & 59913.5 & 59923.5 & -16.32 & 0.12 & 0.00 \\
ZTF22abyokkf/SN2022acri & 14:34:19.17 & +25:52:55.4 & 0.022 & 59915.0 & 59923.5 & -17.22 & 0.08 & 0.00 \\
ZTF22abzdzek/SN2022adtt & 01:14:05.28 & +38:07:05.2 & 0.027 & 59930.2 & 59932.1 & -16.80 & 0.13 & 0.00 \\
ZTF22abzqwmp/SN2022adth & 10:15:39.94 & +45:56:25.2 & 0.031 & 59932.8 & 59933.3 & -16.84 & 0.02 & 0.00 \\
ZTF22acahbko/SN2022advr & 10:22:48.99 & +03:45:18.7 & 0.033 & 59934.5 & 59935.4 & -16.61 & 0.10 & 0.00 \\
ZTF23aaaatjn/SN2023cf & 04:26:49.49 & +29:56:59.8 & 0.018 & 59945.8 & 59951.2 & -18.51 & 1.26 & 0.00 \\
ZTF23aaaigqy/SN2023fu & 03:06:21.26 & +36:01:11.1 & 0.016 & 59948.7 & 59957.1 & -17.97 & 0.66 & 0.00 \\
ZTF23aaarmtb/SN2023qh & 09:07:15.44 & +37:12:54.8 & 0.024 & 59947.4 & 59957.4 & -15.89 & 0.06 & 0.00 \\
ZTF23aaavxye/SN2023abq & 15:34:33.34 & +41:08:11.6 & 0.032 & 59962.7 & 59967.5 & -17.63 & 0.07 & 0.00 \\
ZTF23aaazdla/SN2023wn & 13:36:04.57 & -01:35:39.7 & 0.015 & 59962.5 & 59968.5 & -16.79 & 0.09 & 0.00 \\
ZTF23aabngtm/SN2023axu & 06:45:55.32 & -18:13:53.5 & 0.003 & 59970.3 & 59972.3 & -17.27 & 1.06 & 0.00 \\
ZTF23aabtmzm/SN2023blw & 07:28:09.44 & +52:28:17.5 & 0.022 & 59975.7 & 59979.2 & -18.35 & 0.16 & 2.97 \\
ZTF23aacdlsh/SN2023bmd & 15:10:08.83 & +46:06:23.2 & 0.020 & 59982.5 & 59984.5 & -16.69 & 0.07 & 0.00 \\
ZTF23aacjetk/SN2023buy & 08:20:53.69 & +39:14:29.7 & 0.029 & 59988.3 & 59991.2 & -17.87 & 0.12 & 0.00 \\
ZTF23aackdba/SN2023bql & 08:11:27.82 & +08:56:24.7 & 0.019 & 59984.8 & 59985.3 & -16.51 & 0.07 & 0.28 \\
ZTF23aackjhs/SN1995al & 09:50:56.03 & +33:33:11.0 & 0.005 & 59989.8 & 59992.3 & -14.88 & 0.04 & 0.07 \\
ZTF23aaflnok/SN2023fub & 07:40:26.79 & +25:08:00.2 & 0.029 & 60047.2 & 60049.2 & -17.22 & 0.12 & 0.00 \\
ZTF23aafumlg/SN2023fou & 12:40:20.46 & -10:02:55.1 & 0.026 & 60047.8 & 60050.3 & -17.85 & 0.10 & 0.00 \\
ZTF23aagkajy/SN2023gdt & 10:30:10.64 & +43:21:23.4 & 0.014 & 60050.2 & 60051.2 & -15.42 & 0.03 & 0.71 \\
ZTF23aagkutf/SN2023ghl & 11:09:11.73 & +53:21:43.9 & 0.027 & 60052.2 & 60053.2 & -17.24 & 0.03 & 0.00 \\
ZTF23aagqyym/SN2023gjg & 09:07:17.96 & +37:30:12.6 & 0.030 & 60054.2 & 60055.2 & -17.08 & 0.05 & 0.00 \\
ZTF23aahqvtz/SN2023gss & 14:04:23.54 & -27:08:58.1 & 0.021 & 60058.8 & 60059.3 & -17.07 & 0.19 & 0.00 \\
ZTF23aaiecnn/SN2023gxq & 10:18:12.24 & +34:40:19.7 & 0.029 & 60061.2 & 60062.2 & -16.51 & 0.04 & 0.00 \\
ZTF23aailjjs/SN2023hcp & 16:48:42.72 & +35:56:57.4 & 0.031 & 60062.4 & 60063.4 & -17.91 & 0.05 & 0.00 \\
ZTF23aaitpjv/SN2023hlf & 12:26:26.17 & +31:13:32.2 & 0.002 & 60053.3 & 60054.4 & -16.53 & 0.05 & 5.09 \\
ZTF23aajrmfh/SN2023ijd & 12:36:32.47 & +11:13:19.7 & 0.007 & 60078.2 & 60079.0 & -15.39 & 0.09 & 0.00 \\
ZTF23aajsjon/SN2023hzt & 13:30:01.55 & +75:34:09.0 & 0.030 & 60071.9 & 60076.4 & -17.05 & 0.10 & 0.00 \\
ZTF23aakirso/SN2023jid & 22:40:43.23 & +36:38:39.5 & 0.027 & 59994.8 & 60054.5 & -16.12 & 0.42 & 0.00 \\
ZTF23aamfqxc/SN2023jri & 23:30:27.09 & +30:13:11.2 & 0.015 & 60094.2 & 60097.4 & -18.81 & 0.40 & 0.28 \\
ZTF23aamqycj/SN2023jvi & 13:31:22.27 & +25:37:01.0 & 0.025 & 60094.8 & 60097.3 & -17.03 & 0.03 & 0.00 \\
ZTF23aamzlzc/SN2023kne & 17:25:19.11 & +58:49:02.6 & 0.028 & 60095.4 & 60097.3 & -15.81 & 0.10 & 0.00 \\
ZTF23aanymcl/SN2023kzz & 17:17:06.29 & -14:54:05.1 & 0.028 & 60106.8 & 60110.3 & -17.24 & 1.26 & 0.00 \\
ZTF23aanzmoz/SN2023kyi & 19:16:43.95 & -17:30:35.0 & 0.030 & 60105.9 & 60108.4 & -17.00 & 0.33 & 0.00 \\
ZTF23aaomzth/SN2023rpu & 09:23:47.48 & +42:11:12.1 & 0.014 & 60105.2 & 60113.2 & -16.56 & 0.05 & 0.00 \\
ZTF23aaphnyz/SN2023lkw & 16:48:36.21 & +41:36:02.6 & 0.031 & 60117.8 & 60118.2 & -18.07 & 0.06 & 0.00 \\
ZTF23aaqknaw/SN2023lzn & 00:55:07.58 & +31:32:47.6 & 0.018 & 60124.4 & 60128.4 & -17.37 & 0.17 & 0.85 \\
ZTF23aaqtckr/SN2023mpj & 14:34:24.79 & +02:53:04.9 & 0.030 & 60120.2 & 60120.3 & -16.62 & 0.10 & 0.00 \\
ZTF23aaqwpio/SN2023nca & 16:39:26.36 & +11:12:45.3 & 0.023 & 60129.3 & 60135.2 & -15.61 & 0.14 & 0.00 \\
ZTF23aasbvab/SN2023ngy & 22:18:30.18 & +29:14:41.0 & 0.021 & 60139.9 & 60140.3 & -16.80 & 0.22 & 0.00 \\
ZTF23aasrcyv/SN2023nlu & 00:45:09.04 & -09:37:38.3 & 0.020 & 60142.5 & 60143.4 & -16.78 & 0.10 & 0.00 \\
ZTF23aasyvbf/SN2023nmh & 00:37:38.73 & -04:16:53.2 & 0.020 & 60142.4 & 60144.5 & -15.93 & 0.10 & 0.00 \\
ZTF23aaxadel/SN2023pbg & 00:14:54.90 & +26:20:00.4 & 0.025 & 60166.9 & 60168.4 & -16.88 & 0.11 & 0.00 \\
ZTF23aazprcc/SN2023vhb & 12:19:13.19 & +22:25:42.4 & 0.022 & 60180.1 & 60181.1 & -17.20 & 0.07 & 0.21 \\
ZTF23abadrow/SN2023qxp & 21:56:14.96 & +02:10:27.8 & 0.028 & 60175.9 & 60178.4 & -16.65 & 0.15 & 0.00 \\

\hline
\end{tabular}
\end{center}
\end{table*}

\begin{table*}
\ContinuedFloat
\caption{Continued.}
\small
\begin{center}
\begin{tabular}{ccccccccc}
\hline
Name & RA & Dec & Redshift & $t_{\rm expl}$ & 1st detection & Peak mag$_{\textrm{r}}$ & $A_{V,MW}$ & $A_{V,\text{host}}$ \\
 & (hh:mm:ss) & (dd:mm:ss) & & (MJD) & (MJD) & (mag) & (mag) & (mag) \\
\hline

ZTF23abascqa/SN2023rbk & 03:15:20.60 & +41:36:53.5 & 0.020 & 60186.4 & 60187.4 & -17.96 & 0.38 & 0.00 \\
ZTF23abaxtlq/SN2023rix & 02:47:56.84 & +41:14:48.3 & 0.013 & 60191.9 & 60192.4 & -16.37 & 0.23 & 0.00 \\
ZTF23abbsxzs/SN2023rtq & 04:51:46.59 & +38:56:10.0 & 0.013 & 60193.4 & 60195.4 & -17.56 & 2.68 & 0.57 \\
ZTF23abbtkrv/SN2023rvo & 08:49:16.37 & +36:07:14.8 & 0.025 & 60193.0 & 60194.5 & -17.05 & 0.09 & 0.00 \\
ZTF23aberpzw/SN2023swf & 21:05:58.52 & -14:50:30.3 & 0.023 & 60201.7 & 60203.2 & -16.66 & 0.17 & 0.00 \\
ZTF23abhruov/SN2023ucx & 03:59:54.88 & +32:36:41.1 & 0.018 & 60210.4 & 60215.4 & -17.08 & 0.65 & 1.91 \\
ZTF23abhyroo/SN2023udb & 04:25:05.77 & -10:18:51.9 & 0.033 & 60219.0 & 60221.5 & -17.25 & 0.25 & 0.00 \\
ZTF23abhzfww/SN2023twg & 08:45:54.52 & +12:47:12.2 & 0.030 & 60214.5 & 60220.5 & -17.62 & 0.09 & 0.00 \\
ZTF23abiewbt/SN2023ujp & 15:57:33.01 & +20:02:54.2 & 0.033 & 60222.1 & 60223.1 & -17.36 & 0.12 & 0.00 \\
ZTF23abjwgre/SN2023vcg & 23:56:05.87 & +29:22:40.5 & 0.023 & 60231.3 & 60231.3 & -16.70 & 0.17 & 0.00 \\
ZTF23abkhajf/SN2023vcj & 07:55:15.92 & +53:44:47.7 & 0.025 & 60228.5 & 60232.5 & -17.38 & 0.09 & 0.07 \\
ZTF23abmoxlu/SN2023vog & 09:45:09.63 & +68:35:11.8 & 0.015 & 60237.4 & 60238.4 & -17.59 & 0.28 & 0.00 \\
ZTF23abndgbw/SN2023way & 21:24:11.15 & +15:59:22.9 & 0.018 & 60240.7 & 60242.2 & -16.68 & 0.31 & 0.00 \\
ZTF23abnogui/SN2023wcr & 12:23:31.29 & +74:57:01.3 & 0.005 & 60240.5 & 60244.5 & -15.58 & 0.09 & 0.00 \\
ZTF23abonlit/SN2023wuj & 08:26:18.36 & +02:55:28.6 & 0.031 & 60247.0 & 60248.5 & -16.79 & 0.14 & 0.00 \\
ZTF23abphqjk/SN2023xgn & 03:02:48.82 & -15:42:21.6 & 0.031 & 60253.9 & 60254.4 & -17.42 & 0.15 & 0.00 \\
ZTF23abqwald/SN2023xvo & 11:20:24.29 & +28:17:55.4 & 0.033 & 60234.0 & 60234.5 & -16.90 & 0.05 & 0.21 \\
ZTF23absdcgi/SN2023zcu & 06:01:06.82 & -23:40:29.3 & 0.006 & 60288.2 & 60289.3 & -16.73 & 0.09 & 0.00 \\
ZTF23abvommm/SN2023acbr & 02:27:03.18 & -09:25:02.3 & 0.016 & 60300.7 & 60305.2 & -15.57 & 0.08 & 0.00 \\
ZTF24aaabbse/SN2023achj & 08:36:07.46 & +25:06:47.0 & 0.023 & 60303.9 & 60311.3 & -17.79 & 0.10 & 0.00 \\
ZTF24aaarlvj/SN2024V & 11:52:01.50 & +57:41:37.4 & 0.031 & 60300.0 & 60308.5 & -16.49 & 0.04 & 0.00 \\
ZTF24aaasazz/SN2024ov & 11:50:28.79 & -18:34:42.8 & 0.023 & 60291.0 & 60291.6 & -16.03 & 0.10 & 0.00 \\
ZTF24aabppgn/SN2024wp & 11:24:39.25 & +14:56:52.9 & 0.014 & 60320.5 & 60325.5 & -15.52 & 0.10 & 0.00 \\
ZTF24aabpzuz/SN2024vs & 07:51:07.27 & +72:57:57.4 & 0.010 & 60317.9 & 60321.5 & -16.70 & 0.08 & 0.00 \\
ZTF24aabsmvc/SN2024ws & 08:28:46.69 & +73:45:08.6 & 0.012 & 60318.9 & 60322.3 & -16.15 & 0.07 & 0.00 \\
ZTF24aadkwni/SN2024aul & 10:21:53.24 & +00:17:44.3 & 0.021 & 60330.7 & 60335.4 & -17.77 & 0.13 & 0.28 \\
ZTF24aaejehf/SN2024bzq & 11:44:33.94 & +36:26:41.2 & 0.033 & 60346.4 & 60351.4 & -17.97 & 0.05 & 0.00 \\
ZTF24aaejjps/SN2024btx & 11:54:55.45 & +29:20:34.5 & 0.021 & 60346.4 & 60351.4 & -16.04 & 0.06 & 0.00 \\
ZTF24aaejvcx/SN2024atk & 13:18:31.13 & -14:36:39.1 & 0.010 & 60343.2 & 60351.4 & -16.54 & 0.22 & 0.00 \\
ZTF24aaemydm/SN2024chx & 09:54:28.51 & -18:38:10.8 & 0.013 & 60352.8 & 60354.2 & -18.01 & 0.13 & 0.00 \\
ZTF24aafqzur/SN2024daa & 14:25:57.20 & -02:23:32.2 & 0.031 & 60355.4 & 60355.5 & -16.72 & 0.15 & 0.00 \\
ZTF24aagiwoi/SN2024dhi & 11:13:08.67 & +05:04:28.3 & 0.027 & 60360.9 & 60363.4 & -16.65 & 0.20 & 0.00 \\
ZTF24aagupsf/SN2024egd & 16:30:41.48 & +44:30:40.0 & 0.032 & 60372.0 & 60374.5 & -17.47 & 0.04 & 0.00 \\
ZTF24aahalmb/SN2024ees & 12:32:42.19 & +14:32:27.4 & 0.024 & 60376.8 & 60379.4 & -16.52 & 0.11 & 0.00 \\
ZTF24aahewml/SN2024etq & 18:13:41.23 & +10:56:03.8 & 0.022 & 60381.0 & 60387.5 & -16.60 & 0.51 & 0.00 \\
ZTF24aahgmyj/SN2024epy & 09:46:46.54 & +13:31:53.4 & 0.024 & 60384.2 & 60388.2 & -16.77 & 0.12 & 0.64 \\
ZTF24aahmgck/SN2024faf & 11:28:57.38 & +73:02:11.9 & 0.021 & 60385.8 & 60388.3 & -16.85 & 0.13 & 1.27 \\
ZTF24aahwfsa/SN2024fas & 10:51:55.36 & +37:35:23.9 & 0.026 & 60393.4 & 60396.4 & -15.80 & 0.04 & 0.00 \\
ZTF24aajxppf/SN2024grw & 17:58:21.63 & +09:40:53.5 & 0.021 & 60415.5 & 60416.5 & -18.01 & 0.48 & 0.00 \\
ZTF24aakzive/SN2024hpg & 21:31:43.20 & +0:21:43.71 & 0.029 & 60427.7 & 60430.4 & -17.53 & 0.14 & 0.00 \\
ZTF24aalceob/SN2024hme & 16:05:57.74 & +27:11:21.00 & 0.031 & 60418.9 & 60423.4 & -16.09 & 0.10 & 0.00 \\
ZTF24aamzqsv/SN2024izq & 9:24:57.24 & +40:23:58.49 & 0.028 & 60440.2 & 60441.2 & -16.29 & 0.04 & 0.00 \\
ZTF24aaozxhx/SN2024jlf & 14:37:42.32 & +2:17:04.17 & 0.006 & 60457.3 & 60458.2 & -16.99 & 0.11 & 0.00 \\
ZTF24aaplfjd/SN2024jxm & 0:58:01.36 & +30:42:23.84 & 0.016 & 60460.0 & 60460.5 & -16.00 & 0.18 & 0.00 \\
ZTF24aarajmv/SN2024ldu & 19:54:05.17 & +49:56:47.25 & 0.025 & 60466.4 & 60469.4 & -16.26 & 0.50 & 0.00 \\
ZTF24aarvbxj/SN2024lby & 20:22:40.78 & -8:10:41.95 & 0.020 & 60472.4 & 60473.4 & -17.56 & 0.17 & 0.00 \\
ZTF24aasktmr/SN2024lss & 16:31:22.18 & +22:42:08.73 & 0.024 & 60479.8 & 60480.2 & -17.19 & 0.13 & 0.00 \\

\hline
\end{tabular}
\end{center}
\end{table*}

\begin{table*}
\ContinuedFloat
\caption{Continued.}
\small
\begin{center}
\begin{tabular}{ccccccccc}
\hline
Name & RA & Dec & Redshift & $t_{\rm expl}$ & 1st detection & Peak mag$_{\textrm{r}}$ & $A_{V,MW}$ & $A_{V,\text{host}}$ \\
 & (hh:mm:ss) & (dd:mm:ss) & & (MJD) & (MJD) & (mag) & (mag) & (mag) \\
\hline

ZTF24aatifzm/SN2024mxq & 14:44:05.84 & +9:16:47.52 & 0.031 & 60480.8 & 60484.2 & -16.83 & 0.08 & 0.00 \\

\hline
\end{tabular}
\end{center}
\end{table*}